\documentclass[aps,prfluids,preprint,linenumbers,amsfonts,amssymb,amsmath]{revtex4-2}

\usepackage[table,xcdraw]{xcolor}
\usepackage{newtxtext}
\usepackage{newtxmath}
\usepackage{makecell}
\usepackage{soul} 

\usepackage{graphicx}
\usepackage{subcaption}
\usepackage{booktabs}
\usepackage{multirow}
\usepackage{threeparttable}
\usepackage{placeins}
\usepackage{tabularx}
\usepackage{hhline}
\usepackage{enumitem}

\usepackage{natbib}
\usepackage{url}
\usepackage{hyperref}
\hypersetup{
    colorlinks = true,
    urlcolor   = blue,
    citecolor  = blue, 
    linkcolor = blue
}
\usepackage[capitalise]{cleveref} 
\crefname{figure}{Figure}{Figures}
\Crefname{figure}{Figure}{Figures}
\crefname{table}{Table}{Tables}
\Crefname{table}{Table}{Tables}

\begin{document}

\preprint{FC10303/Hunt}

\title{Experimental validation of a linear momentum and bluff-body model for high-blockage cross-flow turbine arrays}

\author{Aidan Hunt}
\email{ahunt94@uw.edu}
\affiliation{Department of Mechanical Engineering\\
             University of Washington\\
             Seattle, Washington, USA}%

\author{Ari Athair}
\affiliation{Department of Aeronautics and Astronautics\\
             University of Washington\\
             Seattle, Washington, USA}%
             
\author{Owen Williams}
\affiliation{Department of Aeronautics and Astronautics\\
             University of Washington\\
             Seattle, Washington, USA}%

\author{Brian Polagye}%
\affiliation{Department of Mechanical Engineering\\
             University of Washington\\
             Seattle, Washington, USA}%

\date{\today}

\begin{abstract}
    In a confined flow, the performance of a turbine and its near-wake fluid dynamics depend on the blockage ratio, defined as the ratio of the turbine projected area to the channel cross-sectional area.
    While blockage is understood to increase the power coefficient for turbine ``fences'' spanning a channel, most investigations at the upper range of practically-achievable blockage ratios have been theoretical or numerical in nature.
    Furthermore, while linear momentum actuator disk theory is frequently used to model turbines in confined flows, as confinement increases, the ability of this idealized model to describe performance and flow fields has not been established.
    In this work, the performance and near-wake flow field of a pair of cross-flow turbines are experimentally evaluated at blockage ratios from $30\%$ to $55\%$.
    The fluid velocity measured in the bypass region is found to be well-predicted by the open-channel linear momentum model developed by \citet{houlsby_application_2008}, while the wake velocity is not.
    Additionally, self-similar power and thrust coefficients are identified across this range of blockage ratios when array performance is scaled by the modeled bypass velocity following Whelan \textit{et al.}'s \citep{whelan_freesurface_2009} adaptation of the bluff-body theory of \citet{maskell_ec_theory_1963}.
    This result demonstrates that, despite multiple non-idealities, relatively simple models can quantitatively describe highly confined turbines.
    From this, an analytical method for predicting array performance as a function of blockage is presented. 
    Overall, this work illustrates turbine performance at relatively high confinement and demonstrates the suitability of analytical models for predicting and interpreting their hydrodynamics. 
\end{abstract}

\nolinenumbers
\maketitle


\section{Introduction}
\label{sec:intro}

Flowing water in river and tidal channels is a promising source of renewable power.
The efficiency of individual turbines or arrays deployed in these channels is influenced by how much of the channel cross-section they occupy. For a row of turbines, this is represented by the blockage ratio, defined as the ratio between the turbine projected area and the channel cross-sectional area: 
\begin{equation}
    \beta = \frac{A_{\mathrm{turbines}}}{A_{\mathrm{channel}}} .
\end{equation}
As the blockage ratio increases, the turbines present greater flow resistance, but confinement from the channel boundaries leads to increased flow through the turbines relative to unconfined conditions.
Consequently, as confinement increases, so does turbine efficiency \citep{garrett_efficiency_2007, whelan_freesurface_2009, nishino_efficiency_2012, vogel_effect_2016}.
While blockage augments the performance of all turbine designs, cross-flow turbines---which have an axis of rotation perpendicular to the direction of inflow---are particularly well-suited to exploit these physics since their rectangular projected area matches the profile of natural and constructed channels.

\citet{garrett_efficiency_2007} were the first to establish the theoretical basis for turbine performance in confined flows by applying one-dimensional linear momentum actuator disk theory (LMADT).
For relatively low Froude numbers, they showed that the maximum efficiency in confined flow was augmented by a factor of $(1-\beta)^{-2}$.
Subsequent studies by \citet{houlsby_application_2008} and \citet{whelan_freesurface_2009} relaxed the assumption of low Froude number by allowing the free surface to deform in response to momentum extraction. In parallel, \citet{nishino_efficiency_2012, nishino_twoscale_2013} expanded the conceptual framework with a two-scale LMADT model for turbine arrays that considered confinement effects from both the channel boundaries (``global'' blockage) and the presence of neighboring turbines within the array (``local'' blockage).
\citet{vogel_effect_2016} later combined these approaches into a single model that incorporated both free surface deformation and inter-turbine proximity effects, and \citet{dehtyriov_fractallike_2021} extended the two-scale model of \citet{nishino_efficiency_2012} to an arbitrary number of scales (i.e., arrays of turbine sub-arrays) using a fractal-like approach.

Blockage effects have been the focus of multiple experimental and numerical studies (\cref{tab:priorWork}).
Across these, turbine dynamics are generally consistent and aligned with theory: for all types of turbines, as blockage increases, so do the rotor thrust loading, maximum efficiency, and optimal operating speed (i.e., tip-speed ratio or reduced frequency).
Studies that examine the flow field near confined turbines \citep{battisti_aerodynamic_2011, nishino_effects_2012, mctavish_experimental_2014, dossena_experimental_2015, schluntz_effect_2015, sarlak_assessment_2016, ross_experimental_2020a} find that an increase in blockage accelerates the flow through and around the turbine as predicted by theory, as well as narrows the turbine wake.
\citet{gauthier_impact_2016} and \citet{kinsey_impact_2017} examined how blockage effects change for various turbine types in asymmetric confinement (i.e., different combinations of lateral blockage and vertical blockage for a given $\beta$).
The authors parameterized confinement asymmetry as the ratio between lateral blockage and vertical blockage (or its reciprocal, if a larger value), and found turbine performance to be invariant for values of confinement asymmetry less than three.
The interplay between rotor geometry and blockage effects has also been explored for both cross-flow turbines \citep{goude_simulations_2014, kong_correction_2025, hunt_performance_2025} and axial-flow turbines \citep{schluntz_effect_2015, abutunis_comprehensive_2022}. These studies found that higher-solidity rotors benefit more from blockage -- a consequence of how their higher thrust coefficients couple with channel-level energetics.

Several groups have investigated analytical blockage corrections---which adjust observed turbine performance in confined flow to predict the performance of the same turbine in unconfined flow---and evaluated those corrections through experiments \citep{whelan_freesurface_2009, battisti_aerodynamic_2011, chen_blockage_2011, ross_wind_2011, dossena_experimental_2015, jeong_blockage_2018, ross_experimental_2020,steiros_analytical_2022} or simulations \citep{gauthier_impact_2016, kinsey_impact_2017, zilic_de_arcos_numerical_2020,steiros_analytical_2022,zhang_analysis_2023}.
Such methods generally proceed from \citet{glauert_airplane_1935} (who originally proposed a blockage correction for airplane propellers) by assuming that there exists some freestream velocity in unconfined flow that yields the same thrust and velocity through the rotor as in the confined condition.
This unconfined freestream velocity is found through an analytical model, often 
LMADT \citep[e.g.,][]{barnsley_final_1990, whelan_freesurface_2009, ross_experimental_2020, dehtyriov_twoscale_2025}.
However, blockage corrections based on LMADT can yield non-physical results for ``highly-loaded'' turbines since sufficiently high thrust coefficients necessitate reversed flow in the turbine wake \citep{steiros_drag_2018, liew_unified_2024a}.
This limitation has prompted investigation into blockage corrections that are applicable to a broader range of turbine solidities and operating conditions.
For example, \citet{whelan_freesurface_2009} proposed a LMADT-based correction inspired by the bluff-body theory of \citet{maskell_ec_theory_1963}, which, unlike Glauert's theory, assumes that the thrust on a turbine depends on the velocity of the accelerated flow that bypasses it (i.e., the rotor thrust and bypass velocity are unchanged between corresponding confined and unconfined conditions).
\citeauthor{whelan_freesurface_2009} found that such a correction more accurately predicted unconfined thrust coefficients for highly-loaded turbines.
Recently, \citet{steiros_analytical_2022} developed an alternative blockage correction for high-solidity turbines that proceeds from Glauert's theory, but replaces LMADT with a two-dimensional potential flow model following \citeauthor{taylor_air_1944}'s \citep{taylor_air_1944} theory for drag on a porous plate, with adjustments for mass conservation \citep{koo_fluid_1973} and wake pressure \citep{steiros_drag_2018}.
\citet{steiros_analytical_2022} demonstrated that this approach overcame the limitations of LMADT for higher thrust coefficients, while maintaining good agreement with the more conventional method at lower confinements and solidities.
However, when validating their model with drag measurements on porous plates, \citet{steiros_analytical_2022} observed increasing disagreement between their theory and experiments with increasing confinement and solidity, which leaves open the question of whether a bluff-body correction could be more accurate in this regime.

\begin{table*}
    \begin{threeparttable}
        \centering
        \resizebox{\textwidth}{!}{      
        \begin{tabular}{@{}lll|lll@{}}
            \toprule
            \multicolumn{3}{c|}{\textbf{Experimental}} & \multicolumn{3}{c}{\textbf{Numerical}} \\ \midrule
            Author & Turbine Type & $\beta$ tested [\%] & Author & Turbine Type & $\beta$ tested [\%] \\ \midrule
            \textbf{\citet{takamatsu_study_1985}} & \textbf{CFT} & \textbf{75, 95}\tnote{\dag} & \textbf{\citet{nishino_effects_2012}} & \textbf{Actuator disk} & 3.1 - \textbf{50.3} \\
            \textbf{\citet{whelan_freesurface_2009}} & \textbf{AFT} & 5, \textbf{64} & \textbf{\citet{consul_blockage_2013}} & \textbf{CFT} & 12.5, 25, \textbf{50} \\
            \textbf{\citet{mcadam_experimental_2013,   mcadam_experimental_2013a}} & \textbf{CFT} & \textbf{47, 59} & \textbf{\citet{goude_simulations_2014}} & \textbf{CFT} & 0, 12.5, 25, \textbf{50} \\
            \citet{battisti_aerodynamic_2011} & CFT & 2.8\tnote{\dag}, 10 & \textbf{\citet{kolekar_performance_2015}} & \textbf{AFT} & 4.2, 8.2, 11, 17, 20, \textbf{33, 42} \\
            \citet{chen_blockage_2011} & AFT & 10.2, 20.2, 28.3 & \textbf{\citet{schluntz_effect_2015}} & \textbf{AFT} & 0, 7.9, 19.6, \textbf{31.4} \\
            \citet{ross_wind_2011} & CFT (Savonius) & 2, 3.5, 8 & \textbf{\citet{gauthier_impact_2016}} & \textbf{Oscillating foil} & 0.2 - \textbf{60.5} \\
            \textbf{\citet{birjandi_power_2013}} & \textbf{CFT} & 24.5 - \textbf{49.2}\tnote{\dag} & \citet{sarlak_assessment_2016} & AFT & 2, 5, 9, 20 \\
            \citet{mctavish_experimental_2014} & AFT & 6.3, 9.9, 14.3, 19.4, 25.4 & \textbf{\citet{kinsey_impact_2017}} & \textbf{AFT} & 0, 10, 20, \textbf{50, 60} \\
            \citet{gaurier_tidal_2015} & AFT & 1.2, 3.3, 4.8 &  & \textbf{CFT} & 0, 13, 26, \textbf{51} \\
            \textbf{\citet{ryi_blockage_2015}} & \textbf{AFT} & 8.1, 18.0, \textbf{48.1} & \citet{badshah_cfd_2019} & AFT & 2 - 19 \\
            \citet{dossena_experimental_2015} & CFT & 2.8\tnote{\dag}, 10 & \textbf{\citet{gauvin-tremblay_twoway_2020}} & \textbf{CFT} & 5, 20, \textbf{40} \\
            \citet{jeong_blockage_2018} & CFT & 3.5, 13.4, 24.7 & \textbf{\citet{zilic_de_arcos_numerical_2020}} & \textbf{AFT} & 1, 5, 10, 20, \textbf{40} \\
            \textbf{\citet{ross_experimental_2020}} & \textbf{AFT} & 2, \textbf{35} & \textbf{\citet{abutunis_comprehensive_2022}} & \textbf{AFT} & 4.2, 20.8, \textbf{41.6, 62.5} \\
            \textbf{} & \textbf{CFT} & 3, \textbf{36} & \citet{steiros_analytical_2022} & \textbf{AFT} & 3.1, 19.6 \\
            \textbf{\citet{ross_experimental_2020a}} & \textbf{CFT} & 14, \textbf{36} & \citet{zhang_analysis_2023} & AFT & 2, 2.8, 4.4, 7.9, 17.7 \\
            \textbf{\citet{hunt_performance_2025}} & \textbf{CFT} & \textbf{35, 45, 55} &  & CFT (Savonius) & 2.5, 3.6, 5.6, 10, 22.5 \\
             &  &  & \textbf{\citet{kong_correction_2025}} & \textbf{CFT} & 8.3, 16, 25, \textbf{30.8, 36.4, 44.4, 57.1} \\ \bottomrule
        \end{tabular}
        }
        \footnotesize{
        \begin{tablenotes}
             \item[\dag] Value(s) estimated from provided turbine and channel dimensions.
        \end{tablenotes}
        }
    \end{threeparttable}
    \caption{Summary of experimental and numerical studies that consider the effects of varying the blockage ratio for different types of turbines (axial-flow turbines are indicated by ``AFT'' and cross-flow turbines are indicated by ``CFT''). Studies that consider $\beta \geq 30\%$ are bolded.}
    \label{tab:priorWork}
\end{table*}

Although the studies in \cref{tab:priorWork} may appear to comprehensively explore the effects of confinement on turbine performance, the upper end of blockage ratios that are achievable in a realistic river or tidal channel ($\beta = 30\% - 60\%$) have been primarily explored through numerical simulation.
While the general trends in turbine performance and wake characteristics observed in these simulations are in agreement with theory, the computational methods employed are often validated only at lower blockage \citep{consul_blockage_2013, kolekar_performance_2015, gauthier_impact_2016, kinsey_impact_2017}---or not at all---before investigating higher blockages.
This is likely due to a lack of equivalent experimental data at high blockage ratios; only a handful of the experimental studies have considered blockage ratios greater than $30\%$.
Experimental studies that do explore the upper-end of blockage ratios consider only a sparse set of $\beta$, likely due to difficulty of maintaining dynamic similarity across a wider range of blockage ratios \citep{hunt_experimental_2023}.
Furthermore, meaningful cross-comparisons between experimental studies are complicated by differences in rotor geometry, and several studies simultaneously vary blockage along with Reynolds and Froude numbers \citep[e.g.,][]{mcadam_experimental_2013, mcadam_experimental_2013a, birjandi_power_2013}.
Consequently, although the general effects of blockage on turbine performance are well-known, there remains uncertainty as to how these effects evolve at higher confinement. 
Additionally, despite the widespread application of analytical blockage corrections in prior work, their suitability for higher confinements has not been fully established, nor has their ability to describe the flow field around confined turbines been thoroughly explored.

 \begin{figure*}
    \centering
    \includegraphics[width=\textwidth]{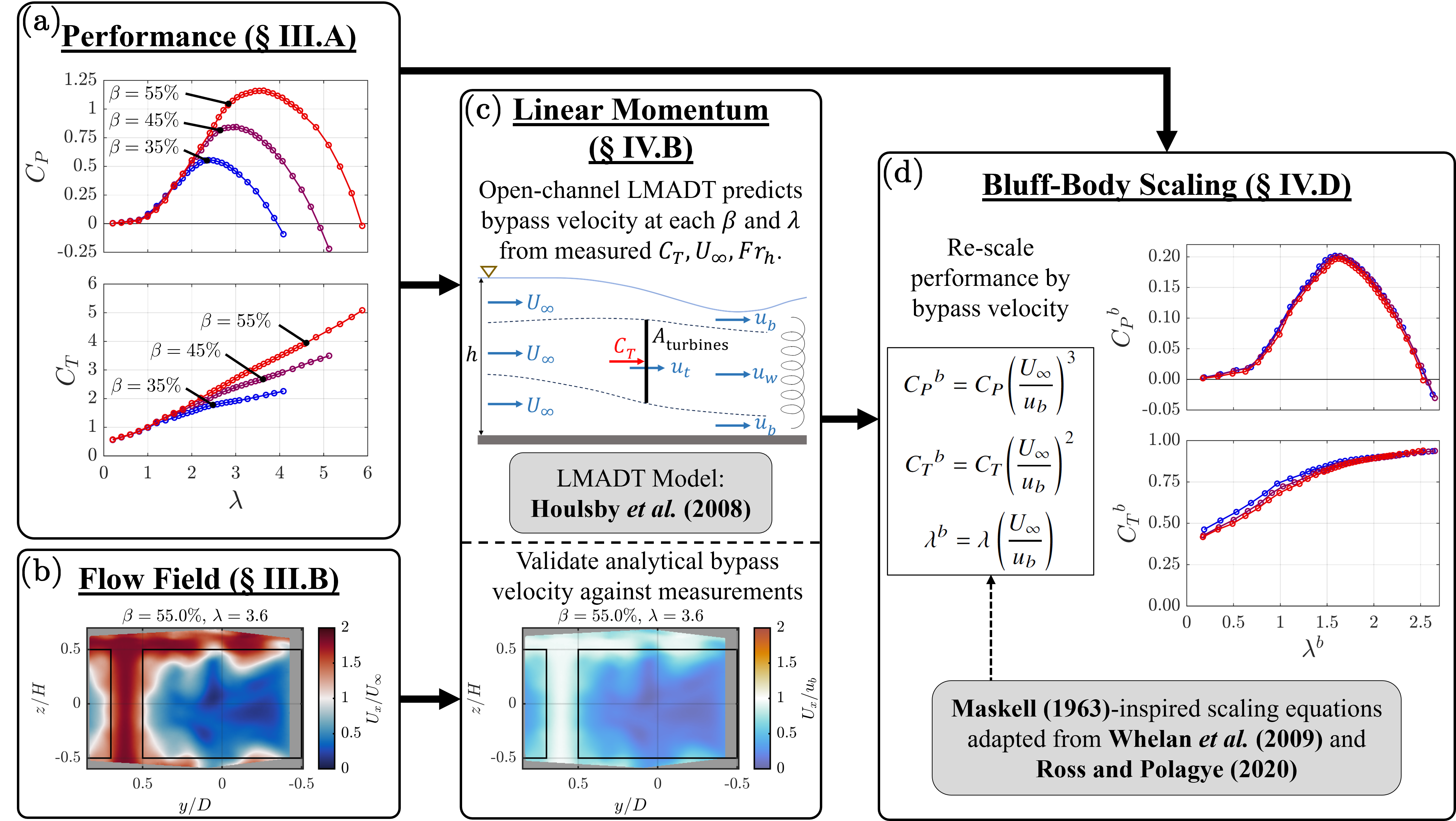}

    {\phantomsubcaption\label{fig:procedure_perf}
     \phantomsubcaption\label{fig:procedure_piv}
     \phantomsubcaption\label{fig:procedure_lmad}
     \phantomsubcaption\label{fig:procedure_scaling}
     }
     \caption{Outline of the procedure in \Cref{sec:results,sec:analyticalModels} to interpret high blockage cross-flow turbine dynamics using \subref{fig:procedure_perf} experimental performance data, \subref{fig:procedure_piv} flow fields, and \subref{fig:procedure_lmad} linear momentum theory, resulting in \subref{fig:procedure_scaling} self-similar array performance across blockage ratios.
     Key methodologies that are adapted from prior work are highlighted in gray boxes.}
     \label{fig:procedureSummary}
 \end{figure*}

The objective of this work is to develop a more comprehensive understanding of performance and flow fields around a cross-flow turbine array at higher blockages than previously considered in most experiments.
Such information can be used to understand the relevance of theory to non-ideal arrays that arguably violate multiple assumptions underpinning analytical models (e.g., treating the turbines as porous, non-rotating disks in LMADT).
To this end, a symmetric two-turbine array is tested in a recirculating water channel at blockage ratios ranging from $30\%$ to $55\%$, and flow fields in the near wake are characterized using particle image velocimetry (PIV).
The Reynolds and Froude numbers are controlled to avoid convolving changes in blockage with other relevant non-dimensional parameters \citep{ross_effects_2022}.
\Cref{sec:methods} describes the experimental methods used to characterize array performance and flow fields, with resulting performance, near-wake velocities, and free surface deformation discussed in \Cref{sec:results}.
In \Cref{sec:analyticalModels}, the open-channel LMADT model developed by \citet{houlsby_application_2008} is applied to the experimental performance data to model the wake velocity, bypass velocity, and free surface deformation in the vicinity of the array.
In addition, comparisons are made between this approach and the recent potential flow model proposed by \citet{steiros_analytical_2022}.
The analytically-modeled velocities are compared to measurements and are used to interpret high-blockage turbine hydrodynamics following \citeauthor{whelan_freesurface_2009}'s \citep{whelan_freesurface_2009} adaptation of the bluff-body theory of \citet{maskell_ec_theory_1963}. 
The procedure followed in \Cref{sec:results,sec:analyticalModels} is graphically outlined in \cref{fig:procedureSummary} and methodologies that are adapted from prior work are highlighted.
Based on the characteristic dynamics observed, an analytical blockage adjustment that leverages LMADT and \citeauthor{maskell_ec_theory_1963}'s \citep{maskell_ec_theory_1963} bluff-body theory to predict turbine performance across blockage ratios (i.e., the reverse of the procedure in \cref{fig:procedureSummary}) is presented in \Cref{sec:lmadForecasting}.

\section{Experimental Methods}
\label{sec:methods}

\subsection{Cross-Flow Turbine Test Setup}
\label{methods:testSetUp}

Experiments are conducted in the Alice C. Tyler recirculating water flume at the University of Washington. The flume has a test section that is 4.88 m long with a width ($w$) of 0.76 m, and can accommodate water depths up to 0.60 m and flow speeds up to ${\sim}\!1.1$ m/s. The flume is equipped with both a heater and chiller for temperature control, and can maintain water temperatures between $10$ $^{\circ} \mathrm{C}$ and $40$ $^{\circ}\mathrm{C}$.

The laboratory-scale array consists of two identical straight-bladed cross-flow turbines.
Experiments with a pair of turbines, rather than a single turbine, are motivated by the research application to confinement benefits for turbine arrays.
The rotors are two-bladed, with each blade consisting of a NACA 0018 profile with a 0.0742 m chord length ($c$) mounted at a $-6^{\circ}$ (i.e., toe-out) preset pitch angle as referenced from the quarter chord. The blade span ($H$) is 0.215 m and the turbine diameter ($D$) is 0.315 m, measured at the outermost point swept by the blades.
The blades are attached to the central driveshaft of each rotor using thin, hydrodynamic blade-end struts (NACA 0008 profile, 0.0742 m chord length).
This arrangement results in an aspect ratio ($H/D$) of 0.68, chord-to-radius ratio ($c/R$) of 0.47, and solidity ($Nc / \pi D$, where $N$ is the blade count) of 0.15.

\begin{figure*}
    \begin{minipage}{0.475\textwidth}
        \centering
        \includegraphics[width=\textwidth]{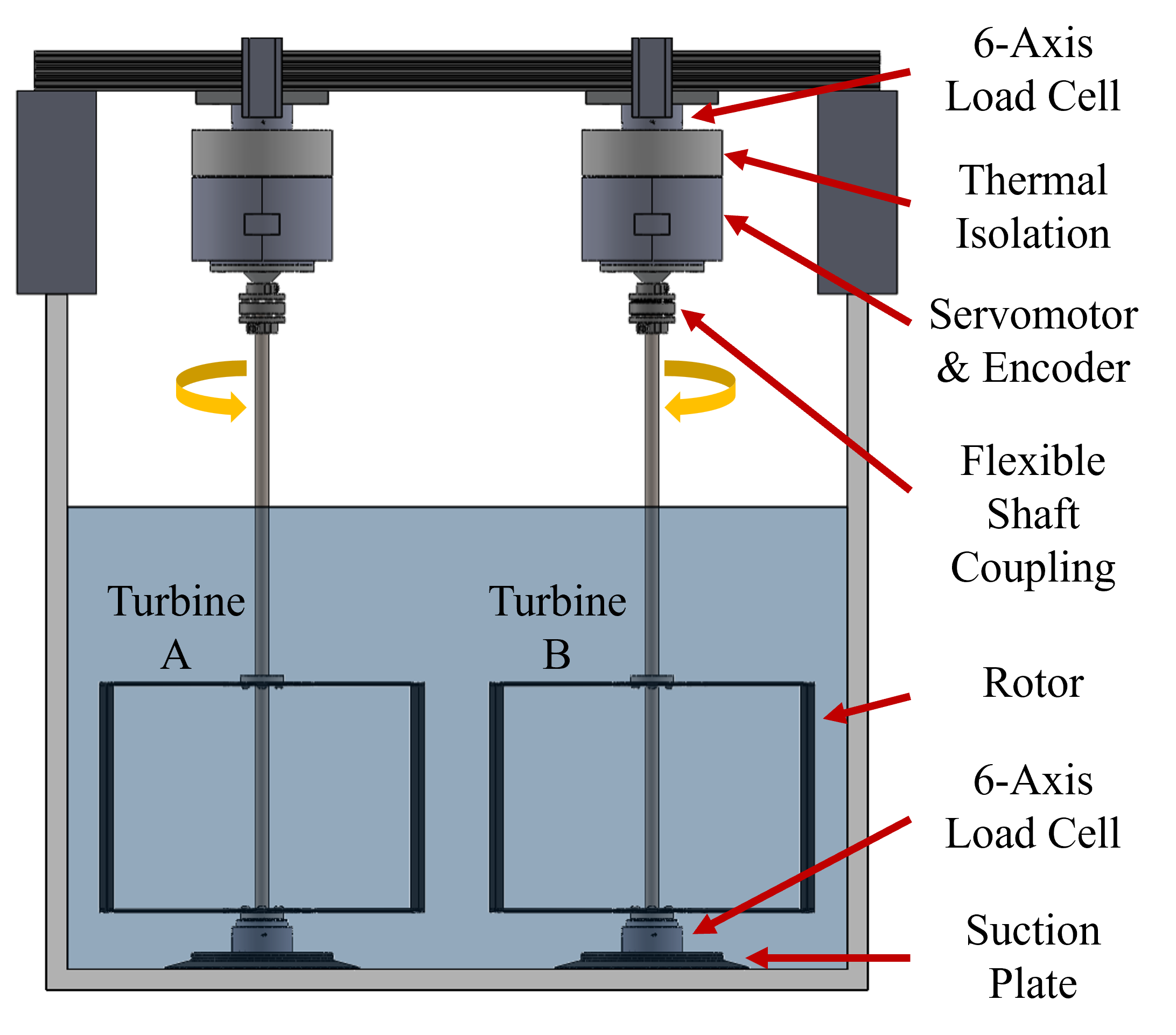}
        \caption{A rendering of the experimental test-rig, as viewed from upstream. The configuration shown corresponds to a blockage ratio of $40.1\%$.}
        \label{fig:testRig}
    \end{minipage} 
    \hfill
    \centering
    \begin{minipage}{0.475\textwidth}
        \centering
        \includegraphics[width=\textwidth]{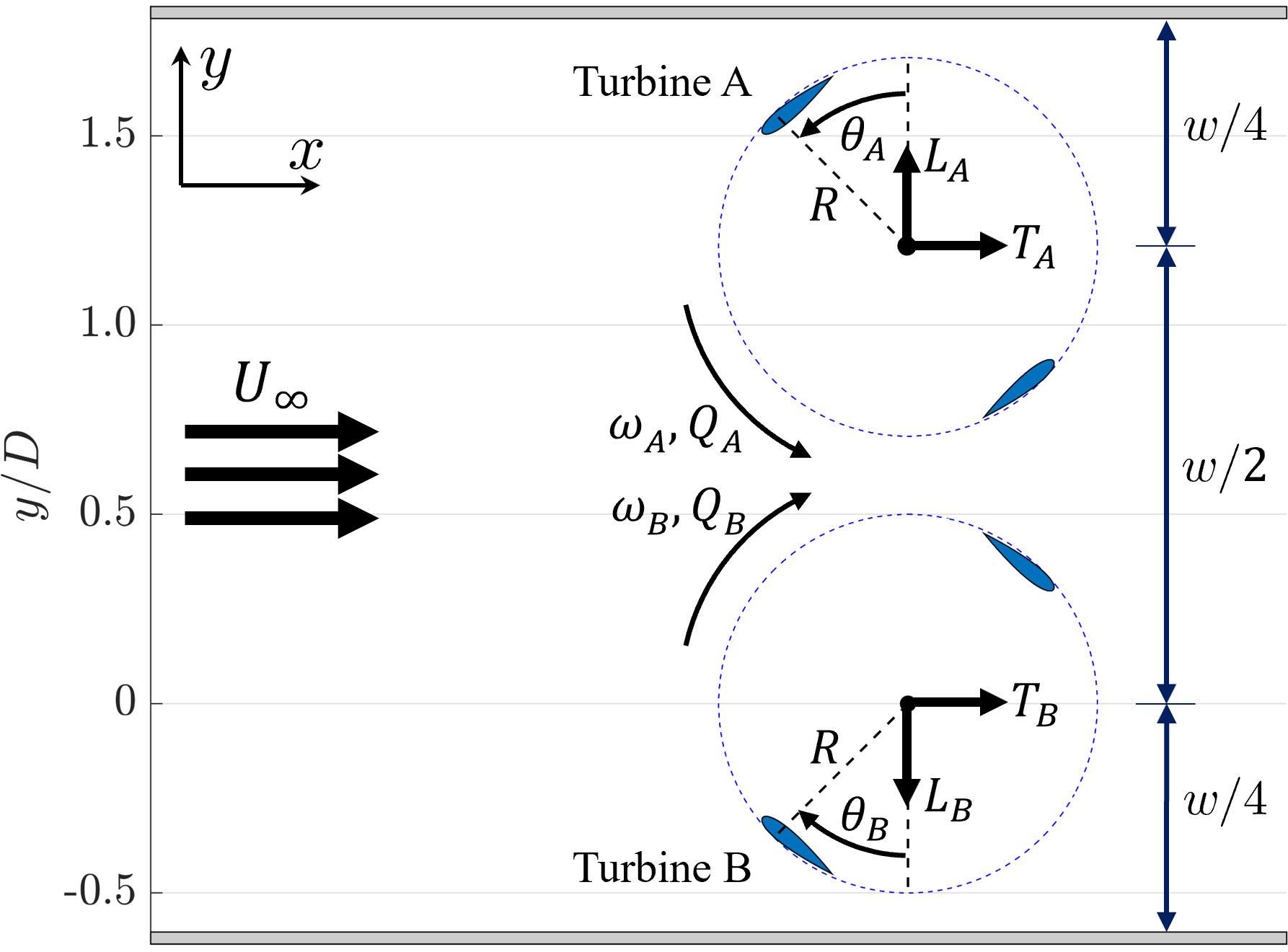}
        \caption{Overhead view of the array layout in the Tyler flume, with key measured quantities and the array layout annotated. The outermost circle swept by each turbine (with diameter $D$) is shown as a dashed blue circle.}
        \label{fig:arrayOverhead}
    \end{minipage} 
\end{figure*}

The two rotors, designated as ``Turbine A'' and ``Turbine B'', are integrated into the experimental setup shown in \cref{fig:testRig}, which consists of two identical test-rigs.
The top of each turbine's central shaft is connected by a flexible shaft coupling (Zero-Max SC040R) to a servomotor (Yaskawa SGMCS-05BC341) which regulates the rotation rate of the turbine.
This control strategy yields similar time-average performance \citep{polagye_comparison_2019} and wake characteristics \citep{araya_comparison_2015} to applying a constant oppositional torque and reduces cycle-to-cycle variability.
The angular position of each turbine is measured via the servomotor encoder ($2^{16}$ counts per revolution), from which the angular velocity is estimated. 
The bottom of each turbine's central shaft sits in a bearing.
The net forces and torques on each turbine are measured by a pair of 6-axis load cells: an upper load cell (ATI Mini45-IP65) connected to the servomotor through a thermal isolation assembly and fixed to a rigid crossbeam, and a lower load cell (ATI Mini45-IP68) mounted to the bottom bearing and fixed to the bottom of the flume via a suction plate.
Measurements from the load cells and servomotor encoders for both turbines are acquired synchronously at 1000 Hz in MATLAB using two National Instruments PCIe-6353 DAQs.

The freestream velocity, $U_{\infty}$, is measured using an acoustic Doppler velocimeter (Nortek Vectrino Profiler) sampling at 16 Hz.
The velocimeter samples a single cell positioned laterally in the center of the flume, vertically at the turbine midspan, and $5D$ upstream of the turbines' axes of rotation.
Velocity measurements are despiked using the method of \citet{goring_despiking_2002}.
The turbulence intensity of the freestream flow measured by the ADV is ${\sim}2\%$ for all experiments.
The water depth far upstream of the array ($5.8D$ upstream of the turbines' axes of rotation, centered laterally in the channel) and depths in vicinity of the array ($0.5D$, $1.0D$, and $2.0D$ directly upstream and downstream of Turbine B) are measured using seven synchronized ultrasonic free surface transducers (Omega LVU32) sampling at 0.5 Hz.
As the array is symmetric about the channel centerline (\cref{fig:arrayOverhead}) the streamwise free surface profile in the vicinity of Turbine B is assumed to be representative of that in the vicinity of Turbine A.
The water temperature is measured using a temperature probe (Omega Ultra-Precise RTD) and maintained within $\pm 0.1^{\circ} \mathrm{C}$ of the target value during each experiment.

\subsection{Non-Dimensional Parameters}
\label{methods:ndParams}

For the two-turbine array, the blockage ratio is defined as:
\begin{equation}
	\beta = \frac{A_{\mathrm{turbines}}}{A_{\mathrm{channel}}} = \frac{2HD}{hw} ,
	\label{eq:arrayBlockage}
\end{equation}
\noindent where $h$ is the dynamic water depth and $w$ is the channel width.
As subsequently discussed, because of the array layout, global and local blockage for this array are identical and can be described by a single value.
The blockage ratio is varied by holding the array geometry constant and changing the water depth in the flume.
To isolate blockage effects, other non-dimensional flow parameters must be controlled as the depth changes. 
For example, decreasing water depth (increasing $\beta$) increases the depth-based Froude number,
\begin{equation}
    Fr_h = \frac{U_{\infty}}{\sqrt{gh}} \ \ , 
    \label{eq:depthFroude}
\end{equation}
\noindent where $g$ is the acceleration due to gravity.
The array's proximity to the free surface also changes as depth decreases, represented here by the normalized submergence, $s/h$, where $s$ is the distance between the free surface and the top of the turbine blades.
Both $Fr_h$ \citep{consul_blockage_2013, hunt_effect_2020, ross_effects_2022} and $s/h$ \citep{birjandi_power_2013, kolekar_performance_2015, kolekar_blockage_2019, ross_effects_2022} have been shown to influence turbine performance.
Therefore, as $h$ decreases, $U_{\infty}$ and $s$ must also decrease to hold $Fr_h$ and $s/h$, respectively, constant.

However, a decrease in $U_{\infty}$ also decreases the Reynolds number, which is defined here with respect to the freestream velocity and turbine diameter as
\begin{equation}
    Re_D =  \frac{ U_{\infty} D}{\nu} \ \ ,
    \label{eq:ReD}
\end{equation}
\noindent where $\nu$ is the kinematic viscosity.
The dependence of turbine performance on the Reynolds number is well documented \citep{miller_verticalaxis_2018, miller_solidity_2021, bachant_effects_2016, ross_effects_2022, hunt_experimental_2024}.
Although turbine performance becomes independent of the Reynolds number above a certain threshold, for cross-flow turbines, this is difficult to achieve at laboratory scale without a compressed-air wind tunnel \citep{miller_verticalaxis_2018, miller_solidity_2021} or relatively large tow tank \citep{bachant_effects_2016}.
To compensate for the decrease in $U_{\infty}$, the kinematic viscosity is decreased by changing the water temperature.
In this way, $\beta$ can be varied while holding $Fr_h$, $s/h$, and $Re_D$ constant.

\begin{table*}
    \centering
    \resizebox{\textwidth}{!}{      
        \begin{tabular}{@{}cccccccc@{}}
        \toprule
        Target Blockage Condition & \multicolumn{4}{c}{Flume Parameters} & \multicolumn{3}{c}{Non-Dimensional Flow Parameters} \\ \midrule
        $\beta$ [\%] & $h$ [m] & $U_{\infty}$ [m/s] & $s$ [m] & Temp [$^{\circ}\mathrm{C}$] & $s/h$ & $Fr_h$ & $Re_D$ \\ \midrule
        30.0 & 0.593 & 0.528 & 0.326 & 21.0 & 0.55 & \multirow{7}{*}{0.22} & \multirow{7}{*}{$1.70\!\times\!10^5$} \\
        33.4 & 0.534 & 0.501 & 0.267 & 23.3 & 0.50 &  &  \\
        36.7 & 0.485 & 0.477 & 0.218 & 25.4 & 0.45 &  &  \\
        40.1 & 0.445 & 0.457 & 0.178 & 27.3 & 0.40 &  &  \\
        45.0 & 0.396 & 0.431 & 0.129 & 30.1 & 0.33 &  &  \\
        50.0 & 0.356 & 0.409 & 0.090 & 32.6 & 0.25 &  &  \\
        55.0 & 0.324 & 0.390 & 0.057 & 35.0 & 0.18 &  &  \\ \bottomrule
        \end{tabular}
    }
    \caption{Experimental parameters for each blockage condition tested.}
    \label{tab:expMatrix}
\end{table*}

Array performance is measured at blockage ratios from $30.0\%$ to $55.0\%$, which is the widest range possible given the size of these turbines and flume capabilities. The inflow conditions required to achieve each blockage while holding $Fr_h$ and $Re_D$ constant are summarized in \cref{tab:expMatrix}.
Across all experiments, the measured $\beta$ are within $1.5\%$ of the target values in \cref{tab:expMatrix}, and the measured $Fr_h$ and $Re_D$ do not deviate more than $5\%$ from the target values.

The value of $Re_D$ in this study is constrained by the maximum $U_{\infty}$ for the highest blockage ratio.
This maximum velocity, in turn, is constrained by the entrainment of air from the free surface into the rotor (i.e., ventilation), which decreases lift on the blades and degrades performance \citep{birjandi_power_2013, young_ventilation_2017}.
Consequently, the maximum $U_{\infty}$ for the highest tested blockage ratio is set such that any ventilation occurs beyond the rotation rate corresponding to maximum turbine performance. 
To further limit the risk of ventilation, $s/h$ is maximized at each blockage rather than held constant across all blockages, since ventilation becomes more likely as $s/h$ decreases.
Array performance at various $s/h$ and $\beta$ is separately evaluated (\Cref{app:submergence}) and confirms that varying $s/h$ has minor effects on performance at these test conditions before the onset of ventilation.

\subsection{Array Layout and Control}
\label{methods:arrayLayout}

An overhead view of the array layout is shown in \cref{fig:arrayOverhead}.
The turbines are positioned laterally such that the center-to-center spacing is ${\sim}1.2D$ and the blade-to-blade spacing between adjacent turbines (${\sim}0.2D$; ${\sim}0.9c$) is twice the wall-to-blade spacing (i.e., the walls notionally correspond to symmetry planes in a larger array).
The turbines are operated under a counter-rotating, phase-locked scheme, wherein both turbines rotate at the same, constant speed, but in opposite directions (i.e., $\omega_A = -\omega_B$), with zero phase offset between them. 
The turbines are counter-rotated such that the blades rotate toward the channel centerline, which has been shown to augment performance relative to other rotation schemes \citep{zanforlin_fluid_2016, scherl_optimization_2022, gauvin-tremblay_hydrokinetic_2022}.

\subsection{Performance Metrics}
\label{methods:perfMet}

Array performance metrics are calculated from the measured quantities shown in \cref{fig:arrayOverhead}.
The rotation rate is non-dimensionalized as the ratio of the blade tangential velocity to the freestream velocity, or the tip-speed ratio, as 
\begin{equation}
    \lambda = \frac{\omega R}{U_{\infty}} \ \ ,
    \label{eq:TSR}
\end{equation}
\noindent where $\omega = |\omega_A| = |\omega_B|$ is the magnitude of the angular velocity. Data are collected at each tip-speed ratio for 60 seconds, and the time series is cropped to an integer number of turbine rotations.

The array efficiency (power coefficient) is the ratio of the net mechanical power produced by the turbines to the kinetic power in the freestream flow that passes through array's projected area
\begin{equation}
    C_{P} = \frac{Q_{A}\omega_{A} + Q_{B}\omega_{B}}{\frac{1}{2}\rho U_{\infty}^3 A_{\mathrm{turbines}}} \ \ ,
    \label{eq:cp}
\end{equation}
\noindent where $Q_{A}$ and $Q_{B}$ are the the hydrodynamic torques on the turbines and $\rho$ is the density of the working fluid.
Structural loads on the array are characterized via the thrust coefficient
\begin{equation}
    C_{T} = \frac{T_A + T_B}{\frac{1}{2} \rho U_{\infty}^2 A_{\mathrm{turbines}}} \ \ ,
    \label{eq:cThrust}
\end{equation}
\noindent where $T_A$ and $T_B$ are the streamwise forces on the turbines.
Cross-flow turbines also experience forces in the cross-stream direction ($L_{A}$ and $L_{B}$ in \cref{fig:arrayOverhead}), which can be similarly non-dimensionalized as an array lateral force coefficient using an analogous equation to \cref{eq:cThrust}. The lateral forces are tangential to this work, but provided as supplemental material \citep{hunt_supplemental}.

\subsection{Flow Visualization}
\label{methods:piv}

\begin{figure*}
    \centering
    \begin{minipage}[t]{0.475\textwidth}
        \vspace{0pt}
        \centering
        \includegraphics[width=0.9\textwidth]{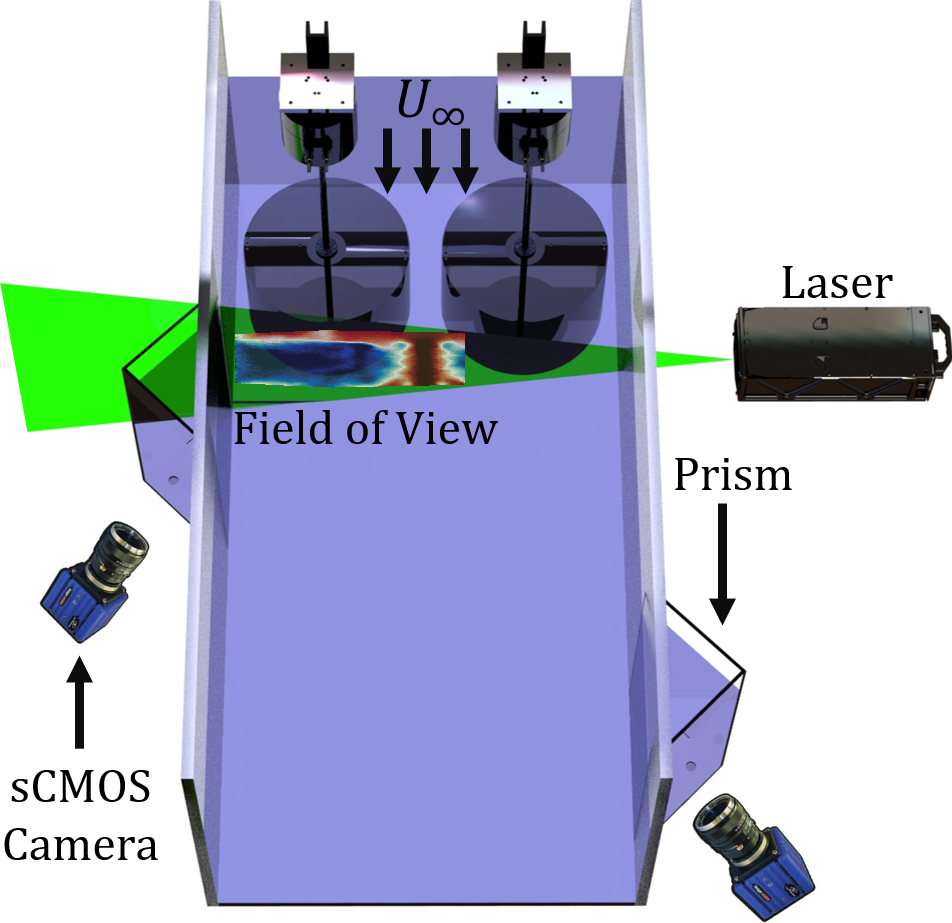}
        \caption{Schematic of the PIV test setup including turbines and example flow field data.}
        \label{fig:PIVRig}
    \end{minipage} \hfill
    \begin{minipage}[t]{0.475\textwidth}
        \vspace{0pt}
        \centering  
        \includegraphics[width=0.9\textwidth]{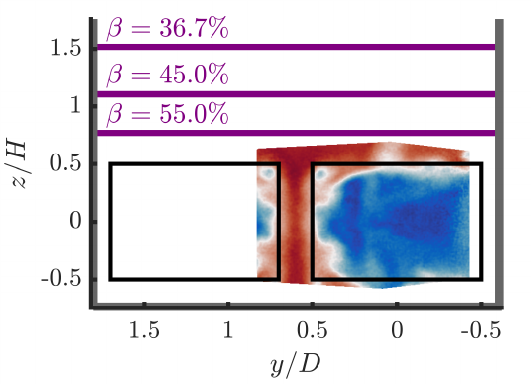}
        \caption{Location of PIV field of view within the flume, looking downstream. The lateral and vertical coordinates for the field of view are defined with respect to the center of Turbine B. The black rectangles indicate the projected areas of the turbines. The purple lines indicate the nominal location of the free surface at each blockage ratio.}
        \label{fig:PIVFoV}
    \end{minipage}
\end{figure*}

Two-camera, three-component particle image velocimetry is used to measure the near-wake flow field in planes perpendicular to the flow (\cref{fig:PIVRig}).
Two cameras (Imager sCMOS 5.5 MPx, maximum 50 fps) with 35 mm and 60 mm lenses, respectively, are positioned on either side of the flume at approximately $45^{\circ}$ incidence to the image plane.
The cameras are oriented perpendicular to water-filled acrylic prisms shown in \cref{fig:PIVRig}, which are used to reduce refractive distortion from the angled camera positioning relative to the flume.
Scheimpflug camera lens adaptors correct any residual focal distortions from angled fields of view (FoV).
The FoV of each camera is focused on a domain centered behind one of the turbines and results in an overlapping 42.3$\times$27.7 cm$^2$ data collection region.
The resulting FoV (\cref{fig:PIVFoV}) is positioned approximately 5.2 cm above the bottom of the flume and approximately 8.75 cm from the side wall to maximize capture of the flow through and around the array while reducing overexposure from laser scattering by solid boundaries.
The FoV encompasses the majority of the projected area of Turbine B, a portion of the bypass region above Turbine B, and the bypass region between the turbines.
Although the FoV does not capture the majority of the flow around Turbine A, the expectation of symmetry is borne out by the available data. 
The entire flume is seeded with neutrally buoyant hollow glass spheres with a diameter of $\sim\!10\mu \mathrm{m}$ (Potters Industries Sphericel 110P8), and illumination is provided by a 532 nm Nd:YAG laser (EverGreen 200), which produces a light sheet approximately 2 mm thick.

LaVision Davis 10.2.1 software is used to acquire and process the PIV data.
A dual-plane calibration target and self-calibration are employed.
The resulting scale factor is 7.97 pix/mm with fitting errors of 0.81 and 0.92 pixels for each camera, respectively.
A minimum filter subtracts background noise and an average polynomial filter using an interrogation window of 11$\times$11 pixels minimizes unsteady reflections from the free surface or near-blade regions.
Each PIV vector field is obtained by post-processing four images (two successive images from each camera).
Vector calculation employs an initial $128\times128$ interrogation window with a 50\% overlap followed by four sequential reducing passes to $32\times32$ pixels at a 75\% overlap.
After vector calculation, three passes of a universal outlier detection median filter \citep{westerweel_universal_2005} removes vectors with a median normalized residual greater than 1.5 on a filter domain of $7\times7$ vectors.
These vectors are reinstated if their normalized residual is less than 2.5 and alternate correlation peaks have a higher residual. Groups with fewer than five vectors are removed.

The near-wake flow field is measured in vertical planes $0.6D$ and $1.5D$ downstream of the turbines' axes of rotation, as well as $1.0D$ upstream.
To change the relative location of the vertical plane, the array is shifted in the streamwise direction.
We focus our analysis on the plane nearest to the array at $0.6D$ downstream, with the flow fields at $1.5D$ downstream and $1.0D$ upstream provided in supplemental material \citep{hunt_supplemental}.
Flow field data in each plane are collected at $\beta = 36.7\%$, $45.0\%$, and $55.0\%$ for tip-speed ratios of $\lambda = 1.6$, $2.6$, $2.9$, and $3.6$ (all $\beta$) and $\lambda = 4.2$ ($\beta = 55.0\%$ only).
These tip-speed ratios include the ``optimal'' $\lambda$ corresponding to maximum $C_{P}$ for $\beta = 36.7\%$, $45.0\%$, and $55.0\%$, as well as $\lambda$ lower (``underspeed'') and higher (``overspeed'') than the optimal $\lambda$ at these blockage ratios, which are relevant in field deployments for regulating turbine power above the rated flow speed \citep{pao_tutorial_2009}.

Data acquisition at each $\beta$ and $\lambda$ is synchronized with turbine phase ($\theta$), with images collected in $3^{\circ}$ increments at the optimal $\lambda$ for each $\beta$ and $12^{\circ}$ increments for all other $\lambda$ cases.
The higher azimuthal resolution at the optimal $\lambda$ was motivated by related work on the three-dimensionality of the array's wake \citep{long_threedimensionality_2024}.
Here, the data with greater phase resolution is used only for secondary data products (e.g., standard deviation, vorticity, and TKE fields) and, for all other analysis, data collected at optimal $\lambda$ are decimated to $12^{\circ}$ increments for consistency. 
This azimuthal discretization results in acquisition at 15 unique turbine phases in the range $0^{\circ} \leq \theta < 180^{\circ}$). 
Because the dynamics of a two-bladed turbine are periodic over a $180^{\circ}$ interval, vector fields collected during the second half of the rotation (i.e., $180^{\circ} \leq \theta < 360^{\circ}$) are pooled with the corresponding upstream position.
At each tip-speed ratio, 40 instantaneous vector fields are obtained at each phase from 20 turbine rotations, which was sufficient for convergence of the mean flow field at each phase (see supplemental material \citep{hunt_supplemental}).
The time-average flow fields are then taken as the mean of the phase-averaged fields over all phases, which is mathematically equivalent to a mean of all instantaneous vector fields.
A phase resolution of $12^\circ$ was sufficient for convergence of the time-averaged field (see supplemental material \citep{hunt_supplemental}.

Standard deviation, turbulent kinetic energy (TKE) and vorticity are calculated at the optimal $\lambda$ for each $\beta$ where data with a higher azimuthal resolution ($3^{\circ}$) are available.
Higher resolution data are employed for these quantities to reduce small-scale noise in the resulting time-average fields, with the large-scale structure obtained comparable to that resulting from $12^{\circ}$ phase separation (see supplemental material \citep{hunt_supplemental}).
TKE is calculated from the flow fields for each phase as
\begin{equation}
    \mathrm{TKE}(\theta) = \frac{1}{2}
    \left( 
    (U_x(\theta) - \overline{U_x(\theta)})^2
    + (U_y(\theta) - \overline{U_y(\theta)})^2
    + (U_z(\theta) - \overline{U_z(\theta)})^2
    \right)
\end{equation}
where the overline denotes the phase-average of each velocity component for a given phase.
A representative time-averaged TKE field is then obtained from the mean of the phase-averaged TKE fields.
Streamwise vorticity is calculated for each phase from the phase-averaged velocity fields as 
\begin{equation}
    \Omega_x(\theta) = \frac{\partial \overline{U_z(\theta)}}{\partial y}- \frac{\partial \overline{U_y(\theta)}}{\partial z}
\end{equation}
where $\partial y$ and $\partial z$ denote the cross stream and vertical spatial resolution.
Velocity gradients were calculated from the phase-averaged velocity fields using a central difference scheme for interior points and a one-sided difference for boundary points.
As for other quantities, the time-averaged streamwise vorticity field is an average across all phases.

\section{Experimental Results}
\label{sec:results}

\subsection{Array Performance}
\label{results:perf}

 \begin{figure}
     \centering
     \includegraphics[width=\textwidth]{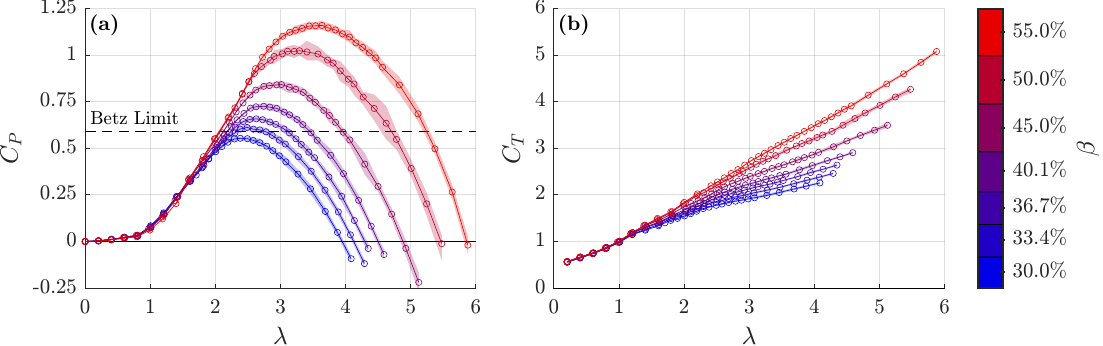}
     {\phantomsubcaption\label{fig:arrayPerf_cp}
      \phantomsubcaption\label{fig:arrayPerf_ct}
      }
     \caption{Time-averaged \subref{fig:arrayPerf_cp} $C_{P}$ and \subref{fig:arrayPerf_ct} $C_{T}$ as a function of $\beta$ and $\lambda$. The shaded regions indicate the interquartile range of the array- and cycle-averaged performance at each $\beta$ and $\lambda$ (the vertical span of the shaded region at each point is similar to the size of the markers for most operating conditions).}
     \label{fig:arrayPerf}
 \end{figure}
 
Time-average $C_{P}$ as a function of $\beta$ and $\lambda$ is shown in \cref{fig:arrayPerf_cp}.
As for prior studies (\cref{tab:priorWork}), as $\beta$ increases, the maximum $C_{P}$ increases, the array produces power over a broader range of tip-speed ratios, and the optimal tip-speed ratio increases.
Similarly, the time-averaged $C_{T}$ (\cref{fig:arrayPerf_ct}) generally increases with $\beta$, particularly at higher blockage ratios.
Above $\beta = 33.4\%$, maximum $C_{P}$ exceeds the Betz limit, and above $\beta = 50.0\%$, maximum $C_{P}$ exceeds unity. 
Such efficiencies are not violations of energy conservation since the definition of $C_{P}$ in \cref{eq:cp} considers only the kinetic power passing through the array projected area.
This neglects the fluid's potential energy, which is appreciably drawn down as $\beta$ and $C_T$ increase.
As $\beta$ increases, a unity-bounded efficiency metric may resemble the hydraulic efficiency of a hydropower turbine (e.g., metrics employed by \citet{takamatsu_study_1985} or \citet{mcadam_experimental_2013}) in which the available power is a function of volumetric flow rate through the channel and the head drop across the array.
This alternative efficiency metric is explored further in the supplemental material \citep{hunt_supplemental}, but is sensitive to the resolution and accuracy of free surface measurements.
Here, the conventional definitions of $C_P$ and $C_T$ facilitate comparison with prior studies, as well as consistency with LMADT---which is formulated using the definition of $C_T$ in \cref{eq:cThrust}.

 \begin{figure}
     \centering
     \includegraphics[width=\textwidth]{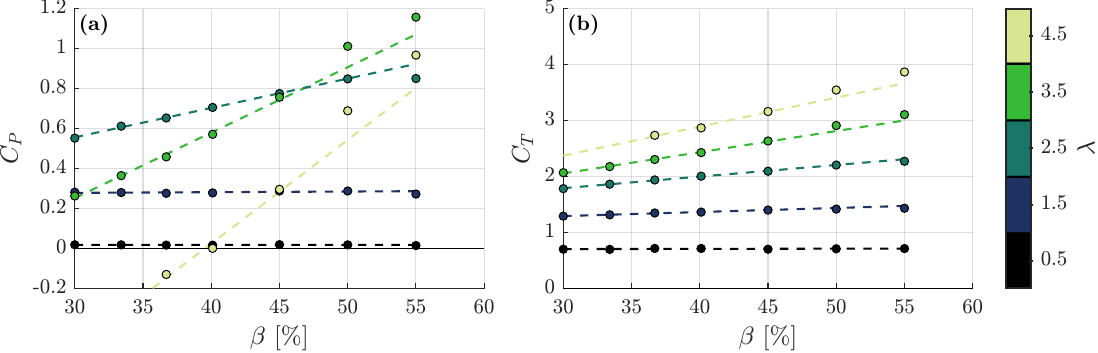}
     {\phantomsubcaption\label{fig:perfVBlockage_cp}
      \phantomsubcaption\label{fig:perfVBlockage_ct}
      }
     \caption{Time-averaged \subref{fig:perfVBlockage_cp} $C_{P}$ and \subref{fig:perfVBlockage_ct} $C_{T}$ versus $\beta$ at constant $\lambda$.
     The data for each $\beta$ are interpolated at each $\lambda$ from the points in \cref{fig:arrayPerf}.
     The dashed lines represent the line of best fit through the data points corresponding to $\beta \leq 45.0\%$ at each $\lambda$.}
     \label{fig:perfVBlockage}
 \end{figure}

When the array power and thrust coefficients are regressed against $\beta$ at constant tip-speed ratio, approximately linear relationships are revealed, which are shown for a subset of $\lambda$ in \cref{fig:perfVBlockage}.
Below a threshold value of $\lambda$, both $C_{P}$ and $C_{T}$ are independent of the blockage ratio and depend only on the tip-speed ratio, which corresponds to the collapsed regions of the $C_{P}-\lambda$ curves ($\lambda < 1.6$) and $C_{T}-\lambda$ curves ($\lambda < 1.2$) in \cref{fig:arrayPerf}.
Similar collapse at low $\lambda$ has been observed in prior work for both cross-flow turbines \citep{consul_blockage_2013} and axial-flow turbines \citep{kolekar_performance_2015, badshah_cfd_2019}.
As $\lambda$ increases further, 
$C_{P}$ and $C_{T}$ vary linearly with the blockage ratio for $\beta \leq 45.0\%$ and moderate $\lambda$, which is emphasized by the linear regressions in \cref{fig:perfVBlockage}.
While similar linear relationships between the blockage ratio, power, and thrust have been previously identified in simulations of oscillating foils \citep{gauthier_impact_2016}, axial-flow turbines \citep{kinsey_impact_2017, abutunis_comprehensive_2022} and cross-flow turbines \citep{kinsey_impact_2017}, this is the first experimental demonstration of this trend.
Finally, as $\beta$ and $\lambda$ increase further, these relationships become non-linear with performance increasingly sensitive to incremental changes in the blockage and tip-speed ratios, which is similar to narrower observations for $C_P$ by \citet{gauthier_impact_2016}.

\subsection{Flow Fields}
\label{results:piv}

\begin{figure*}
     \centering
     \includegraphics[width=\textwidth]{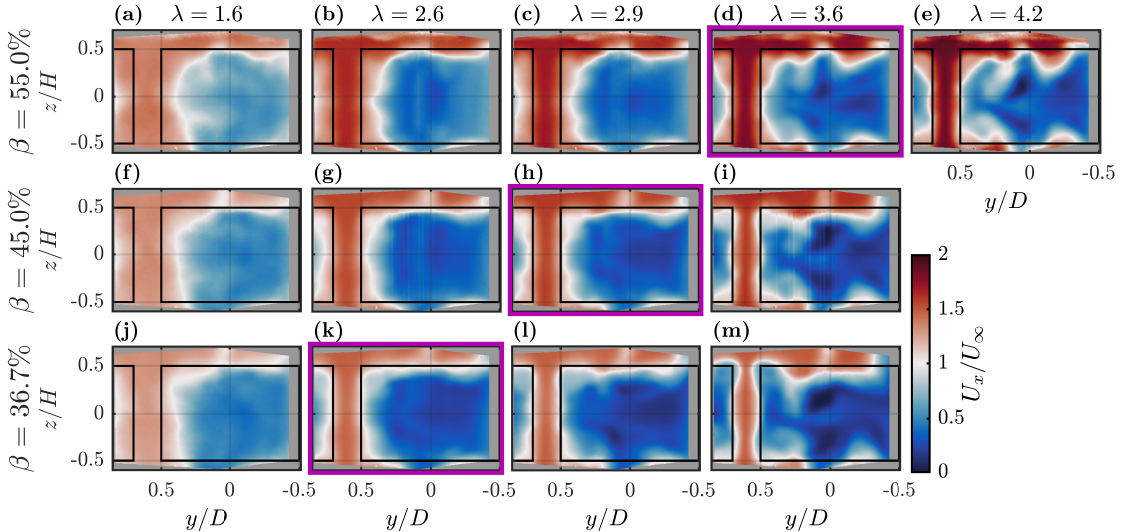}
     {\phantomsubcaption\label{fig:piv_B55_TSR1.6}
      \phantomsubcaption\label{fig:piv_B55_TSR2.6}
      \phantomsubcaption\label{fig:piv_B55_TSR2.9}
      \phantomsubcaption\label{fig:piv_B55_TSR3.6}
      \phantomsubcaption\label{fig:piv_B55_TSR4.2}
      
      \phantomsubcaption\label{fig:piv_B45_TSR1.6}
      \phantomsubcaption\label{fig:piv_B45_TSR2.6}
      \phantomsubcaption\label{fig:piv_B45_TSR2.9}
      \phantomsubcaption\label{fig:piv_B45_TSR3.6}
      
      \phantomsubcaption\label{fig:piv_B36_TSR1.6}
      \phantomsubcaption\label{fig:piv_B36_TSR2.6}
      \phantomsubcaption\label{fig:piv_B36_TSR2.9}
      \phantomsubcaption\label{fig:piv_B36_TSR3.6}
     }
     \caption{Time-averaged streamwise velocity $0.6D$ downstream of the array as function of $\beta$ and $\lambda$. Velocity measurements are normalized by the freestream velocity measured by the ADV. The black rectangles in each tile indicate the projected area of the turbines in the field of view. Flow fields corresponding to the optimal $\lambda$ for each blockage are outlined in purple (tiles \subref{fig:piv_B55_TSR3.6} \subref{fig:piv_B45_TSR2.9}, and \subref{fig:piv_B36_TSR2.6}).}
     \label{fig:pivNearWake}
 \end{figure*} 

\Cref{fig:pivNearWake} shows the normalized time-average streamwise velocity field $0.6D$ downstream of the array (i.e., within the near-wake) as a function of $\beta$ and $\lambda$.
The velocity fields are characterized by a wake region behind the turbines ($U_x/U_{\infty} < 1$, indicated by blue hues) and a bypass region with accelerated flow above, below, and between the turbines ($U_x/U_{\infty} > 1$, indicated by red hues). 
The bypass flow region between Turbine B and the flume wall lies outside of the field of view, and we attribute the persistent region of decelerated flow below Turbine B at all $\lambda$ to the test rig wake (\cref{fig:testRig}).
Qualitatively, the shape of the near-wake resembles that observed in prior work for individual \citep{bachant_characterising_2015, ross_experimental_2020a} and pairs \citep{posa_wake_2022} of cross-flow turbines.
Deformation of the wake at the top and bottom of the turbine projected area is attributable in part to vortices that form at the interface between the blades and the struts, which intensify at higher tip-speeds.

\begin{figure*}
     \centering
     \includegraphics[width=\textwidth]{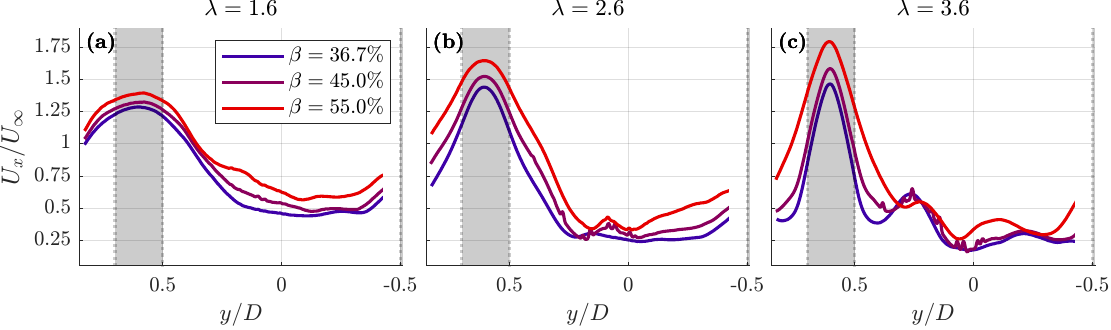}
      {\phantomsubcaption\label{fig:pivlineScaled_infty_TSR1.6}
       \phantomsubcaption\label{fig:pivlineScaled_infty_TSR2.6}
      \phantomsubcaption\label{fig:pivlineScaled_infty_TSR3.6}
      }
     \caption{Time-averaged normalized streamwise velocity profiles measured $0.6D$ downstream of the array for various $\beta$ at \subref{fig:pivlineScaled_infty_TSR1.6} $\lambda = 1.6$, \subref{fig:pivlineScaled_infty_TSR2.6} $\lambda = 2.6$, and \subref{fig:pivlineScaled_infty_TSR3.6} $\lambda = 3.6$.
     The profiles correspond to the average velocity along the middle $70\%$ of the turbine blade span. Vertical gray rectangles correspond to the bypass region.}
     \label{fig:pivlineBetaUinf}
 \end{figure*}

Trends in the evolution of the near-wake with $\beta$ and $\lambda$ are apparent in two-dimensional velocity profiles of the average streamwise velocity at each $y/D$ (\cref{fig:pivlineBetaUinf}).
These profiles are calculated over the middle $70\%$ of the blade span to exclude edge effects from the blade-end struts.
Ripples in the wake region of these profiles are attributed to the passage of vortices shed from the blades during their downstream sweep.
At $\lambda = 1.6$ (\cref{fig:pivlineScaled_infty_TSR1.6}), where array thrust varies weakly with $\beta$ (\cref{fig:perfVBlockage_ct}), $U_x/U_{\infty}$ increases slightly with $\beta$ across the lateral profile.
As $\lambda$ increases (\cref{fig:pivlineScaled_infty_TSR2.6,fig:pivlineScaled_infty_TSR3.6}),  the turbines impart greater resistance on the flow and more fluid is diverted around them, causing $U_x/U_{\infty}$ to increase in the bypass region for all $\beta$.
Similarly, at a given $\lambda$, the bypass velocity increases with $\beta$.
$U_x/U_{\infty}$ in the wake region also tends to increase with $\beta$ at a given $\lambda$ since confinement increases the flow through the turbines.
However, increases in the wake velocity with $\beta$ are less pronounced than the corresponding increases in bypass velocity due to momentum extraction by the array from the core flow through the turbines.
Although the maximum bypass velocity at the array centerline increases with both the $\beta$ and $\lambda$, $U_x/U_{\infty}$ decreases more rapidly away from this midpoint at higher $\lambda$.
This is indicated by the narrower peaks in the velocity profiles as $\lambda$ increases at a given $\beta$.
Conversely, an increase in $\beta$ at constant $\lambda$ results in a more gradual reduction of $U_x/U_{\infty}$ away from the array centerline, which implies increased turbulent mixing between the bypass flow and flow through the turbine.
As discussed by \citet{nishino_effects_2012}, mixing between the bypass and core flows provides a secondary power enhancement with increasing blockage by increasing the pressure drop across the rotor as a consequence of elevated near-wake velocity.

\subsection{Free Surface Deformation}

\begin{figure*}
     \centering
     \includegraphics[width=\textwidth]{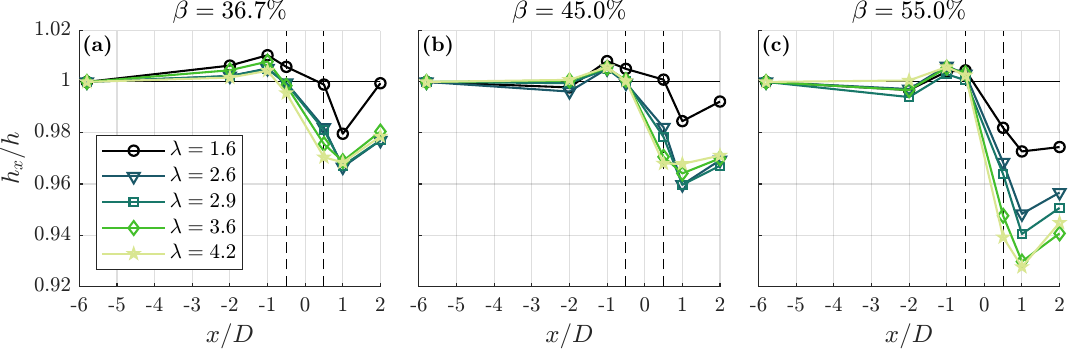}
      {\phantomsubcaption\label{fig:fst_36}
      \phantomsubcaption\label{fig:fst_45}
      \phantomsubcaption\label{fig:fst_55}
      }
     \caption{Time-averaged streamwise water depth profiles for various $\lambda$ at \subref{fig:fst_36} $\beta = 36.7\%$, \subref{fig:fst_45} $\beta = 45.0\%$, and \subref{fig:fst_55} $\beta = 55.0\%$. The farthest upstream measurement is taken at the center of the channel, whereas depth measurements in the vicinity of the array are directly upstream and downstream of the Turbine B (i.e., $y/D = 0$ in the PIV flow fields).
     Depth measurements at each streamwise location ($h_x$) are normalized by the far-upstream water depth ($h$) measured at the same $\beta$ and $\lambda$. The dashed lines indicate the extent of the turbine's swept area.}
     \label{fig:fstProfiles}
 \end{figure*}

The time-average streamwise free surface profile across Turbine B is shown in \cref{fig:fstProfiles} at the same $\beta$ and $\lambda$ as for the flow fields in \cref{fig:pivNearWake}.
For all $\beta$ and $\lambda$, a free surface drop is observed across the turbine, corresponding to the pressure drop across a turbine in an open-channel flow \citep{vogel_effect_2016}.
The magnitude of this drop (in this case, the difference between the measured water depths at $x/D = -1$ and $x/D = 1$) tends to increase with $\beta$ for a given $\lambda$.
However, as shown in \Cref{app:submergence}, the free surface drop behind the array also depends on the normalized submergence depth ($s/h$) which decreases with $\beta$ in these experiments (\Cref{methods:ndParams}).
Consequently, the free surface drop at higher $\beta$ is likely augmented by the closer proximity of the turbines to the free surface.
The free surface drop across the turbine clearly increases with $\lambda$ at $\beta = 55.0\%$ (\cref{fig:fst_55}), consistent with the monotonic increase in $C_T$ with $\lambda$ (\cref{fig:arrayPerf}).
However, even though similar relationships between $C_T$ and $\lambda$ are observed for $\beta = 36.7\%$ and $\beta = 45.0\%$, the free surface drop at these blockages does not vary significantly with the tip-speed ratio for $\lambda \geq 2.6$, although the depth at $x/D = 0.5$ does continuously decrease with $\lambda$ for all $\beta$.
        
\section{Evaluation of LMADT for a Cross-flow Turbine Array at High Blockage}
\label{sec:analyticalModels}

\subsection{Model Overview}
\label{sec:modelOverview}

\begin{figure}
     \centering
     \includegraphics[width=0.75\textwidth]{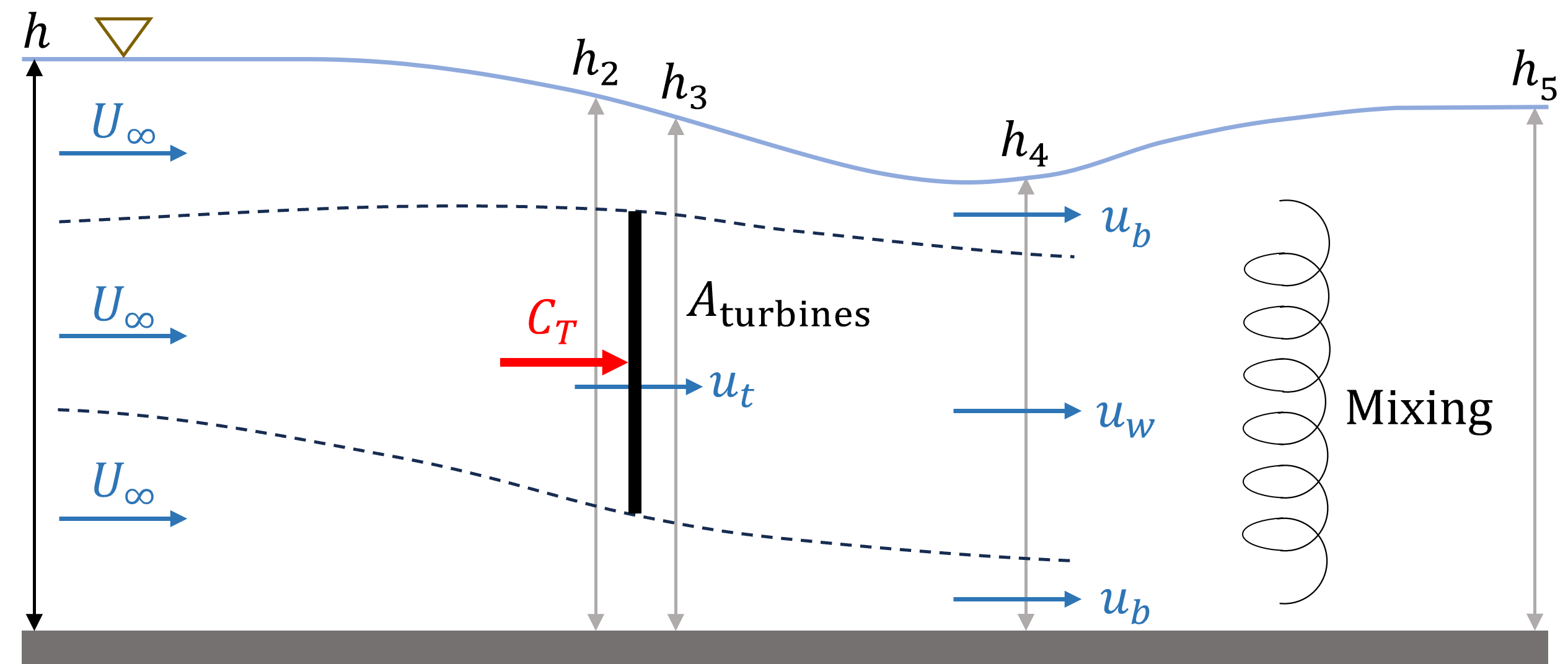}
     \caption{Linear momentum model for an actuator disk in open-channel flow defined by \citet{houlsby_application_2008} and adapted to the cross-flow turbine array nomenclature.}
     \label{fig:lmadDiagram}
\end{figure}

The performance, flow field, and free surface measurements described in \Cref{sec:results} allow us to directly assess the accuracy of a one-dimensional linear momentum actuator disk theory (LMADT) description of the cross-flow turbine array.
Given the appreciable free surface deformation (\cref{fig:fstProfiles}) and the homogeneous lateral distribution of turbines in the channel (i.e., equivalent local and global blockage), the open-channel LMADT model introduced by \citet{houlsby_application_2008} is most relevant.
In the limiting case of negligible free surface drop across the turbine array ($Fr_h^2 \rightarrow 0$), \citeauthor{houlsby_application_2008}'s model reduces to the well-known model of \citet{garrett_efficiency_2007}.
For a non-uniform distribution of turbines along the channel width, a theory that represents the individual turbines in the array as separate actuator disks would be required \citep[e.g.,][]{nishino_efficiency_2012, vogel_effect_2016, dehtyriov_fractallike_2021}.
Although the array's homogeneity does not extend to the vertical direction, the blockage asymmetry in these experiments (1.3 - 2.3) is within the range found by \citet{kinsey_impact_2017} to have limited influence on turbine power and thrust.
We note that LMADT-based models like that of \citet{houlsby_application_2008} are only one of several types of analytical models (e.g., double-multiple streamtube models \citep{paraschivoiu_wind_2002,ayati_doublemultiple_2019}, potential flow models \citep{steiros_analytical_2022}) that could be applied.
While we primarily consider \citeauthor{houlsby_application_2008}'s model here due to the widespread use of LMADT-based blockage corrections, the potential flow model of \citet{steiros_analytical_2022} is separately evaluated (see supplemental material \citep{hunt_supplemental}), with a discussion of key differences in results provided in \cref{sec:lmadScaling}.

In the LMADT model, the array is represented as a single actuator disk with the same blockage ratio and thrust coefficient as the array (\cref{fig:lmadDiagram}).
The flow upstream of the turbine is unidirectional, uniform, and subcritical (i.e., $Fr_h^2 < 1$) with velocity $U_{\infty}$ and depth $h$.
In response to momentum extraction by the turbine, the ``core'' flow in the streamtube that passes through the rotor decelerates from $U_{\infty}$ to $u_t$ just upstream of the rotor plane, and further decelerates to $u_w$ downstream of the turbine in the near-wake.
To satisfy continuity, the fluid outside of this streamtube is accelerated from $U_{\infty}$ to $u_b$.
The core wake velocity ($u_w$) and bypass velocity ($u_b$) are both defined at a location downstream of the turbine where the pressure in the core and bypass flows is equal and hydrostatic. The flow is assumed to be axisymmetric and inviscid up to this location, downstream of which $u_w$ and $u_b$ mix.
Far downstream, the flow is once again uniform, with reduced depth due to the energy extracted by the disk but higher velocity to conserve mass.
This model qualitatively captures aspects of the experimentally observed wake and bypass velocities (\cref{fig:pivNearWake}).
However, although LMADT models are widely used to represent turbine hydrodynamics in confined flows, several of the underlying assumptions are often violated.
For example, the wake is assumed to be axisymmetric and to not mix with the bypass until static pressure has equilibrated between the core and bypass flows, while we observe a non-uniform wake with immediate mixing between core and bypass.

Detailed derivations of the LMADT equation set used here are presented in \citet{houlsby_application_2008} and \citet{houlsby_power_2017}.
The implementation in this study follows \citet{ross_experimental_2020}, who reorganized \citeauthor{houlsby_application_2008}'s original formulation into two equations from which $u_w$ and $u_b$ can be obtained numerically if $U_{\infty}$, $C_{T}$, $Fr_h$, and $\beta$ are known:

\begin{equation}
    u_w = \frac{Fr_h^2 u_b^4 - (4 + 2Fr_h^2) U_{\infty}^2 u_b^2 + 8 U_{\infty}^3 u_b - 4U_{\infty}^4 + 4\beta C_{T} U_{\infty}^4 + Fr_h^2 U_{\infty}^4}
               {-4Fr_h^2 u_b^3 + (4Fr_h^2 + 8) U_{\infty}^2 u_b - 8U_{\infty}^3} \ , 
    \label{eq:uwBlockage}
\end{equation}

\begin{equation}
    u_w = \sqrt{u_b^2 - C_{T}U_{\infty}^2} \ .
    \label{eq:uwThrust}
\end{equation}

\noindent The velocity through the turbine, $u_t$, is then given as

\begin{equation}
    u_t = \frac{u_w (u_b - U_{\infty}) (2gh - u_b^2 - u_b U_{\infty})}
               {2 \beta g h (u_b - u_w)} \ .
    \label{eq:ut}
\end{equation}

\noindent For the solutions to \cref{eq:uwBlockage,eq:uwThrust,eq:ut} to be physically valid, the core and bypass flows must be subcritical and satisfy $u_b > u_t > u_w$ and $u_b > U_{\infty} > u_t$.

The free surface profile in the vicinity of the array can also be calculated from the open-channel LMADT model.
\citeauthor{houlsby_application_2008} define five streamwise stations, $x_1 - x_5$:

\begin{enumerate}[align=right,
                  leftmargin=2em,
                  label=$\mathbf{x_{\arabic*}}$:\ ,
                  topsep=1em]
    \item Far upstream, where the flow is undisturbed by the disk ($h_1$ = $h$; $U_1 = U_{\infty}$);
    \item Just upstream of the actuator disk ($h_2$; $U_2 = u_t$);
    \item Just downstream of the actuator disk ($h_3$; $U_3 = u_t$);
    \item Downstream of the actuator disk, where the static pressure in the core and bypass
    \item[] flows is equal ($h_4$; $U_4$ = $u_w$ in the core flow; $U_4 = u_b$ in the bypass); and
    \item Far downstream of the disk, when the flow has fully mixed ($h_5$; $U_5$),
\end{enumerate}

\noindent where $h_1$ through $h_5$ and $U_1$ through $U_5$ are the water depth and velocity, respectively, at each station.
Once $u_w$, $u_b$, and $u_t$ have been calculated from \cref{eq:uwBlockage,eq:uwThrust,eq:ut}, then $h_4$, $h_3$, and $h_2$ may be calculated following \citet{houlsby_application_2008} as
\begin{equation}
    h_4 = h + \frac{(U_{\infty}^2 - u_b^2)}{2g} \ \ ,
    \label{eq:h4}
\end{equation}
\begin{equation}
    h_3 = h_4 + \frac{(u_w^2 - u_t^2)}{2g} \ \ ,
    \label{eq:h3}
\end{equation}
\begin{equation}
    h_2 = h + \frac{(U_{\infty}^2 - u_t^2)}{2g} \ \ .
    \label{eq:h2}
\end{equation}
\noindent The normalized net free surface drop across the channel from far upstream ($x_1$) to far downstream ($x_5$), $\frac{\Delta h}{h}$, is obtained by solving
\begin{equation}
    \frac{1}{2} \left( \frac{\Delta h}{h} \right)^3 
    - \frac{3}{2} \left( \frac{\Delta h}{h} \right)^2
    + \left( 1 - Fr_{h}^2 + \frac{C_T \beta Fr_{h}^2}{2} \right) \frac{\Delta h}{h}
    - \frac{C_T \beta Fr_{h}^2}{2}
    = 0 \ \ ,
    \label{eq:deltaH}
\end{equation}
\noindent from which $h_5$ is obtained as
\begin{equation}
    h_5 = h \left( \frac{\Delta h}{h} \right) \ \ .
\end{equation}

In our experiments, because the static pressure equilibrium point is unknown, the location of $x_4$ is ambiguous, as are the locations of $x_2$ and $x_3$ due to the non-zero rotor ``thickness''.
In addition, the flume has insufficient length for the flow to fully mix downstream of the array, such that $x_5$ cannot be measured.
Consequently, although $u_b$, $u_w$, and $u_t$ may be estimated from available measurements and $h_2$ through $h_5$ calculated, the corresponding streamwise positions of $x_2$ through $x_5$ in experiments are uncertain.

\subsection{Comparison between Experimental and Modeled Near-Wake Velocities}
\label{sec:lmadVels}

\begin{figure*}
     \centering
     \includegraphics[width=\textwidth]{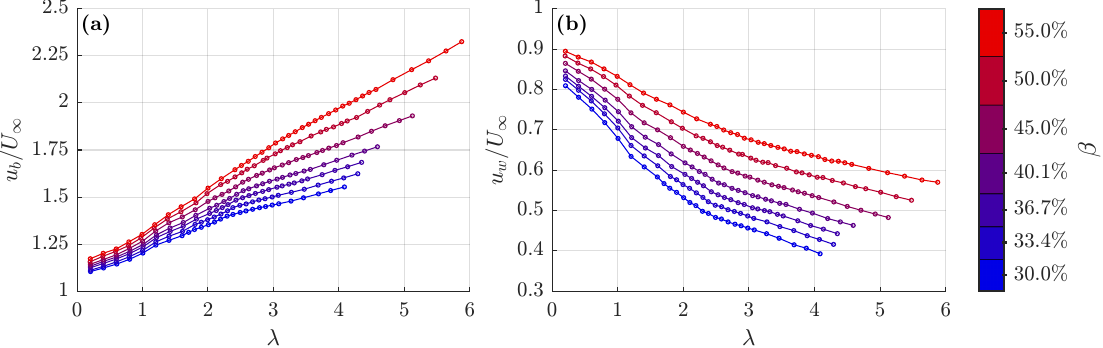}
     {
      \phantomsubcaption\label{fig:lmadVels_bypass}
      \phantomsubcaption\label{fig:lmadVels_wake}
      }
     \caption{The \subref{fig:lmadVels_bypass} bypass velocities, and \subref{fig:lmadVels_wake} wake velocities (all normalized by the freestream velocity) predicted by the open-channel LMADT model of \citet{houlsby_application_2008} from measurements of array thrust and flow conditions.}
     \label{fig:lmadVels}
 \end{figure*}

Using the measured values of $C_{T}$, $U_{\infty}$, $Fr_h$, and $\beta$, $u_b$ and $u_w$ are calculated at each $\lambda$ using \cref{eq:uwBlockage,eq:uwThrust}.
The resulting analytically-predicted velocities are normalized by the freestream velocity in \cref{fig:lmadVels}.
At a given tip-speed ratio, both $u_b$ (\cref{fig:lmadVels_bypass}) and $u_w$ (\cref{fig:lmadVels_wake}) increase with $\beta$ because confinement accelerates the flow through and around the turbine.
$u_b$ increases with $\lambda$ at all $\beta$ since the corresponding increase in thrust diverts more flow around the array, whereas $u_w$ decreases with $\lambda$ at all $\beta$ due to momentum extraction by the array.
In agreement with experimental trends in the velocity profiles in \cref{fig:pivlineBetaUinf}, modeled $u_b$ increases more strongly with tip-speed ratio at higher blockage.
We note that the modeled $u_w$ is always positive, indicating that the upper limit of the thrust coefficient for physically-valid solutions to open-channel LMADT (analogous to $C_T \leq 1$ for LMADT in unconfined flow) is not exceeded in these experiments.

\begin{figure*}
     \centering
     \includegraphics[width=\textwidth]{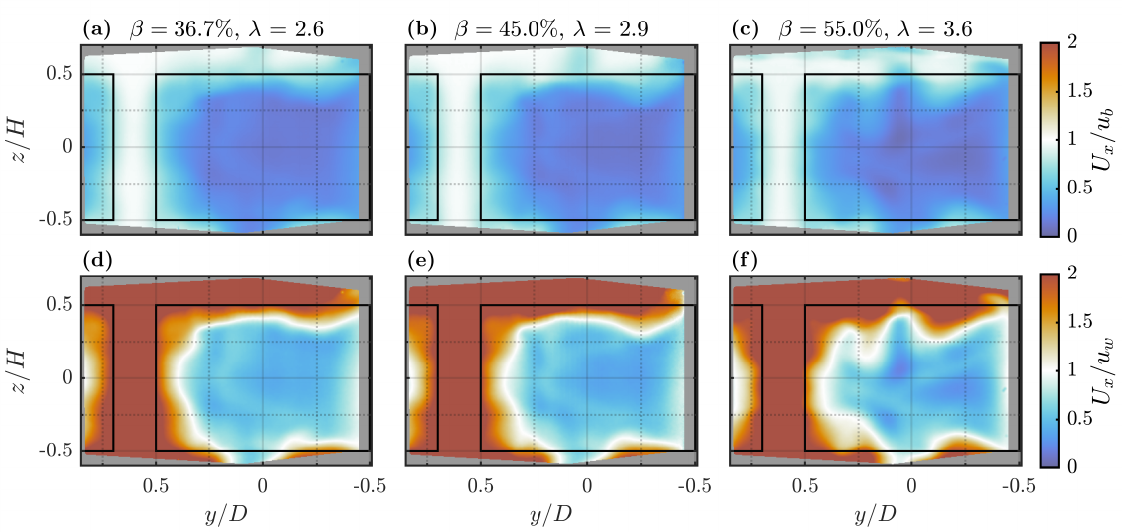}
      {\phantomsubcaption\label{fig:pivScaled_bypass_36}
      \phantomsubcaption\label{fig:pivScaled_bypass_45}
      \phantomsubcaption\label{fig:pivScaled_bypass_55}
      \phantomsubcaption\label{fig:pivScaled_wake_36}
      \phantomsubcaption\label{fig:pivScaled_wake_45}
      \phantomsubcaption\label{fig:pivScaled_wake_55}
      }
     \caption{Time-averaged streamwise velocities measured $0.6D$ downstream of the array at $\beta = 36.7\%$, $45.0\%$, and $55.0\%$ and optimal $\lambda$, normalized by the corresponding \subref{fig:pivScaled_bypass_36}-\subref{fig:pivScaled_bypass_55} modeled bypass velocity, $u_b$, and \subref{fig:pivScaled_wake_36}-\subref{fig:pivScaled_wake_55} modeled wake velocity, $u_w$. The black rectangles indicate the projected area of the turbines in the field of view.}
     \label{fig:pivScaled}
 \end{figure*}

 \begin{figure*}
     \centering
     \includegraphics[width=\textwidth]{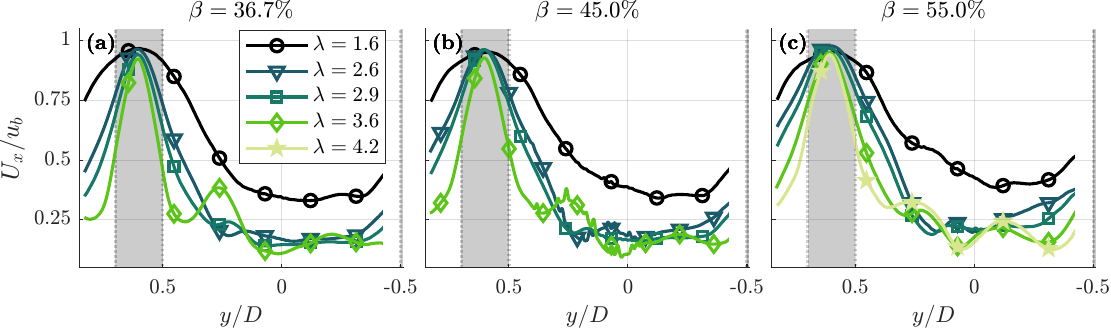}
      {\phantomsubcaption\label{fig:pivlineScaled_bypass_36}
      \phantomsubcaption\label{fig:pivlineScaled_bypass_45}
      \phantomsubcaption\label{fig:pivlineScaled_bypass_55}
      }
     \caption{Time-averaged streamwise velocity profiles measured $0.6D$ downstream of the array normalized by the modeled bypass velocity at \subref{fig:pivlineScaled_bypass_36} $\beta = 36.7\%$, \subref{fig:pivlineScaled_bypass_45} $45.0\%$, and \subref{fig:pivlineScaled_bypass_55} $55.0\%$ and various $\lambda$.
     The profiles shown correspond to the average velocity along the middle $70\%$ of the turbine blade span. Vertical gray rectangles correspond to the bypass region.}
     \label{fig:pivlineScaled}
 \end{figure*}
 
In \cref{fig:pivScaled}, the time-averaged near-wake velocity fields at $\beta = 36.7\%$, $45.0\%$, and $55.0\%$ and optimal $\lambda$ are normalized by $u_b$ and $u_w$, respectively.
Blue hues correspond to experimental velocities lower than the LMADT-modeled value, red hues correspond to experimental velocities higher than the LMADT-modeled value, and white indicates agreement between the experimental and modeled velocities.
The measured bypass velocity is well-predicted by LMADT, as indicated by the substantive white regions in \cref{fig:pivScaled_bypass_36,fig:pivScaled_bypass_45,fig:pivScaled_bypass_55}.
As shown in the bypass-normalized velocity profiles in \cref{fig:pivlineScaled}, the modeled $u_b$ tends to be slightly higher than experimental measurements at the array centerline, but are within approximately $8\%$ of the measured value for all $\beta$ and $\lambda$.
In contrast, the measured wake velocities are lower than modeled wake velocities at all $\lambda$, and there is no substantive region of the flow that is well represented by $u_w$ (\cref{fig:pivScaled_wake_36,fig:pivScaled_wake_45,fig:pivScaled_wake_55}).
Flow fields normalized by $u_b$ and $u_w$ at other $\beta-\lambda$ combinations are provided as supplemental material \citep{hunt_supplemental}.
While it is unknown if the static pressure is equal in the wake and bypass flows at $0.6D$ downstream of the array, $u_b$ is less descriptive of the measured bypass flow further downstream ($1.5D$), as shown in supplemental material \citep{hunt_supplemental}).

 \begin{figure*}
     \centering
     \includegraphics[width=\textwidth]{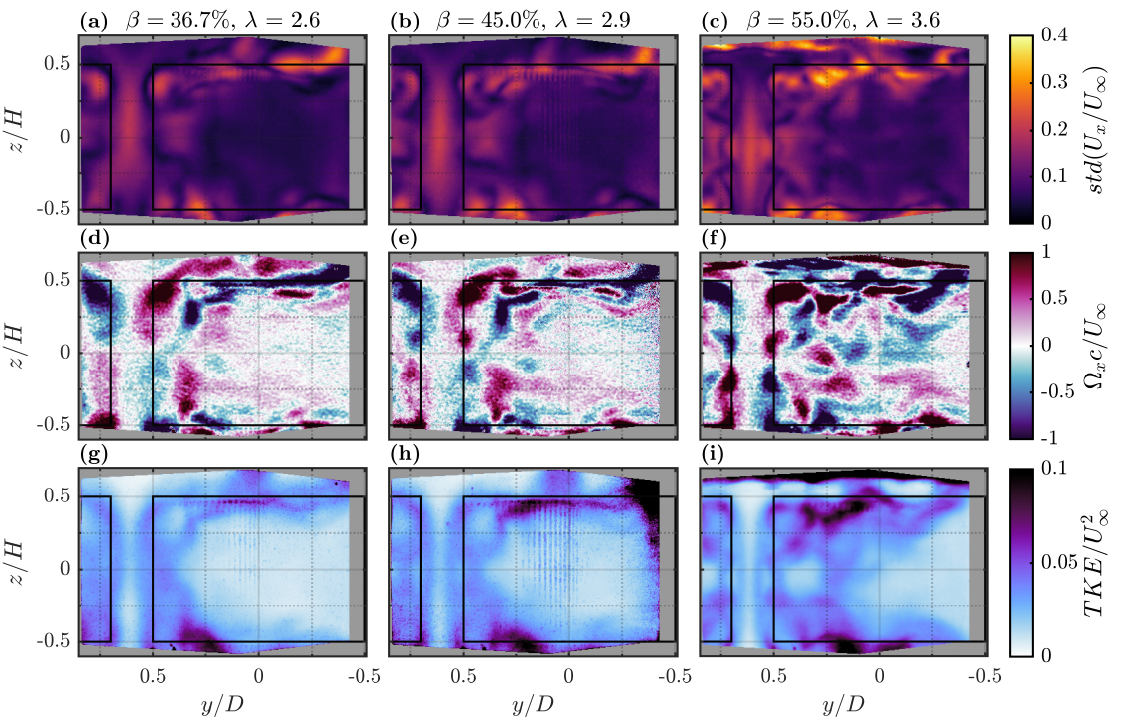}
      {\phantomsubcaption\label{fig:piv_stdev_36}
      \phantomsubcaption\label{fig:piv_stdev_45}
      \phantomsubcaption\label{fig:piv_stdev_55}
      \phantomsubcaption\label{fig:piv_vorticity_36}
      \phantomsubcaption\label{fig:piv_vorticity_45}
      \phantomsubcaption\label{fig:piv_vorticity_55}
      \phantomsubcaption\label{fig:piv_tke_36}
      \phantomsubcaption\label{fig:piv_tke_45}
      \phantomsubcaption\label{fig:piv_tke_55}
      }
     \caption{Examples of non-idealities in the measured flow fields that are not present in the flow fields assumed by linear momentum theory, shown for the streamwise flow through a plane $0.6D$ downstream of the array operating at $\beta = 36.7\%$, $45.0\%$, and $55.0\%$ and optimal $\lambda$.
     \subref{fig:piv_stdev_36}--\subref{fig:piv_stdev_55} Standard deviation of the measured streamwise velocity field, calculated as the standard deviation of the phase-average streamwise velocity field across all phases.
     \subref{fig:piv_vorticity_36}--\subref{fig:piv_vorticity_55} Time-averaged streamwise vorticity at the same location and conditions, calculated from the phase-averaged velocity fields and averaged across all phases.
     \subref{fig:piv_tke_36}--\subref{fig:piv_tke_55} Time-averaged turbulent kinetic energy (TKE) measured at the same location and conditions.
     To account for the inherent periodicity of cross-flow turbine wakes, the TKE field shown is obtained by calculating the TKE at each phase from the individual frames collected at that phase, then averaging the resulting fields across all phases.}
     \label{fig:pivVariability}
 \end{figure*}

The quantitative agreement between the modeled and measured bypass velocities is remarkable given the multiple violations of LMADT in the near-wake flow field.
First, unlike the uniform and one-dimensional flow assumed by the LMADT model, the flow in the near-wake varies in space and time within the bypass and wake regions (\cref{fig:piv_stdev_36,fig:piv_stdev_45,fig:piv_stdev_55}).
Second, much of the flow is rotational (\cref{fig:piv_vorticity_36,fig:piv_vorticity_45,fig:piv_vorticity_55}): dynamic stall vortices are shed into the region between the two turbines and strong tip vortices form along the top and bottom of the turbine projected area. 
The asymmetric vorticity produced by these sources drives flow toward the center of each turbine, contributing to wake asymmetry \citep{bachant_characterising_2015,ryan_threedimensional_2016}. 
Third, as suggested by the velocity profiles in \cref{fig:pivlineBetaUinf} and the turbulent kinetic energy (TKE) in the near-wake (\cref{fig:piv_tke_36,fig:piv_tke_45,fig:piv_tke_55}), mixing occurs between the bypass flow and the turbine wake, consistent with the shear layer in a co-flowing turbulent jet.
In contrast, the LMADT model assumes that the core and bypass flows do not mix until further downstream (\cref{fig:lmadDiagram}).
This near-wake mixing necessarily reduces the bypass velocity, which may explain why the modeled bypass velocity is slightly higher than the measured bypass velocity at the array centerline (\cref{fig:pivlineScaled}).
We note that, as shown in supplemental material \citep{hunt_supplemental}, the flow field $1.0D$ upstream of the array is largely uniform, irrotational, and TKE free, such that the upstream flow does not meaningfully contribute to the variability highlighted in \Cref{fig:pivVariability}.
However, we note that the time-average streamwise vorticity and the TKE in the bypass region are low relative to other areas of the near-wake, such that Bernoulli's equation should be approximately valid along a streamline that connects the upstream flow and the bypass region.
This may explain why the measured bypass velocities are well predicted by LMADT.

\subsection{Comparison between Experimental and Modeled Free Surface Deformation}

\begin{figure}
    \centering
    \includegraphics[width=\textwidth]{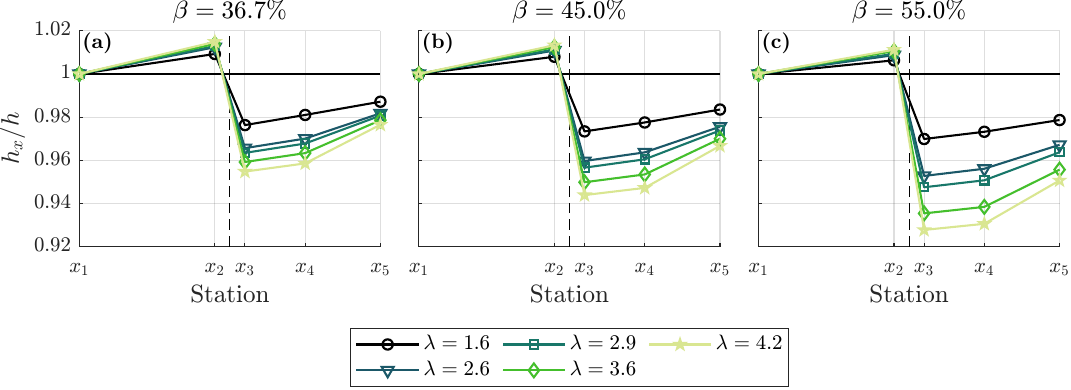}
      {\phantomsubcaption\label{fig:fstTheory_36}
      \phantomsubcaption\label{fig:fstTheory_45}
      \phantomsubcaption\label{fig:fstTheory_55}
      }
    \caption{Modeled free-surface profile at various stations along the channel predicted using the open-channel LMADT model of \citet{houlsby_application_2008} and the measured $C_T$, $h$, and $U_{\infty}$ for \subref{fig:fstTheory_36} $\beta = 36.7\%$, \subref{fig:fstTheory_45} $\beta = 45.0\%$, and \subref{fig:fstTheory_55} $\beta = 55.0\%$ for various $\lambda$. The dashed black line indicates the relative location of the array, represented in the linear momentum model as a thin actuator disk.}
    \label{fig:fstTheory}
\end{figure}

For comparison with the measured free surface profiles in \cref{fig:fstProfiles}, the theoretical free surface profiles are shown for the same $\beta$ and $\lambda$ in \cref{fig:fstTheory}.
As noted in \cref{sec:modelOverview}, we emphasize that the exact locations of $x_2 - x_4$ relative to the physical array are ambiguous and that $x_5$ is likely further downstream than can be measured given the finite flume length.
Consequently, the locations of $x_1$ through $x_5$ relative to the actuator disk model of the array shown in \cref{fig:fstTheory} are qualitative.
With these caveats, we consider the trends in measured free surface deformation relative to model predictions.
For all $\beta$, the modeled water depth just upstream of the actuator disk ($h_2 / h$) increases continuously with $\lambda$.
Similarly, the immediate static head drop across the actuator disk ($(h_3 - h_2)/h$) and the total free surface drop along the channel ($\Delta h / h$) increase continuously with both $\lambda$ and $\beta$.

This analytical result is consistent with a direct relationship between the pressure drop across the array (i.e., the thrust coefficient) and a drop in the free surface across the array, but conflicts with measured free surface profiles in \cref{fig:fstProfiles}.
Although the measured $C_{T}$ increases continuously with $\lambda$ at all $\beta$, the measured free surface drop across the turbines at $\beta = 36.7\%$ (\Cref{fig:fst_36}) and $45.0\%$ (\Cref{fig:fst_45}) does not depend significantly on $\lambda$ for $\lambda \geq 2.6$.
This may suggest that LMADT is not fully descriptive of the evolution of the free surface in the vicinity of high blockage turbines, although the limited spatial and temporal resolution of the instrumentation employed precludes a more comprehensive characterization of the free surface dynamics in this data set.
Experimental trends are consistent with modeled free surface deformation for $\beta = 55.0\%$, which suggests that smaller-scale spatial variability could be masking expected trends at lower $\beta$.

\subsection{Bluff-Body Scaling and Blockage Correction of Performance Data}
\label{sec:lmadScaling}

Given that open-channel LMADT is descriptive of the measured bypass flow across a range of array operating conditions, we turn to the question of whether LMADT also describes the measured array performance.
In prior work, LMADT is commonly used in analytical blockage corrections to estimate unconfined performance from measurements or simulations of confined operation \citep[e.g.,][]{bahaj_power_2007, chen_blockage_2011, kinsey_impact_2017, ross_experimental_2020}.
However, since a perfect blockage correction would be expected to collapse the $C_P$ and $C_T$ measured at different blockage ratios (e.g., \cref{fig:arrayPerf}) onto common unconfined $C_P-\lambda$ and $C_T-\lambda$ curves, such corrections imply not just a connection between confined and unconfined performance, but rather a broader self-similarity in turbine performance across blockage ratios.
Here, we explore such self-similar aspects of array performance with a focus on the significance of the bypass flow.

\begin{figure*}
     \centering
     \includegraphics[width=\textwidth]{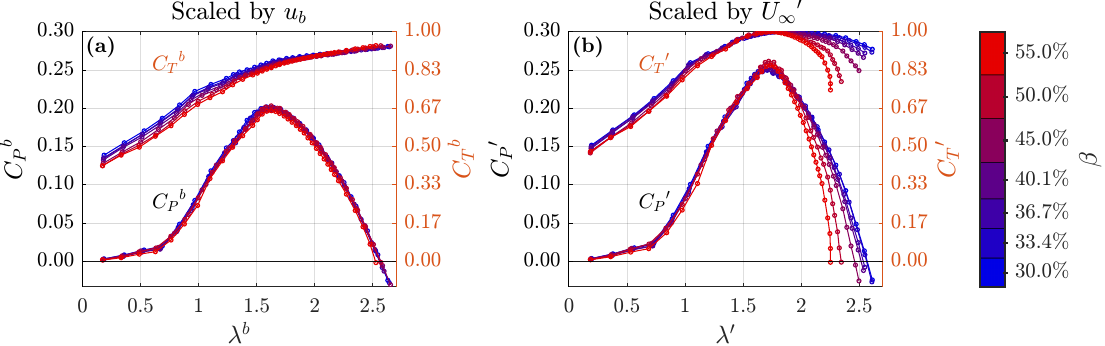}
     {\phantomsubcaption\label{fig:lmadScalings_bypass}
      \phantomsubcaption\label{fig:lmadScalings_UPrime}
    }
    
     \caption{Time-averaged array performance from \cref{fig:arrayPerf} scaled by \subref{fig:lmadScalings_bypass} the analytical bypass velocities obtained using the model of \citet{houlsby_application_2008} and \subref{fig:lmadScalings_UPrime} scaled by the unconfined freestream velocities from the model of \citet{houlsby_application_2008}. \subref{fig:lmadScalings_bypass} can be interpreted as a Maskell-inspired bluff-body blockage correction based on \citeauthor{houlsby_application_2008}'s model, whereas \subref{fig:lmadScalings_UPrime} represents a conventional Glauert-derived blockage correction based on \citeauthor{houlsby_application_2008}'s model, with the colorbar indicating the blockage ratio of the confined data.}
     \label{fig:LMADT_scalings}
 \end{figure*}

To investigate the relationship between the bypass flow and array performance, we apply a blockage correction method introduced by \citet{whelan_freesurface_2009} and inspired by the bluff-body theory of \citet{maskell_ec_theory_1963} to scale the experimental array performance by the corresponding bypass velocities modeled in LMADT:
\begin{equation}
    {C_P}^b = C_{P} {\left( \frac{U_{\infty}}{u_b} \right)}^3 ,
    \label{eq:cpPrime}
\end{equation}
\begin{equation}
    {C_T}^b = C_{T} {\left( \frac{U_{\infty}}{u_b} \right)}^2 ,
    \label{eq:ctPrime}
\end{equation}
\begin{equation}
    {\lambda}^b = \lambda\left( \frac{U_{\infty}}{u_b} \right) .
    \label{eq:TSRPrime}
\end{equation}
\noindent As shown in \cref{fig:lmadScalings_bypass}, the resulting ${C_P}^b-{\lambda}^b$ and ${C_T}^b-{\lambda}^b$ curves collapse across the tested blockage and tip-speed ratios, indicating that array power, thrust, and their relationship to rotation rate scale with the bypass velocity for these experimental conditions.
Consequently, one may interpret $u_b / U_{\infty}$ as a similarity variable for array performance that is well-predicted by LMADT.
Some residual variation in the bypass-scaled thrust curves is observed, particularly at lower $\lambda^b$ (maximum $\Delta {C_T}^b \approx 0.06$), but the bypass-scaled power curves exhibit excellent collapse at all $\lambda^b$ (maximum $\Delta {C_P}^b \approx 0.016$). 
As $u_b$ is calculated using the measured $C_{T}$ in \cref{eq:uwBlockage,eq:uwThrust}, it is not surprising that $C_{T}$ scales with $u_b$. 
However, given that measurements of $C_{P}$ are not involved in the calculation of $u_b$, the collapse in the ${C_P}^b-{\lambda}^b$ curves is notable, especially since the rotation, torque, and lateral forces experienced by the cross-flow turbines are all neglected in LMADT.

The hydrodynamics that underlie the connection between the bypass velocity and array performance are well-described by \citet{nishino_effects_2012}.
To summarize their conclusions, as $\beta$ increases, the velocity of the fluid bypassing the array increases (\cref{fig:pivlineBetaUinf,fig:lmadVels_bypass}).
Since energy is mostly conserved in the bypass flow, this requires an increased pressure drop in the bypass.
To satisfy pressure equilibrium between the core and bypass flows both far upstream and far downstream, the pressure drop in the core flow through the turbines must also increase.
In an open-channel flow, this pressure drop manifests as a free surface drop along the channel. 
As $\beta$ increases, the increased pressure drop across the array accelerates the flow through the turbines, which increases power and thrust at higher blockage.

This connection between the bypass velocity and array performance parallels the bluff-body theory of \citet{maskell_ec_theory_1963}.
Maskell hypothesized that, for a bluff body in a confined flow, there exists an unconfined inflow condition that would result in the same pressure distribution over the body, and therefore, the same drag.
Maskell's subsequent analysis considers a stream surface around the bluff body that relates the undisturbed flow far upstream to the flow just outside the wake, which is treated as an axisymmetric quiescent ``bubble'' with uniform pressure bounded by a uniform and axisymmetric flow.
Since the pressure and velocity on this stream surface are related via conservation of energy, then invariant pressure with blockage requires that the velocity outside the wake must also be constant between the confined and unconfined conditions.
Therefore, Maskell's hypothesis stipulates that the drag force on a bluff body in confined flow is invariant with the blockage ratio if the bypass velocity remains constant, although Maskell does not explicitly use the term ``bypass velocity'' to describe this relationship.
Maskell demonstrated through experiments that this model was descriptive of the drag on square flat plates at relatively low blockage ratios ($\beta \leq 10\%$), and leveraged this model as the basis of a blockage correction for predicting the drag on a bluff body in unconfined flow given thrust measurements on the same body in a confined flow.
As the wake characteristics of wind and water turbines resemble those of a bluff body under certain conditions \citep{medici_measurements_2006, chamorro_reynolds_2012, araya_transition_2017}, several subsequent studies have extended Maskell's theory to correct turbine performance to unconfined conditions.
While some have directly applied Maskell's flat plate blockage correction to turbine performance data \citep{zilic_de_arcos_numerical_2020, zhang_analysis_2023}, others have adapted this theory to specific turbine designs through analogous empirical relationships \citep{alexander_wind_1978, ross_wind_2011, jeong_blockage_2018, zhang_analysis_2023}.

\citet{whelan_freesurface_2009} were the first to combine the core idea of Maskell's theory---that the thrust on a bluff body in confined flow scales with the bypass flow---with LMADT to develop a blockage correction for ``highly loaded'' turbines.
Rather than directly implementing Maskell's blockage correction, \citeauthor{whelan_freesurface_2009} utilized an open-channel LMADT model to calculate the expected $u_b$ from axial-flow turbine thrust measurements at high blockage ($\beta = 64\%$), and utilized \cref{eq:ctPrime,eq:TSRPrime} to scale the measured thrust coefficients and tip-speed ratios by this bypass velocity.
The resulting ${C_T}^b-{\lambda}^b$ data agreed closely with the unconfined thrust coefficients predicted for the same turbine simulated using blade element momentum theory.
\citet{kinsey_impact_2017} applied the same correction to CFD-simulated thrust data for a cross-flow turbine at $\beta = 13\%$, $25\%$, and $51\%$ with $u_b$ calculated via closed-channel LMADT.
The resulting ${C_T}^b-\lambda^b$ curves yielded better agreement with the simulated unconfined thrust coefficients than was obtained with the popular \citet{barnsley_final_1990} blockage correction.
While \citeauthor{whelan_freesurface_2009} left open the question of whether power could be similarly corrected, \citet{ross_experimental_2020} introduced \cref{eq:cpPrime} as an analogous scaling for $C_P$.
Using the LMADT model of \citet{houlsby_application_2008} to calculate $u_b$, \citeauthor{ross_experimental_2020} scaled the corresponding performance data for both an axial-flow turbine ($\beta = 2\%$ and $35\%$) and cross-flow turbine ($\beta = 3\%$ and $36\%$).
However, in contrast to \citeauthor{whelan_freesurface_2009} and \citeauthor{kinsey_impact_2017}, at tip-speed ratios beyond optimal ($\lambda > 5$ for axial-flow, $\lambda > 1.85$ for cross-flow) \citeauthor{ross_experimental_2020} observed moderately poor agreement between the measured low-blockage turbine performance and ${C_P}^b$ and ${C_T}^b$ calculated from high-blockage turbine performance.
Relative to prior work, the bluff-body corrected power and thrust coefficients in \cref{fig:lmadScalings_bypass} exhibit the strongest collapse across blockages observed to date, and highlight similarities between array hydrodynamics and bluff-body dynamics for a wide range of high-blockage operating conditions.

Since the bypass scalings in \cref{eq:cpPrime,eq:ctPrime,eq:TSRPrime} are typically interpreted as a Maskell-inspired bluff-body blockage correction, it is relevant to compare the bluff-body corrected performance (\cref{fig:lmadScalings_bypass}) to that obtained via a ``standard'' LMADT blockage correction based on the theory of \citet{glauert_airplane_1935}, which assumes that the thrust on the turbine responds to the velocity through the turbine ($u_t$) rather than the bypass velocity.
Using the values of $u_t$ obtained from \citeauthor{houlsby_application_2008}'s \citep{houlsby_application_2008} model (\cref{eq:ut}), the unconfined freestream velocity, ${U_{\infty}}'$, that corresponds to the same $u_t$ and dimensional thrust in unconfined flow as for the confined system is given as

\begin{equation}
    {U_{\infty}}' = \frac{U_{\infty} \left( (u_t/U_{\infty})^2 + C_{T}/4 \right)}{u_t / U_{\infty}} \ .
    \label{eq:UPrime}
\end{equation}

\noindent The unconfined power coefficient (${C_P}'$), thrust coefficient (${C_T}'$), and tip-speed ratio ($\lambda'$) are then calculated using \cref{eq:cpPrime,eq:ctPrime,eq:TSRPrime} with ${U_{\infty}}'$ in place of $u_b$.
As shown in \cref{fig:lmadScalings_UPrime}, up to ${C_T}' \approx 1$, the ${C_P}'-\lambda'$ and ${C_T}'-\lambda'$ curves exhibit similar collapse across blockages to the bluff-body correction (maximum $\Delta {C_P}' \approx 0.018$; maximum $\Delta {C_T}' \approx 0.04$), with the values of ${C_P}'$ and ${C_T}'$ generally exceeding those of ${C_P}^b$ and ${C_T}^b$ at comparable re-scaled tip-speed.
However, for LMADT in unconfined flow, the thrust coefficient cannot exceed unity as this requires reversed flow in the turbine wake, which violates conservation of mass in the streamtube of flow through the turbine.
Thus, beyond $\lambda'$ corresponding to ${C_T}' \approx 1$, ${C_T}'$ decreases non-physically and the unconfined efficiency and thrust coefficient curves no longer collapse.
The inability of LMADT to model turbines in unconfined flow with high thrust coefficients is a well-documented limitation of this theory \citep{whelan_freesurface_2009, ross_experimental_2020, liew_unified_2024a} and indicates that a Glauert-derived blockage correction based on \citeauthor{houlsby_application_2008}'s model is not applicable over the full range of operating conditions in these experiments.
Given this constraint, a key advantage of the Maskell-inspired bluff-body blockage correction over a conventional Glauert-derived correction is the ability to describe the performance of high-blockage turbines over a wider range of confined $C_{T}$---and thus a wider range of $\beta$ and $\lambda$---since this scaling does not rely on the relationship between $C_T$ and $u_w$ in unconfined flow.

\begin{figure*}
     \centering
     \includegraphics[width=\textwidth]{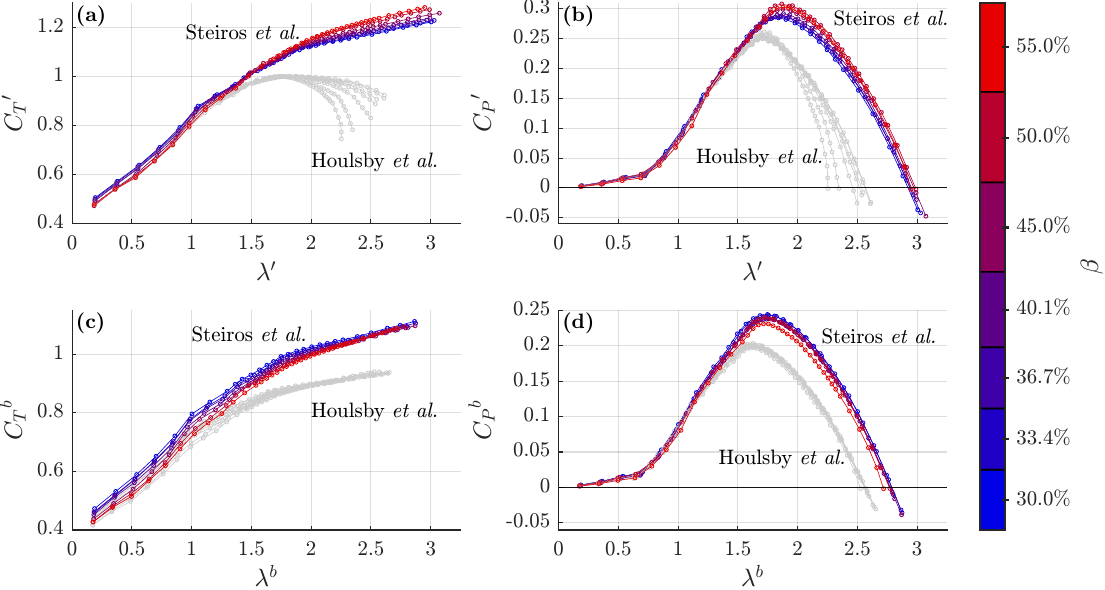}
     {\phantomsubcaption\label{fig:compare_ct_UPrime}
      \phantomsubcaption\label{fig:compare_cp_UPrime}
      \phantomsubcaption\label{fig:compare_ct_bypass}
      \phantomsubcaption\label{fig:compare_cp_bypass}
    }
    
     \caption{Comparison between the model of \citet{steiros_analytical_2022} (curves shown in color) and the model of \citet{houlsby_application_2008} (curves shown in gray, reproduced from \cref{fig:LMADT_scalings}) for Glauert-derived and bluff-body blockage corrections to unconfined conditions for the time-averaged array performance in \cref{fig:arrayPerf}.
     \subref{fig:compare_ct_UPrime} and \subref{fig:compare_cp_UPrime} show  Glauert-derived blockage corrections of $C_T$ and $C_P$, respectively, using the values of ${U_{\infty}}'$ calculated using each each model.
     \subref{fig:compare_ct_bypass} and \subref{fig:compare_cp_bypass} show Maskell-inspired bluff body blockage corrections of $C_T$ and $C_P$, respectively, using the values of $u_b$ calculated using each model.}
     \label{fig:houlsby_steiros_compare}
 \end{figure*}

Since the potential flow model proposed by \citet{steiros_analytical_2022} also relaxes the constraint on ${C_T}'$, the question naturally arises as to whether this model---either coupled with Glauert's theory as presented in \citet{steiros_analytical_2022} or with an adaptation of Maskell's theory following \citet{whelan_freesurface_2009}---can provide a better prediction of flow physics and performance self-similarity for these experiments.
A summary of \citeauthor{steiros_analytical_2022}'s model and associated blockage correction as applied to the experimental data is provided in supplemental material \citep{hunt_supplemental}.
As shown in \cref{fig:compare_ct_UPrime,fig:compare_cp_UPrime}, \citeauthor{steiros_analytical_2022}'s blockage correction improves on the weaknesses of traditional LMADT-based blockage corrections at high $\lambda'$ while providing similar ${C_P}'$ and ${C_T}'$ to that obtained with \citeauthor{houlsby_application_2008}'s model at lower $\lambda'$.
\citeauthor{steiros_analytical_2022}'s model allows ${C_T}'$ to exceed unity, such that physically meaningful solutions for ${C_T}'$ and ${C_P'}$ are obtained at all tip-speed ratios.
While the ${C_T}' - \lambda'$ and ${C_P}' - \lambda'$ curves obtained using the \citeauthor{steiros_analytical_2022}'s model generally collapse across $\beta$, the curves diverge slightly at higher $\lambda'$.
This is consistent with observations by \citeauthor{steiros_analytical_2022} in their validation experiments involving drag on flat plates with various porosities and blockage ratios.
Specifically, \citeauthor{steiros_analytical_2022} noted that the model accuracy decreased with increasing blockage ratio and decreasing plate porosity (analogous to an increase in tip-speed ratio here).

As the model of \citet{steiros_analytical_2022} also provides an estimate for the bypass velocity, a Maskell-inspired bluff-body blockage correction can also be applied from this model's results.
As shown in \cref{fig:compare_ct_bypass,fig:compare_cp_bypass}, \citeauthor{steiros_analytical_2022}'s model yields qualitatively similar bypass scalings to that of \citeauthor{houlsby_application_2008}'s model, and results in higher values of ${C_T}^b$ and ${C_P}^b$ as $\lambda^b$ increases due to slightly lower estimates for $u_b$.
However, the ${C_P}^b-\lambda^b$ curves collapse more tightly for \citeauthor{houlsby_application_2008}'s model. 
Overall, the scalings in \cref{fig:LMADT_scalings} demonstrate the descriptiveness of a Maskell-inspired bluff-body blockage correction for array performance is not exclusive to a particular analytical model type.
Notably, despite neglecting free surface deformation, \citeauthor{steiros_analytical_2022}'s model is still descriptive of the experimental bypass velocities, although the measured wake velocity for this cross-flow turbine array remains poorly predicted (see supplemental material \citep{hunt_supplemental}).
Because \citeauthor{steiros_analytical_2022}'s model does not provide a materially better bluff-body scaling for performance (\cref{fig:compare_ct_bypass,fig:compare_cp_bypass}), we continue with the open-channel LMADT model of \citeauthor{houlsby_application_2008} for the remainder of this study, but wish the emphasize the former's utility for situations where a relatively wide range of thrust coefficients are expected.

In the present work, measurements of array performance at $\beta \approx 0\%$ are not available, such that the accuracy of the bypass scaling in \cref{eq:cpPrime,eq:ctPrime,eq:TSRPrime} as a bluff-body correction to unconfined performance cannot be assessed directly.
Nonetheless, we expect this Maskell-inspired scaling relationship to describe self-similar array performance primarily under conditions where turbine dynamics resemble those of a bluff body.
While the results in this study demonstrate this is likely the case for this particular rotor geometry at $\beta = 30 - 55\%$, \citet{araya_transition_2017} showed that cross-flow turbine wakes at even lower blockage ($\beta = 20\%$) resemble those of a solid cylinder as the ``dynamic solidity'' of the turbine increases via increasing tip-speed ratio or rotor solidity.
Furthermore, related work by the authors of the present study \citep{hunt_performance_2025} evaluated the performance of this experimental array for 90 different rotor geometries (consisting of different combinations of blade count, chord-to-radius ratio, and preset pitch angle) at $\beta = 35\%-55\%$, and found that ${C_T}^b$ becomes increasingly self-similar across blockage ratios and geometries as dynamic solidity increases \citep{hunt_performance_2025}.
Consequently, we hypothesize that the range of blockage ratios over which bluff-body scaling can capture self-similar aspects of array performance depends on the dynamic solidity of the turbine or array.
This may explain why the ${C_T}^b - \lambda^b$ curves in \cref{fig:compare_ct_bypass} increasingly collapse as $\lambda$ (and thus, $C_T$) increases, which is consistent with \citeauthor{whelan_freesurface_2009}'s \citep{whelan_freesurface_2009} original framing of \cref{eq:ctPrime,eq:TSRPrime} as a blockage correction for highly-loaded turbines.
Similarly, the poorer agreement between bluff-body scaled performance at $\beta = 36\%$ and low-blockage performance at $\beta = 3\%$ observed by \citet{ross_experimental_2020} (who used a cross-flow turbine with similar solidity to this study) suggests limitations of bluff-body scaling for describing low-blockage or unconfined performance.
In these regimes, the blockage correction proposed by \citet{steiros_analytical_2022} (\cref{fig:compare_ct_UPrime,fig:compare_cp_UPrime}) has clear advantages for high-solidity turbines.


\section{Generalized Blockage Adjustment}
\label{sec:lmadForecasting}

The self-similar array performance yielded by Maskell-inspired bluff-body scaling provides a pathway for predicting performance at a given blockage ratio using measurements or simulations at another.
For example, such information would be useful for describing time-varying performance in a tidal channel with changing water depth and, therefore, blockage.
However, only a limited number of prior studies have ``corrected'' performance to non-zero blockages.
\citet{whelan_freesurface_2009} applied their Maskell-inspired bluff-body blockage correction in reverse to predict the thrust coefficients of a turbine at $\beta = 64\%$ from unconfined thrust calculated from BEM, but did not detail the procedure or quantify prediction accuracy.
\citet{kinsey_impact_2017} proposed an empirical method for predicting confined performance based on their observation that both the power and thrust coefficients exhibit approximately linear relationships with $\beta$ at constant $\lambda$ (similar to that in \cref{fig:perfVBlockage}).
Specifically, at several $\lambda$, \citeauthor{kinsey_impact_2017} developed a linear regression for confined performance based on simulations at $\beta = 20\%$ and estimated performance at $\beta = 0\%$ using the LMADT blockage correction of \citet{barnsley_final_1990}.
Performance at other $\beta$ was then obtained by interpolation or extrapolation.
Although this method was able to accurately predict the power and thrust coefficients at $\beta = 10\%$, predictions were less accurate for extrapolation to $\beta \geq 50\%$ where the $C_T-\beta$ and $C_P-\beta$ relationships are less linear.
Finally, the blockage correction method developed by \citet{steiros_analytical_2022} can be used to ``correct'' turbine performance to arbitrary blockage by simply specifying a non-zero target value of $\beta$ in their equation set.
\citeauthor{steiros_analytical_2022} evaluated their method using numerical simulations of an axial flow-turbine by correcting simulated performance at $\beta = 20\%$ to $\beta = 3\%$, and comparing to the performance of the same turbine directly simulated at $\beta = 3\%$.

Here, a method for adjusting performance across blockage ratios using \citeauthor{houlsby_application_2008}'s LMADT model and a Maskell-inspired bluff-body blockage correction is described, and its efficacy evaluated over a range of $\beta$ and $\lambda$.
We suspect that this approach is similar to that employed by \citet{whelan_freesurface_2009} and may be interpreted as a generalized bluff-body blockage correction that allows for transformation of known turbine performance at one blockage ratio ($\beta_1$) to some other arbitrary blockage ratio ($\beta_2$).
A separate evaluation of the generalized blockage correction of \citet{steiros_analytical_2022} as applied to the experimental data is provided as supplemental material \citep{hunt_supplemental}.
Consider a turbine in a channel (or an array of equally-spaced turbines spanning the channel width) with known performance at blockage ratio $\beta_1$ and Froude number $Fr_{h}$.
At a given operating condition (e.g., $\lambda_1$) the turbine's thrust coefficient is $C_{T,1}$ and bypass velocity is $u_{b,1}$.
As for Glauert-derived blockage corrections, for the same turbine or array operating at a different blockage ratio, $\beta_2$, assume that there is a freestream velocity, $U_{\infty,2}$, that yields the same dimensional thrust on the turbine as at $\beta_1$ and $U_{\infty,1}$:
\begin{equation}
    T_1  = T_2 \ .
    \label{eq:equalThrust}
\end{equation}
%
\noindent Therefore, 
 \begin{equation}
    C_{T,2} = C_{T,1}{\left( \frac{U_{\infty,1}}{U_{\infty,2}} \right)}^2 \ .
    \label{eq:ct2From1}
\end{equation}
\noindent Since the thrust on the turbine scales with the bypass velocity in a Maskell-inspired approach, equal thrust at $\beta_1$ and $\beta_2$ requires equal bypass velocity:
\begin{equation}
    u_{b,1} = u_{b,2} = u_b \ .
    \label{eq:equalBypass}
\end{equation}
\noindent 
Therefore, $C_{T,2}$ can be calculated from \cref{eq:ct2From1} if the value of $U_{\infty,2}$ that yields $u_b$ at $\beta_2$ can be determined. In other words, $C_{T,1}$ and $C_{T,2}$ are connected through a common value of ${C_T}^b$ when a Maskell-inspired bluff-body blockage correction is applied as
\begin{equation}
    {C_T}^b
    = C_{T,1}{\left( \frac{U_{\infty,1}}{u_b} \right)}^2
    = C_{T,2}{\left( \frac{U_{\infty,2}}{u_b} \right)}^2
    \ ,
\end{equation}
\noindent which is analogous to Equation (1) in \citet{maskell_ec_theory_1963}.
Since rotor power also scales with the bypass velocity, $C_{P,2}$ is given as
\begin{equation}
    C_{P,2} = C_{P,1}{\left( \frac{U_{\infty,1}}{U_{\infty,2}} \right)}^3 \ .
    \label{eq:cp2From1}
\end{equation}
\noindent Finally, assuming constant rotation rate (i.e., $\omega_1 = \omega_2$),
\begin{equation}
    \lambda_{2} = \lambda_{1}{\left( \frac{U_{\infty,1}}{U_{\infty,2}} \right)} \ .
    \label{eq:TSR2From1}
\end{equation}
\noindent In \cref{eq:ct2From1,eq:cp2From1}, it is implicitly assumed that $\rho_1$ = $\rho_2$ (and thus, $\nu_1$ = $\nu_2$), which implies a difference in the freestream $Re_D$ (\cref{eq:ReD}) between $\beta_1$ and $\beta_2$.
However, the Reynolds number based on the bypass velocity ($u_b D / \nu$) remains constant.
We note that the inability of LMADT to describe $u_w$ and, presumably, $u_t$ for relatively high confinement is not a limitation here, since LMADT is able to accurately model the relationship between $\beta$, $C_T$, and $u_b / U_{\infty}$.

\begin{figure}
     \centering
     \includegraphics[width=\textwidth]{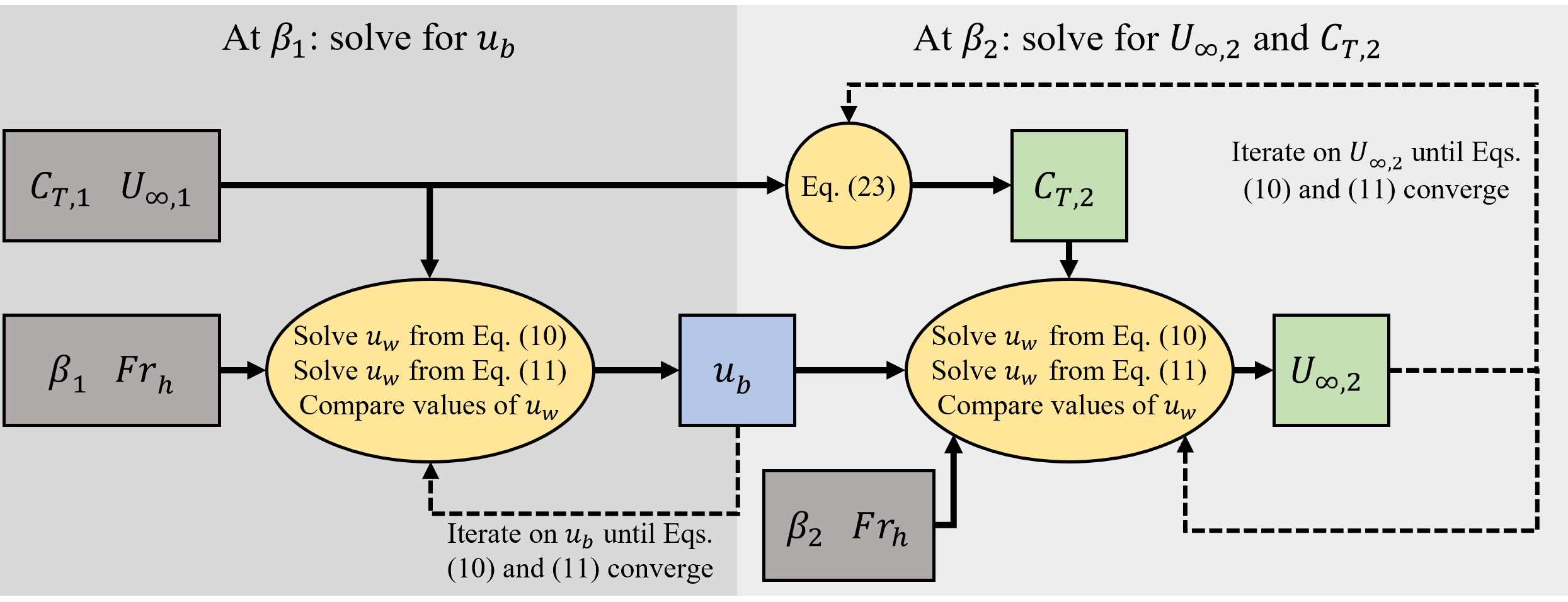}
     \caption{Flowchart describing the procedure for predicting performance at $\beta_2$ based on performance at $\beta_1$, assuming constant Froude number between $\beta_1$ and $\beta_2$. Gray boxes indicate known or specified inputs, yellow circles/ovals indicate equation solving, blue boxes indicate quantities associated with bluff-body scaling, and green boxes indicate the outputs at $\beta_2$. The dashed arrows indicate steps where iteration is required.}
     \label{fig:forecastingDiagram}
\end{figure}

Using $u_b$ as a bridge between $\beta_1$ and $\beta_2$, LMADT can be used to calculate $U_{\infty,2}$ from known quantities at $\beta_1$.
The procedure for calculating $U_{\infty,2}$ using the LMADT model of \citet{houlsby_application_2008} is outlined in \cref{fig:forecastingDiagram}.
Here, we assume that the Froude numbers at $\beta_1$ and $\beta_2$ are constant and subcritical as for the experimental data, although constant Froude number is not required.
First, $u_b$ is determined at $\beta_1$ by numerically solving \cref{eq:uwBlockage,eq:uwThrust}, using $C_{T,1}$, $U_{\infty,1}$, and $Fr_h$ as inputs.
Next, a reasonable value of $U_{\infty,2}$ is guessed (e.g., $U_{\infty,2} = U_{\infty,1}$), from which a corresponding $C_{T,2}$ is calculated from \cref{eq:ct2From1}.
\Cref{eq:uwBlockage,eq:uwThrust} are then solved separately for $u_w$ at $\beta_2$ using $u_b$, $Fr_h$, and the guesses for $U_{\infty,2}$ and $C_{T,2}$.
The resulting values of $u_w$ are compared, and this process is iterated with a new guess for $U_{\infty,2}$ until the value of $U_{\infty,2}$ minimizes the difference between \cref{eq:uwBlockage} and \cref{eq:uwThrust}.
With $U_{\infty,2}$ known, $C_{T,2}$, $C_{P,2}$ and $\lambda_2$ may be calculated from $C_{T,1}$, $C_{P,1}$, and $\lambda_1$ via \cref{eq:ct2From1,eq:cp2From1,eq:TSR2From1}.
Note, however, that $C_{T,1}$ is the only performance metric at $\beta_1$ that influences the value of $U_{\infty,2}$.

In the limiting case of $\beta_2 \rightarrow 0$, $U_{\infty,2} \rightarrow u_b$ since the freestream velocity and bypass velocity are equal in unconfined flow.
Consequently, \cref{eq:ct2From1,eq:cp2From1,eq:TSR2From1} revert to \cref{eq:cpPrime,eq:ctPrime,eq:TSRPrime} corresponding to a Maskell-inspired blockage correction (\cref{fig:lmadScalings_bypass}) with the implicit assumption that a bluff-body model is still representative of turbine dynamics at low blockage.
Conversely, as $\beta_2 \rightarrow 1$, the solution space is constrained by the requirements of equal bypass velocity and $Fr_h < 1$ at $\beta_1$ and $\beta_2$, such that physically meaningful subcritical solutions at $\beta_2$ do not exist for all $\beta_1$ and $C_{T,1}$.
The maximum value of $\beta_2$ at which a physical solution can be obtained depends on $C_{T,1}$, $\beta_1$, and the Froude numbers at $\beta_1$ and $\beta_2$; a visual representation of this limit for the experimental data is provided as supplemental material \citep{hunt_supplemental}.
However, for the performance data in this study, physically meaningful solutions exist for all practically-achievable blockages (i.e., $\beta_2 \leq 65\%$).

 \begin{figure*}
     \centering
     \includegraphics[width=\textwidth]{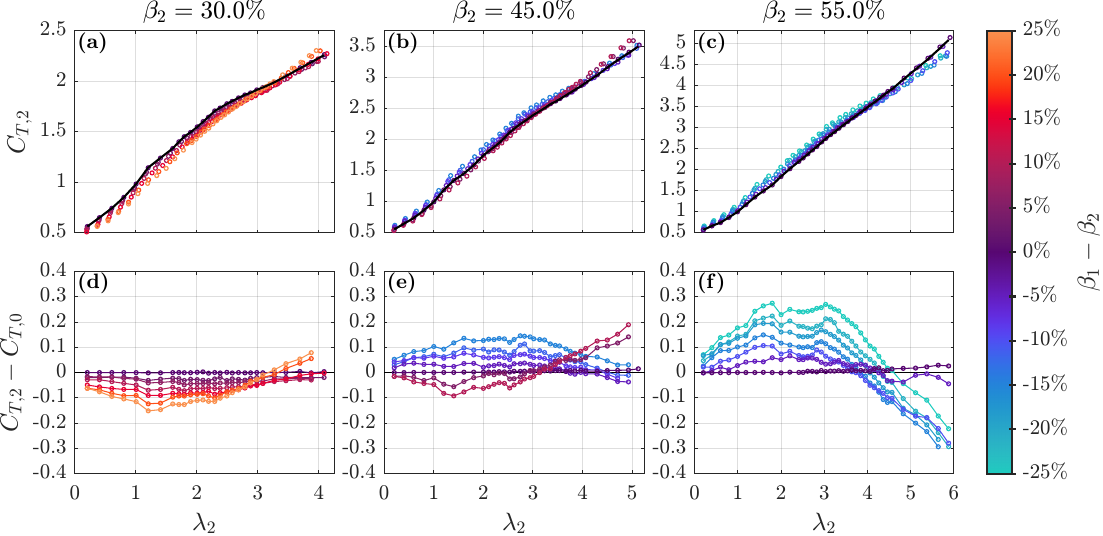}
     {\phantomsubcaption\label{fig:forecastCT_30}
      \phantomsubcaption\label{fig:forecastCT_45}
      \phantomsubcaption\label{fig:forecastCT_55}
      \phantomsubcaption\label{fig:forecastCT_30_error}
      \phantomsubcaption\label{fig:forecastCT_45_error}
      \phantomsubcaption\label{fig:forecastCT_55_error}
     }
     \caption{\subref{fig:forecastCT_30}-\subref{fig:forecastCT_55} Predicted thrust coefficient and \subref{fig:forecastCT_30_error}-\subref{fig:forecastCT_55_error} error between the predicted thrust coefficient and measured thrust coefficient ($C_{T,0}$, shown as the gray lines in \subref{fig:forecastCT_30}-\subref{fig:forecastCT_55}) for $\beta_2 = 30.0\%$, $45.0\%$, and $55.0\%$. Color indicates the difference between the blockage ratio of the performance data that was used to make each prediction ($\beta_1$) and the target blockage ratio ($\beta_2$). Note the different axes limits in \subref{fig:forecastCT_30}-\subref{fig:forecastCT_55}.}
     \label{fig:forecastCT}
 \end{figure*}

 \begin{figure*}
     \centering
     \includegraphics[width=\textwidth]{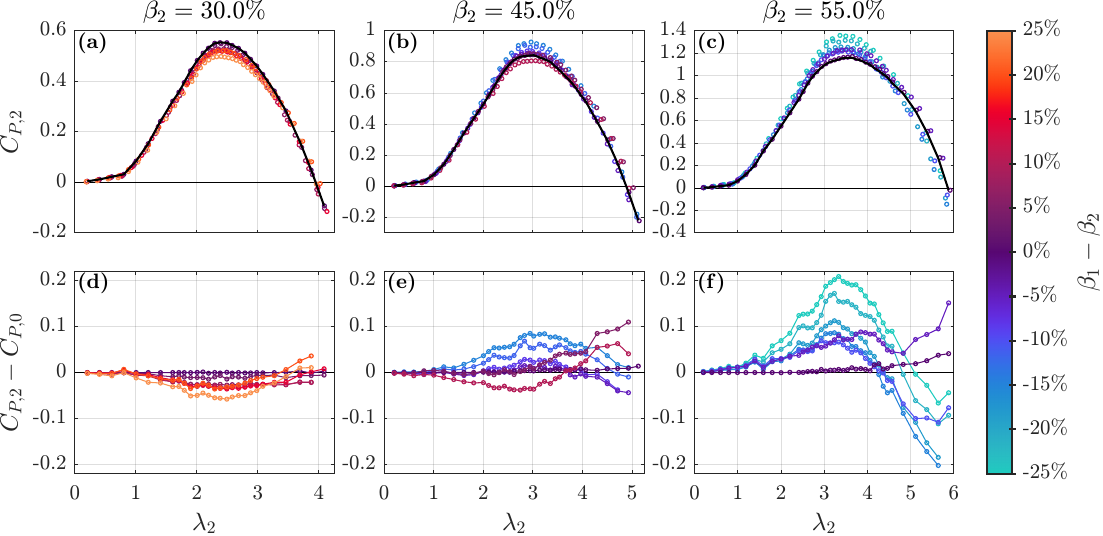}
     {\phantomsubcaption\label{fig:forecastCP_30}
      \phantomsubcaption\label{fig:forecastCP_45}
      \phantomsubcaption\label{fig:forecastCP_55}
      \phantomsubcaption\label{fig:forecastCP_30_error}
      \phantomsubcaption\label{fig:forecastCP_45_error}
      \phantomsubcaption\label{fig:forecastCP_55_error}
     }
     \caption{\subref{fig:forecastCP_30}-\subref{fig:forecastCP_55} Predicted efficiency and \subref{fig:forecastCP_30_error}-\subref{fig:forecastCP_55_error} error between the predicted efficiency and measured efficiency ($C_{P,0}$, shown as the black lines in \subref{fig:forecastCP_30}-\subref{fig:forecastCP_55}) for $\beta_2 = 30.0\%$, $45.0\%$, and $55.0\%$. Color indicates the difference between the nominal blockage ratio of the performance data that was used to make each prediction ($\beta_1$) and the target blockage ratio ($\beta_2$). Note the different axes limits for each $\beta_2$ in \subref{fig:forecastCP_30}-\subref{fig:forecastCP_55}.}
     \label{fig:forecastCP}
 \end{figure*}

To quantify the effectiveness of this blockage adjustment method, the measured array performance at $\beta_1 = 30.0\% - 55.0\%$ (\cref{fig:arrayPerf}) is used to predict array performance at $\beta_2 = 30.0\%$, $45.0\%$, and $55.0\%$, with the resulting $C_{T,2}$ and $C_{P,2}$ at each operating point compared to the experimentally measured performance (denoted as $C_{T,0}$ and $C_{P,0}$).
The predicted thrust coefficients at each $\beta_2$ and the corresponding error between the predicted and measured thrust coefficients are given in \cref{fig:forecastCT} as a function of the difference between $\beta_1$ and $\beta_2$.
The predictions are within $\sim\!0.2$ of measurements at $\beta_2 = 30.0\%$ and $45.0\%$ (\cref{fig:forecastCT_30_error,fig:forecastCT_45_error}) and within $\sim\!0.3$ at $\beta_2 = 55.0\%$ (\cref{fig:forecastCT_55_error}).
In general, the magnitude of the thrust prediction error increases with the difference between $\beta_1$ and $\beta_2$ (i.e., predictions further from the starting blockage are less accurate), and the absolute prediction error tends to grow as $\beta_2$ increases for the same difference between $\beta_1$ and $\beta_2$. In relative terms, these errors are appreciable at lower $\lambda_2$ where $C_T$ is relatively low, but decline for higher $\lambda_2$.
Since the foundation of this blockage adjustment method is a Maskell-inspired bluff-body blockage correction (\cref{eq:cpPrime,eq:ctPrime,eq:TSRPrime}), the differences in the predicted $C_{T,2}$ across $\beta_1$ are driven by differences in ${C_{T}}^b$ across $\beta_1$.
Thus, the shapes and relative positions of the ${C_{T}}^b-\lambda^b$ curves across $\beta_1$ (\cref{fig:lmadScalings_bypass}) dictate the shapes and relative positions of the resulting $C_{T,2}-\lambda_2$ curves when scaled to $\beta_2$ (\cref{fig:forecastCT_30,fig:forecastCT_45,fig:forecastCT_55}).
For the current data set, at low to moderate tip-speed ratios this corresponds to underpredicted $C_{T,2}$ if $\beta_1 > \beta_2$ and overpredicted $C_{T,2}$ if $\beta_1 < \beta_2$, except at relatively high $\lambda_2$ where this trend inverts.

The corresponding predicted power coefficients and prediction error are shown in \cref{fig:forecastCP}.
The predicted power coefficients are within $\sim\!0.05$ of measurements at $\beta_2 = 30.0\%$ (\cref{fig:forecastCP_30_error}), $\sim\!0.1$ at $45.0\%$ (\cref{fig:forecastCP_45_error}), and $\sim\!0.2$ at $\beta_2 = 55.0\%$ (\cref{fig:forecastCP_55_error}).
The greatest prediction error tends to occur near the performance peak and at tip-speed ratios well beyond the optimal operating point.
However, the tip-speed ratio corresponding to maximum performance is well-predicted (\cref{fig:forecastCP_30,fig:forecastCP_45,fig:forecastCP_55}).
Since only the thrust influences the outcome of the blockage adjustment procedure, trends in the $C_{P,2}-\lambda_2$ curves with $\beta_1$ resemble those for the ${C_{T}}^b-\lambda^b$ curves.

The proposed generalized blockage adjustment method combines linear momentum theory and the Maskell-inspired bluff-body scaling to enable predictions for confined turbine performance from experimental or simulation data.
However, since this method utilizes $u_b$ as a link between $\beta_1$ and $\beta_2$, the prediction accuracy depends on whether the bypass velocity is descriptive of array performance at both $\beta_1$ and $\beta_2$.
Consequently, the effectiveness of this blockage adjustment scheme outside of the tested range of $\beta$ is unknown, and the method is likely less accurate when turbine dynamics depart from a bluff-body model.
If there exists a transitional condition at which both a Maskell-inspired LMADT scaling (\cref{fig:lmadScalings_bypass}) and traditional Glauert-derived LMADT scaling (\cref{fig:lmadScalings_UPrime}) provide similar results, then a hybrid method that translates between these flow regimes might allow for predicting performance across a wider ranges of blockages.
However, the existence of such a transitional condition is speculative.

\section{Conclusions}
\label{sec:conclusions}

This research investigates the performance and near-wake flow fields of a cross-flow turbine array at the upper end of practically achievable blockage ratios and demonstrates the efficacy of theoretical frameworks for describing the dynamics of real turbines under these conditions. 
The performance and near-wake flow field of a two-turbine array were experimentally characterized over a wide range of operating conditions at array blockage ratios from $30\%$ to $55\%$, holding other non-dimensional flow parameters constant to isolate blockage effects.
The observed array performance is consistent with trends observed in prior work while providing a uniquely high-resolution view of the manner in which performance evolves with the blockage ratio at relatively high confinements.
This highlights relatively subtle trends, such as linear behavior of both the power and thrust coefficients with blockage for low-to-moderate $\beta$ and $\lambda$.
Similarly, the measured near-wake velocity fields quantitatively illustrate how the wake and bypass flows change with $\beta$ and $\lambda$.
This work also provides an openly-available set of experimental performance and flow-field data for cross-flow turbines at high blockage, which will aid in the validation of numerical simulations and reduced-order models that seek to examine confinement effects and inform the design of full scale systems that can exploit these physics.

A key outcome of this work is the demonstration that linear momentum actuator disk models can  quantitatively describe aspects of real turbine arrays in highly confined flows.
It is shown that, at a location $0.6$ turbine diameters downstream of the array, the velocity of the flow bypassing the array measured using PIV is well represented by the bypass velocity predicted from the open-channel LMADT model of \citet{houlsby_application_2008} using the measured array thrust coefficients and inflow conditions as inputs.
This result was confirmed for a wide range of blockages and tip-speed ratios, with the bypass velocity calculated from LMADT within approximately $8\%$ of the measured velocity between the turbines at the array centerline at all conditions.
The residual deviation between the modeled and measured bypass velocities may be a consequence of near-wake mixing that is neglected in LMADT.
Following \citet{whelan_freesurface_2009} and inspired by the bluff-body theory of \citet{maskell_ec_theory_1963}, scaling the array power and thrust coefficients by this analytical bypass velocity collapses these coefficients across the tested blockage and tip-speed ratios.
Similar results are obtained when the potential flow model of \citet{steiros_analytical_2022} is used in place of \citeauthor{houlsby_application_2008}'s LMADT model.
These results highlight similar dynamics between high-blockage turbine arrays and bluff bodies, and demonstrate that, despite their inherent simplicity, analytical approaches can effectively model aspects of the flow around non-ideal turbines in confined flows and provide insights into the salient hydrodynamics that govern turbine performance.
Leveraging the connection between array performance and the bypass velocity, a generalized analytical blockage adjustment method based on LMADT is presented by which known turbine performance at one confinement ($\beta_1$) can be used to predict turbine performance at a different confinement ($\beta_2$) with moderate accuracy.
This method provides a pathway for estimating turbine performance under confined conditions based on the dominant dynamics of high-blockage arrays.

Although this work focuses on cross-flow turbines, the characteristic fluid dynamics of confinement are not exclusive to this archetype, and the analytical models employed are agnostic to turbine geometry.
However, the design of the rotor is expected to influence how these characteristic dynamics change with blockage and the tip-speed ratio \citep{schluntz_effect_2015, hunt_performance_2025}, and the effectiveness of Maskell-inspired bluff-body scaling and the generalized blockage adjustment are likely dependent on turbine geometry and operating condition.
It is recommended that future work explore the broader application of these analytical methods to turbines at lower confinement, as well as to different turbine designs.
Additionally, although this work focuses on the time-averaged array performance and near-wake flow fields, the unsteady fluid dynamics of cross-flow turbines also inform their performance and the properties of the near-wake.
Future research should examine how the phase-resolved performance of cross-flow turbines develops with confinement, and how these dynamics influence the temporal and spatial evolution of the flow field in the near- and far-wake regions.

\section*{Acknowledgments}
This work was supported by the United States Advanced Research Projects Agency – Energy (ARPA-E) under award number DE-AR0001441. 
The authors would like to thank Gregory Talpey, Gemma Calandra, and Corey Crisp for their assistance in commissioning the high-confinement test-rig, as well as help with data collection.
Abigale Snortland, Hannah Ross, Jennifer Franck, and Mukul Dave are acknowledged for insightful conversations regarding the results in this work. We are grateful to the Alice C. Tyler Charitable Trust for upgrades to the experimental flume, including closed loop heating and cooling.
Finally, we appreciate the substantive and insightful comments by a pair of anonymous reviewers on the initial version of this manuscript.

\section*{Data Availability}

The data that support the findings of this study are openly available in the Dryad data repository at \url{https://doi.org/10.5061/dryad.dfn2z35b5} \citep{hunt_data_2024a}.
The MATLAB code used to implement the models of \citet{houlsby_application_2008} and \citet{steiros_analytical_2022} is available on GitHub at \url{https://github.com/aidan-hunt/turbine-confinement-models}.

\section*{Supplemental Material}

See supplemental material at \textcolor{red}{[URL will be inserted by publisher]} for lateral force measurements, estimates of hydraulic efficiency, application of Steiros \textit{et al.}'s model to the experimental data, a visualization of the solution space for the generalized blockage adjustment method, sensitivity analysis of the phase-averaged and time-averaged PIV flow fields, and additional figures of PIV flow fields at $0.6D$ and $1.5D$ downstream of the array and $1.0D$ upstream of the array.

\section*{Author contributions (CRediT)}

\textbf{A. Hunt:} Conceptualization, Investigation, Validation, Methodology, Formal Analysis, Visualization, Software, Writing - Original Draft and Review \& Editing.
\textbf{A. Athair:} Investigation, Validation, Methodology, Formal Analysis, Visualization, Software, Writing - Original Draft and Review \& Editing.
\textbf{O. Williams:} Conceptualization, Methodology, Supervision, Writing - Review \& Editing
\textbf{B. Polagye:} Conceptualization, Methodology, Supervision, Writing - Original Draft and Review \& Editing, Resources, Funding Acquisition.

\FloatBarrier
\appendix

\section{Influence of Submergence Depth on Performance}
\label{app:submergence}

\begin{figure}
     \centering
     \includegraphics[width=\textwidth]{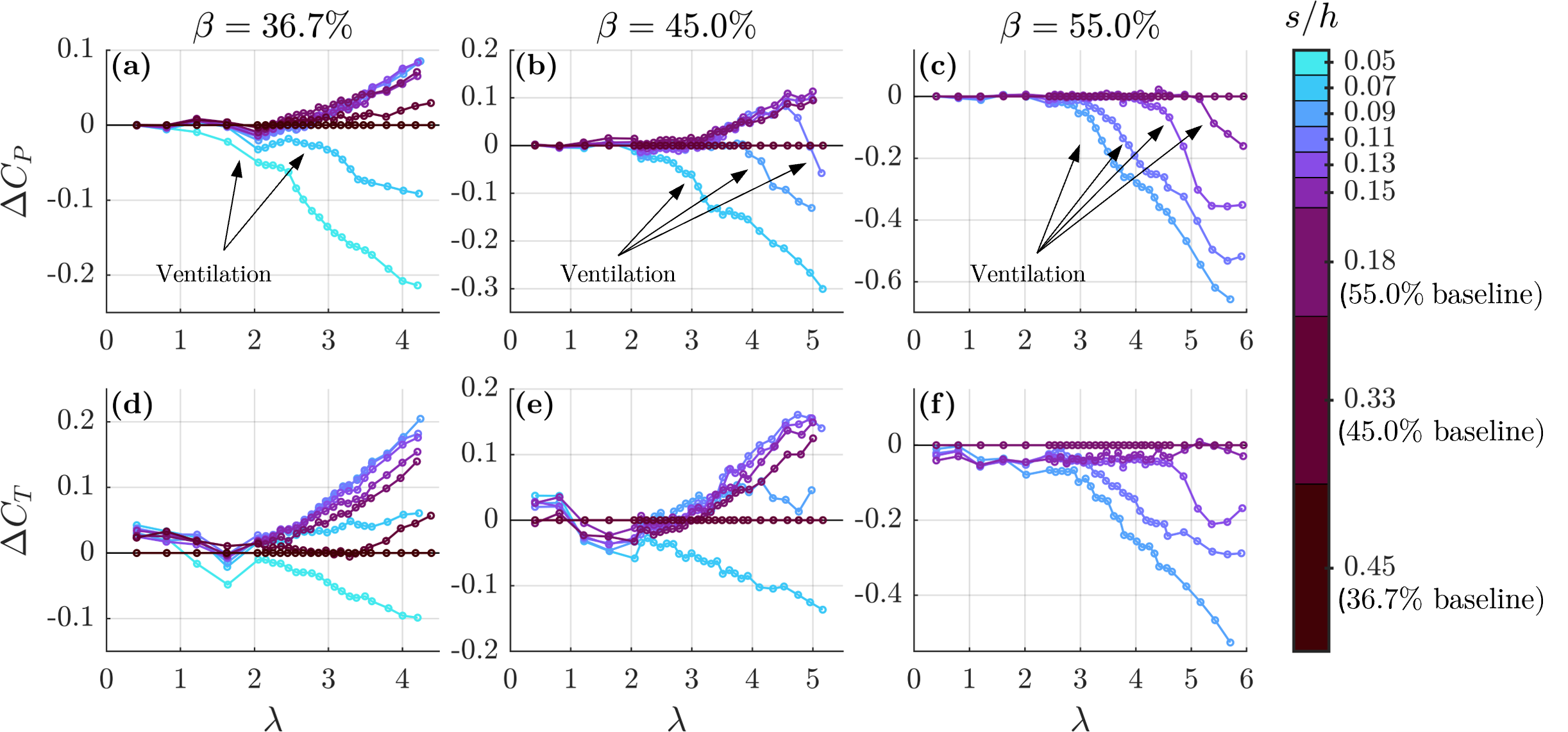}
     {\phantomsubcaption\label{fig:subPerf_cp_36}
      \phantomsubcaption\label{fig:subPerf_cp_45}
      \phantomsubcaption\label{fig:subPerf_cp_55}
      \phantomsubcaption\label{fig:subPerf_ct_36}
      \phantomsubcaption\label{fig:subPerf_ct_45}
      \phantomsubcaption\label{fig:subPerf_ct_55}
      }
     \caption{Change in the \subref{fig:subPerf_cp_36}-\subref{fig:subPerf_cp_55} $C_{P}-\lambda$ and \subref{fig:subPerf_ct_36}-\subref{fig:subPerf_ct_55} $C_{T}-\lambda$ curves as a function of $s/h$ and $\lambda$ relative to the baseline performance curves in \cref{fig:arrayPerf} for $\beta = 36.7\%$, $45.0\%$, and $55.0\%$, with general regions of significant ventilation annotated.
     The value of $s/h$ at each $\beta$ that corresponds to baseline experimental results is indicated in the colorbar and corresponds to the zero line in each tile.}
     \label{fig:subPerf}
\end{figure}

As mentioned in \Cref{methods:ndParams}, $s/h$ was maximized at each blockage to minimize the impact of ventilation on array performance, particularly at the highest blockage ratios.
However, this resulted in variation of $s/h$ across the tested $\beta$ (\cref{tab:priorWork}).
To assess the sensitivity of array performance to submergence depth, the array was also tested at a range of $s/h$ at $\beta = 36.7\%$, $45.0\%$, and $55.0\%$ through variation of the rotors' vertical position in the water column.
The ADV position was likewise adjusted such that $U_\infty$ was always sampled at the turbine midplane.

The difference between the $C_{P}\!-\!\lambda$ and $C_{T}\!-\!\lambda$ curves presented in \cref{fig:arrayPerf} and those obtained at different $s/h$ are shown in \cref{fig:subPerf}.
At $\beta = 36.7\%$ and $\beta = 45.0\%$, both $C_{P}$ and $C_{T}$ increase slightly as $s/h$ decreases (i.e., the turbines are moved closer to the free surface), particularly for $\lambda$ beyond the performance peak.
However, at all $\beta$, once $s/h$ drops below a critical value, ventilation begins, resulting in decreased $C_{P}$ and $C_{T}$ relative to deeper submergence depths.
As $\beta$ increases, the onset of ventilation occurs at larger $s/h$ (i.e., deeper submergence), and at a given $\beta$, as $s/h$ decreases, ventilation begins at lower $\lambda$.
However, other than ventilation, the effects of submergence depth on both $C_{P}$ and $C_{T}$ are minor relative to the influence of $\beta$ and the associated effects are mainly limited to $\lambda$ beyond the performance peak.
Consequently, variation in $s/h$ across the tested $\beta$ is unlikely to  significantly affect the trends presented in \cref{fig:arrayPerf} or the flow field structure.

\begin{figure}
    \centering
    \includegraphics[width=\textwidth]{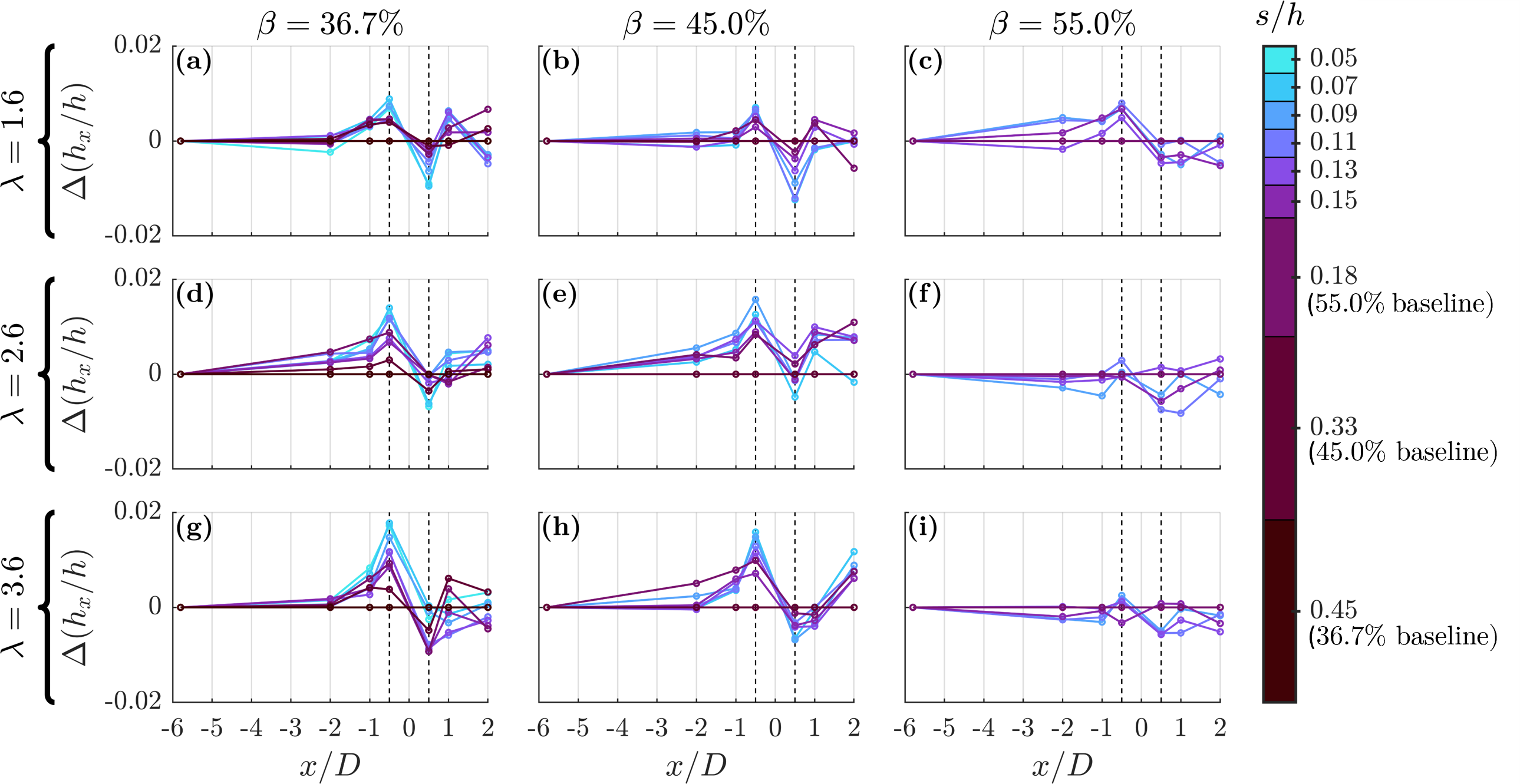}
    {\phantomsubcaption\label{fig:subFST_36_1.6}
     \phantomsubcaption\label{fig:subFST_45_1.6}
     \phantomsubcaption\label{fig:subFST_55_1.6}
     \phantomsubcaption\label{fig:subFST_36_2.6}
     \phantomsubcaption\label{fig:subFST_45_2.6}
     \phantomsubcaption\label{fig:subFST_55_2.6}
     \phantomsubcaption\label{fig:subFST_36_3.6}
     \phantomsubcaption\label{fig:subFST_45_3.6}
     \phantomsubcaption\label{fig:subFST_55_3.6}
    }
    \caption{Change in the time-averaged normalized streamwise water depth profiles as a function of $s/h$ at \subref{fig:fst_36} $\beta = 36.7\%$, $\beta = 45.0\%$, and \subref{fig:fst_55} $\beta = 55.0\%$ for \subref{fig:subFST_36_1.6}--\subref{fig:subFST_55_1.6} $\lambda = 1.6$, \subref{fig:subFST_36_2.6}--\subref{fig:subFST_55_2.6} $\lambda = 2.6$, and \subref{fig:subFST_36_3.6}--\subref{fig:subFST_55_3.6} $\lambda = 3.6$.
    The farthest upstream measurement is taken at the center of the channel, whereas depth measurements in the vicinity of the array are directly upstream and downstream of Turbine B (\Cref{fig:arrayOverhead}).
    For each case, depth measurements at each streamwise location ($h_x$) are normalized by the far-upstream water depth ($h$). The dashed lines indicate the swept area of Turbine B.
    The value of $s/h$ for each $\beta$ that corresponds to baseline experimental results (\cref{fig:fstProfiles}) is indicated in the colorbar and corresponds to the zero line in each tile.}
    \label{fig:subFST}
\end{figure}

The normalized submergence depth also slightly affects free surface deformation in the vicinity of the array, as shown by the change in the water depth profiles in \Cref{fig:subFST} for $\lambda = 1.6$, $2.6$, and $3.6$ at $\beta = 36.7\%$, $45.0\%$, and $55.0\%$.
Particularly for $\beta = 36.7\%$ and $45.0\%$, decreasing $s/h$ somewhat elevates the free surface upstream of Turbine B and lowers the free surface directly downstream at $x/D = 0.5$.
In other words, the magnitude of the free surface drop across Turbine B somewhat increases with greater proximity of the rotor to the free surface, and this drop becomes more pronounced as $\lambda$ and $\beta$ increase.
This is consistent with the non-ventilated power and thrust trends shown in \cref{fig:subPerf}.
Trends in free surface deformation with $s/h$ at $\beta = 55.0\%$ are less clear, likely due to the significant ventilation that occurs this blockage ratio.
Similarly, for all cases trends in the free surface with $s/h$ further downstream ($x/D = 1 - 2$) are more ambiguous, and may be obscured by the relatively low spatial and temporal resolution of the free surface transducers utilized in this study.

\FloatBarrier
\bibliography{bv_refs}

\begin{thebibliography}{76}%
\makeatletter
\providecommand \@ifxundefined [1]{%
 \@ifx{#1\undefined}
}%
\providecommand \@ifnum [1]{%
 \ifnum #1\expandafter \@firstoftwo
 \else \expandafter \@secondoftwo
 \fi
}%
\providecommand \@ifx [1]{%
 \ifx #1\expandafter \@firstoftwo
 \else \expandafter \@secondoftwo
 \fi
}%
\providecommand \natexlab [1]{#1}%
\providecommand \enquote  [1]{``#1''}%
\providecommand \bibnamefont  [1]{#1}%
\providecommand \bibfnamefont [1]{#1}%
\providecommand \citenamefont [1]{#1}%
\providecommand \href@noop [0]{\@secondoftwo}%
\providecommand \href [0]{\begingroup \@sanitize@url \@href}%
\providecommand \@href[1]{\@@startlink{#1}\@@href}%
\providecommand \@@href[1]{\endgroup#1\@@endlink}%
\providecommand \@sanitize@url [0]{\catcode `\\12\catcode `\$12\catcode
  `\&12\catcode `\#12\catcode `\^12\catcode `\_12\catcode `\%12\relax}%
\providecommand \@@startlink[1]{}%
\providecommand \@@endlink[0]{}%
\providecommand \url  [0]{\begingroup\@sanitize@url \@url }%
\providecommand \@url [1]{\endgroup\@href {#1}{\urlprefix }}%
\providecommand \urlprefix  [0]{URL }%
\providecommand \Eprint [0]{\href }%
\providecommand \doibase [0]{https://doi.org/}%
\providecommand \selectlanguage [0]{\@gobble}%
\providecommand \bibinfo  [0]{\@secondoftwo}%
\providecommand \bibfield  [0]{\@secondoftwo}%
\providecommand \translation [1]{[#1]}%
\providecommand \BibitemOpen [0]{}%
\providecommand \bibitemStop [0]{}%
\providecommand \bibitemNoStop [0]{.\EOS\space}%
\providecommand \EOS [0]{\spacefactor3000\relax}%
\providecommand \BibitemShut  [1]{\csname bibitem#1\endcsname}%
\let\auto@bib@innerbib\@empty
\bibitem [{\citenamefont {Houlsby}\ \emph {et~al.}(2008)\citenamefont
  {Houlsby}, \citenamefont {Draper},\ and\ \citenamefont
  {Oldfield}}]{houlsby_application_2008}%
  \BibitemOpen
  \bibfield  {author} {\bibinfo {author} {\bibfnamefont {G.}~\bibnamefont
  {Houlsby}}, \bibinfo {author} {\bibfnamefont {S.}~\bibnamefont {Draper}},\
  and\ \bibinfo {author} {\bibfnamefont {M.}~\bibnamefont {Oldfield}},\
  }\href@noop {} {\emph {\bibinfo {title} {Application of Linear Momentum
  Actuator Disc Theory to Open Channel Flow}}},\ \bibinfo {type} {Technical
  {{Report}}}\ \bibinfo {number} {OUEL 2296/08}\ (\bibinfo  {institution}
  {University of Oxford},\ \bibinfo {year} {2008})\BibitemShut {NoStop}%
\bibitem [{\citenamefont {Whelan}\ \emph {et~al.}(2009)\citenamefont {Whelan},
  \citenamefont {Graham},\ and\ \citenamefont
  {Peir{\'o}}}]{whelan_freesurface_2009}%
  \BibitemOpen
  \bibfield  {author} {\bibinfo {author} {\bibfnamefont {J.~I.}\ \bibnamefont
  {Whelan}}, \bibinfo {author} {\bibfnamefont {J.~M.~R.}\ \bibnamefont
  {Graham}},\ and\ \bibinfo {author} {\bibfnamefont {J.}~\bibnamefont
  {Peir{\'o}}},\ }\bibfield  {title} {\bibinfo {title} {A free-surface and
  blockage correction for tidal turbines},\ }\href
  {https://doi.org/10.1017/S0022112009005916} {\bibfield  {journal} {\bibinfo
  {journal} {Journal of Fluid Mechanics}\ }\textbf {\bibinfo {volume} {624}},\
  \bibinfo {pages} {281} (\bibinfo {year} {2009})}\BibitemShut {NoStop}%
\bibitem [{\citenamefont {Maskell}(1963)}]{maskell_ec_theory_1963}%
  \BibitemOpen
  \bibfield  {author} {\bibinfo {author} {\bibfnamefont {E.}~\bibnamefont
  {Maskell}},\ }\href@noop {} {\emph {\bibinfo {title} {A {{Theory}} of the
  {{Blockage Effects}} on {{Bluff Bodies}} and {{Stalled Wings}} in a {{Closed
  Wind Tunnel}}}}},\ \bibinfo {type} {Reports and {{Memoranda}}}\ \bibinfo
  {number} {3400}\ (\bibinfo  {institution} {Ministry of Aviation},\ \bibinfo
  {year} {1963})\BibitemShut {NoStop}%
\bibitem [{\citenamefont {Garrett}\ and\ \citenamefont
  {Cummins}(2007)}]{garrett_efficiency_2007}%
  \BibitemOpen
  \bibfield  {author} {\bibinfo {author} {\bibfnamefont {C.}~\bibnamefont
  {Garrett}}\ and\ \bibinfo {author} {\bibfnamefont {P.}~\bibnamefont
  {Cummins}},\ }\bibfield  {title} {\bibinfo {title} {The efficiency of a
  turbine in a tidal channel},\ }\href
  {https://doi.org/10.1017/S0022112007007781} {\bibfield  {journal} {\bibinfo
  {journal} {Journal of Fluid Mechanics}\ }\textbf {\bibinfo {volume} {588}},\
  \bibinfo {pages} {243} (\bibinfo {year} {2007})}\BibitemShut {NoStop}%
\bibitem [{\citenamefont {Nishino}\ and\ \citenamefont
  {Willden}(2012{\natexlab{a}})}]{nishino_efficiency_2012}%
  \BibitemOpen
  \bibfield  {author} {\bibinfo {author} {\bibfnamefont {T.}~\bibnamefont
  {Nishino}}\ and\ \bibinfo {author} {\bibfnamefont {R.~H.~J.}\ \bibnamefont
  {Willden}},\ }\bibfield  {title} {\bibinfo {title} {The efficiency of an
  array of tidal turbines partially blocking a wide channel},\ }\href
  {https://doi.org/10.1017/jfm.2012.349} {\bibfield  {journal} {\bibinfo
  {journal} {Journal of Fluid Mechanics}\ }\textbf {\bibinfo {volume} {708}},\
  \bibinfo {pages} {596} (\bibinfo {year} {2012}{\natexlab{a}})}\BibitemShut
  {NoStop}%
\bibitem [{\citenamefont {Vogel}\ \emph {et~al.}(2016)\citenamefont {Vogel},
  \citenamefont {Houlsby},\ and\ \citenamefont {Willden}}]{vogel_effect_2016}%
  \BibitemOpen
  \bibfield  {author} {\bibinfo {author} {\bibfnamefont {C.~R.}\ \bibnamefont
  {Vogel}}, \bibinfo {author} {\bibfnamefont {G.~T.}\ \bibnamefont {Houlsby}},\
  and\ \bibinfo {author} {\bibfnamefont {R.~H.~J.}\ \bibnamefont {Willden}},\
  }\bibfield  {title} {\bibinfo {title} {Effect of free surface deformation on
  the extractable power of a finite width turbine array},\ }\href
  {https://doi.org/10.1016/j.renene.2015.11.050} {\bibfield  {journal}
  {\bibinfo  {journal} {Renewable Energy}\ }\textbf {\bibinfo {volume} {88}},\
  \bibinfo {pages} {317} (\bibinfo {year} {2016})}\BibitemShut {NoStop}%
\bibitem [{\citenamefont {Nishino}\ and\ \citenamefont
  {Willden}(2013)}]{nishino_twoscale_2013}%
  \BibitemOpen
  \bibfield  {author} {\bibinfo {author} {\bibfnamefont {T.}~\bibnamefont
  {Nishino}}\ and\ \bibinfo {author} {\bibfnamefont {R.~H.~J.}\ \bibnamefont
  {Willden}},\ }\bibfield  {title} {\bibinfo {title} {Two-scale dynamics of
  flow past a partial cross-stream array of tidal turbines},\ }\href
  {https://doi.org/10.1017/jfm.2013.340} {\bibfield  {journal} {\bibinfo
  {journal} {Journal of Fluid Mechanics}\ }\textbf {\bibinfo {volume} {730}},\
  \bibinfo {pages} {220} (\bibinfo {year} {2013})}\BibitemShut {NoStop}%
\bibitem [{\citenamefont {Dehtyriov}\ \emph {et~al.}(2021)\citenamefont
  {Dehtyriov}, \citenamefont {Schnabl}, \citenamefont {Vogel}, \citenamefont
  {Draper}, \citenamefont {Adcock},\ and\ \citenamefont
  {Willden}}]{dehtyriov_fractallike_2021}%
  \BibitemOpen
  \bibfield  {author} {\bibinfo {author} {\bibfnamefont {D.}~\bibnamefont
  {Dehtyriov}}, \bibinfo {author} {\bibfnamefont {A.~M.}\ \bibnamefont
  {Schnabl}}, \bibinfo {author} {\bibfnamefont {C.~R.}\ \bibnamefont {Vogel}},
  \bibinfo {author} {\bibfnamefont {S.}~\bibnamefont {Draper}}, \bibinfo
  {author} {\bibfnamefont {T.~a.~A.}\ \bibnamefont {Adcock}},\ and\ \bibinfo
  {author} {\bibfnamefont {R.~H.~J.}\ \bibnamefont {Willden}},\ }\bibfield
  {title} {\bibinfo {title} {Fractal-like actuator disc theory for optimal
  energy extraction},\ }\href {https://doi.org/10.1017/jfm.2021.766} {\bibfield
   {journal} {\bibinfo  {journal} {Journal of Fluid Mechanics}\ }\textbf
  {\bibinfo {volume} {927}},\ \bibinfo {pages} {A40} (\bibinfo {year}
  {2021})}\BibitemShut {NoStop}%
\bibitem [{\citenamefont {Battisti}\ \emph {et~al.}(2011)\citenamefont
  {Battisti}, \citenamefont {Zanne}, \citenamefont {Dell'Anna}, \citenamefont
  {Dossena}, \citenamefont {Persico},\ and\ \citenamefont
  {Paradiso}}]{battisti_aerodynamic_2011}%
  \BibitemOpen
  \bibfield  {author} {\bibinfo {author} {\bibfnamefont {L.}~\bibnamefont
  {Battisti}}, \bibinfo {author} {\bibfnamefont {L.}~\bibnamefont {Zanne}},
  \bibinfo {author} {\bibfnamefont {S.}~\bibnamefont {Dell'Anna}}, \bibinfo
  {author} {\bibfnamefont {V.}~\bibnamefont {Dossena}}, \bibinfo {author}
  {\bibfnamefont {G.}~\bibnamefont {Persico}},\ and\ \bibinfo {author}
  {\bibfnamefont {B.}~\bibnamefont {Paradiso}},\ }\bibfield  {title} {\bibinfo
  {title} {Aerodynamic {{Measurements}} on a {{Vertical Axis Wind Turbine}} in
  a {{Large Scale Wind Tunnel}}},\ }\bibfield  {journal} {\bibinfo  {journal}
  {Journal of Energy Resources Technology}\ }\textbf {\bibinfo {volume}
  {133}},\ \href {https://doi.org/10.1115/1.4004360} {10.1115/1.4004360}
  (\bibinfo {year} {2011})\BibitemShut {NoStop}%
\bibitem [{\citenamefont {Nishino}\ and\ \citenamefont
  {Willden}(2012{\natexlab{b}})}]{nishino_effects_2012}%
  \BibitemOpen
  \bibfield  {author} {\bibinfo {author} {\bibfnamefont {T.}~\bibnamefont
  {Nishino}}\ and\ \bibinfo {author} {\bibfnamefont {R.~H.~J.}\ \bibnamefont
  {Willden}},\ }\bibfield  {title} {\bibinfo {title} {Effects of 3-{{D}}
  channel blockage and turbulent wake mixing on the limit of power extraction
  by tidal turbines},\ }\href
  {https://doi.org/10.1016/j.ijheatfluidflow.2012.05.002} {\bibfield  {journal}
  {\bibinfo  {journal} {International Journal of Heat and Fluid Flow}\ }\textbf
  {\bibinfo {volume} {37}},\ \bibinfo {pages} {123} (\bibinfo {year}
  {2012}{\natexlab{b}})}\BibitemShut {NoStop}%
\bibitem [{\citenamefont {McTavish}\ \emph {et~al.}(2014)\citenamefont
  {McTavish}, \citenamefont {Feszty},\ and\ \citenamefont
  {Nitzsche}}]{mctavish_experimental_2014}%
  \BibitemOpen
  \bibfield  {author} {\bibinfo {author} {\bibfnamefont {S.}~\bibnamefont
  {McTavish}}, \bibinfo {author} {\bibfnamefont {D.}~\bibnamefont {Feszty}},\
  and\ \bibinfo {author} {\bibfnamefont {F.}~\bibnamefont {Nitzsche}},\
  }\bibfield  {title} {\bibinfo {title} {An experimental and computational
  assessment of blockage effects on wind turbine wake development},\ }\href
  {https://doi.org/10.1002/we.1648} {\bibfield  {journal} {\bibinfo  {journal}
  {Wind Energy}\ }\textbf {\bibinfo {volume} {17}},\ \bibinfo {pages} {1515}
  (\bibinfo {year} {2014})}\BibitemShut {NoStop}%
\bibitem [{\citenamefont {Dossena}\ \emph {et~al.}(2015)\citenamefont
  {Dossena}, \citenamefont {Persico}, \citenamefont {Paradiso}, \citenamefont
  {Battisti}, \citenamefont {Dell'Anna}, \citenamefont {Brighenti},\ and\
  \citenamefont {Benini}}]{dossena_experimental_2015}%
  \BibitemOpen
  \bibfield  {author} {\bibinfo {author} {\bibfnamefont {V.}~\bibnamefont
  {Dossena}}, \bibinfo {author} {\bibfnamefont {G.}~\bibnamefont {Persico}},
  \bibinfo {author} {\bibfnamefont {B.}~\bibnamefont {Paradiso}}, \bibinfo
  {author} {\bibfnamefont {L.}~\bibnamefont {Battisti}}, \bibinfo {author}
  {\bibfnamefont {S.}~\bibnamefont {Dell'Anna}}, \bibinfo {author}
  {\bibfnamefont {A.}~\bibnamefont {Brighenti}},\ and\ \bibinfo {author}
  {\bibfnamefont {E.}~\bibnamefont {Benini}},\ }\bibfield  {title} {\bibinfo
  {title} {An {{Experimental Study}} of the {{Aerodynamics}} and
  {{Performance}} of a {{Vertical Axis Wind Turbine}} in a {{Confined}} and
  {{Unconfined Environment}}},\ }\bibfield  {journal} {\bibinfo  {journal}
  {Journal of Energy Resources Technology}\ }\textbf {\bibinfo {volume}
  {137}},\ \href {https://doi.org/10.1115/1.4030448} {10.1115/1.4030448}
  (\bibinfo {year} {2015})\BibitemShut {NoStop}%
\bibitem [{\citenamefont {Schluntz}\ and\ \citenamefont
  {Willden}(2015)}]{schluntz_effect_2015}%
  \BibitemOpen
  \bibfield  {author} {\bibinfo {author} {\bibfnamefont {J.}~\bibnamefont
  {Schluntz}}\ and\ \bibinfo {author} {\bibfnamefont {R.~H.~J.}\ \bibnamefont
  {Willden}},\ }\bibfield  {title} {\bibinfo {title} {The effect of blockage on
  tidal turbine rotor design and performance},\ }\href
  {https://doi.org/10.1016/j.renene.2015.02.050} {\bibfield  {journal}
  {\bibinfo  {journal} {Renewable Energy}\ }\textbf {\bibinfo {volume} {81}},\
  \bibinfo {pages} {432} (\bibinfo {year} {2015})}\BibitemShut {NoStop}%
\bibitem [{\citenamefont {Sarlak}\ \emph {et~al.}(2016)\citenamefont {Sarlak},
  \citenamefont {Nishino}, \citenamefont {{Mart{\'i}nez-Tossas}}, \citenamefont
  {Meneveau},\ and\ \citenamefont {S{\o}rensen}}]{sarlak_assessment_2016}%
  \BibitemOpen
  \bibfield  {author} {\bibinfo {author} {\bibfnamefont {H.}~\bibnamefont
  {Sarlak}}, \bibinfo {author} {\bibfnamefont {T.}~\bibnamefont {Nishino}},
  \bibinfo {author} {\bibfnamefont {L.~A.}\ \bibnamefont
  {{Mart{\'i}nez-Tossas}}}, \bibinfo {author} {\bibfnamefont {C.}~\bibnamefont
  {Meneveau}},\ and\ \bibinfo {author} {\bibfnamefont {J.~N.}\ \bibnamefont
  {S{\o}rensen}},\ }\bibfield  {title} {\bibinfo {title} {Assessment of
  blockage effects on the wake characteristics and power of wind turbines},\
  }\href {https://doi.org/10.1016/j.renene.2016.01.101} {\bibfield  {journal}
  {\bibinfo  {journal} {Renewable Energy}\ }\textbf {\bibinfo {volume} {93}},\
  \bibinfo {pages} {340} (\bibinfo {year} {2016})}\BibitemShut {NoStop}%
\bibitem [{\citenamefont {Ross}\ and\ \citenamefont
  {Polagye}(2020{\natexlab{a}})}]{ross_experimental_2020a}%
  \BibitemOpen
  \bibfield  {author} {\bibinfo {author} {\bibfnamefont {H.}~\bibnamefont
  {Ross}}\ and\ \bibinfo {author} {\bibfnamefont {B.}~\bibnamefont {Polagye}},\
  }\bibfield  {title} {\bibinfo {title} {An experimental evaluation of blockage
  effects on the wake of a cross-flow current turbine},\ }\href
  {https://doi.org/10.1007/s40722-020-00172-w} {\bibfield  {journal} {\bibinfo
  {journal} {Journal of Ocean Engineering and Marine Energy}\ }\textbf
  {\bibinfo {volume} {6}},\ \bibinfo {pages} {263} (\bibinfo {year}
  {2020}{\natexlab{a}})}\BibitemShut {NoStop}%
\bibitem [{\citenamefont {Gauthier}\ \emph {et~al.}(2016)\citenamefont
  {Gauthier}, \citenamefont {Kinsey},\ and\ \citenamefont
  {Dumas}}]{gauthier_impact_2016}%
  \BibitemOpen
  \bibfield  {author} {\bibinfo {author} {\bibfnamefont {E.}~\bibnamefont
  {Gauthier}}, \bibinfo {author} {\bibfnamefont {T.}~\bibnamefont {Kinsey}},\
  and\ \bibinfo {author} {\bibfnamefont {G.}~\bibnamefont {Dumas}},\ }\bibfield
   {title} {\bibinfo {title} {Impact of {{Blockage}} on the {{Hydrodynamic
  Performance}} of {{Oscillating-Foils Hydrokinetic Turbines}}},\ }\bibfield
  {journal} {\bibinfo  {journal} {Journal of Fluids Engineering}\ }\textbf
  {\bibinfo {volume} {138}},\ \href {https://doi.org/10.1115/1.4033298}
  {10.1115/1.4033298} (\bibinfo {year} {2016})\BibitemShut {NoStop}%
\bibitem [{\citenamefont {Kinsey}\ and\ \citenamefont
  {Dumas}(2017)}]{kinsey_impact_2017}%
  \BibitemOpen
  \bibfield  {author} {\bibinfo {author} {\bibfnamefont {T.}~\bibnamefont
  {Kinsey}}\ and\ \bibinfo {author} {\bibfnamefont {G.}~\bibnamefont {Dumas}},\
  }\bibfield  {title} {\bibinfo {title} {Impact of channel blockage on the
  performance of axial and cross-flow hydrokinetic turbines},\ }\href
  {https://doi.org/10.1016/j.renene.2016.11.021} {\bibfield  {journal}
  {\bibinfo  {journal} {Renewable Energy}\ }\textbf {\bibinfo {volume} {103}},\
  \bibinfo {pages} {239} (\bibinfo {year} {2017})}\BibitemShut {NoStop}%
\bibitem [{\citenamefont {Goude}\ and\ \citenamefont
  {{\AA}gren}(2014)}]{goude_simulations_2014}%
  \BibitemOpen
  \bibfield  {author} {\bibinfo {author} {\bibfnamefont {A.}~\bibnamefont
  {Goude}}\ and\ \bibinfo {author} {\bibfnamefont {O.}~\bibnamefont
  {{\AA}gren}},\ }\bibfield  {title} {\bibinfo {title} {Simulations of a
  vertical axis turbine in a channel},\ }\href
  {https://doi.org/10.1016/j.renene.2013.09.038} {\bibfield  {journal}
  {\bibinfo  {journal} {Renewable Energy}\ }\textbf {\bibinfo {volume} {63}},\
  \bibinfo {pages} {477} (\bibinfo {year} {2014})}\BibitemShut {NoStop}%
\bibitem [{\citenamefont {Kong}\ \emph {et~al.}(2025)\citenamefont {Kong},
  \citenamefont {Ji}, \citenamefont {Wu}, \citenamefont {Sun}, \citenamefont
  {Zhang}, \citenamefont {Zhang}, \citenamefont {Zhu},\ and\ \citenamefont
  {Reabroy}}]{kong_correction_2025}%
  \BibitemOpen
  \bibfield  {author} {\bibinfo {author} {\bibfnamefont {M.}~\bibnamefont
  {Kong}}, \bibinfo {author} {\bibfnamefont {R.}~\bibnamefont {Ji}}, \bibinfo
  {author} {\bibfnamefont {M.}~\bibnamefont {Wu}}, \bibinfo {author}
  {\bibfnamefont {K.}~\bibnamefont {Sun}}, \bibinfo {author} {\bibfnamefont
  {J.}~\bibnamefont {Zhang}}, \bibinfo {author} {\bibfnamefont
  {Y.}~\bibnamefont {Zhang}}, \bibinfo {author} {\bibfnamefont
  {R.}~\bibnamefont {Zhu}},\ and\ \bibinfo {author} {\bibfnamefont
  {R.}~\bibnamefont {Reabroy}},\ }\bibfield  {title} {\bibinfo {title}
  {Correction of sidewall blockage effect for twin-rotor vertical axis tidal
  stream turbine subjected to different solidities},\ }\href
  {https://doi.org/10.1063/5.0263049} {\bibfield  {journal} {\bibinfo
  {journal} {Physics of Fluids}\ }\textbf {\bibinfo {volume} {37}},\ \bibinfo
  {pages} {035157} (\bibinfo {year} {2025})}\BibitemShut {NoStop}%
\bibitem [{\citenamefont {Hunt}\ \emph {et~al.}(2025)\citenamefont {Hunt},
  \citenamefont {Talpey}, \citenamefont {Calandra},\ and\ \citenamefont
  {Polagye}}]{hunt_performance_2025}%
  \BibitemOpen
  \bibfield  {author} {\bibinfo {author} {\bibfnamefont {A.}~\bibnamefont
  {Hunt}}, \bibinfo {author} {\bibfnamefont {G.}~\bibnamefont {Talpey}},
  \bibinfo {author} {\bibfnamefont {G.}~\bibnamefont {Calandra}},\ and\
  \bibinfo {author} {\bibfnamefont {B.}~\bibnamefont {Polagye}},\ }\bibfield
  {title} {\bibinfo {title} {Performance characteristics and bluff-body
  modeling of high-blockage cross-flow turbine arrays with varying rotor
  geometry},\ }\href {https://doi.org/10.1063/5.0272110} {\bibfield  {journal}
  {\bibinfo  {journal} {Journal of Renewable and Sustainable Energy}\ }\textbf
  {\bibinfo {volume} {17}},\ \bibinfo {pages} {034501} (\bibinfo {year}
  {2025})}\BibitemShut {NoStop}%
\bibitem [{\citenamefont {Abutunis}\ and\ \citenamefont
  {Menta}(2022)}]{abutunis_comprehensive_2022}%
  \BibitemOpen
  \bibfield  {author} {\bibinfo {author} {\bibfnamefont {A.}~\bibnamefont
  {Abutunis}}\ and\ \bibinfo {author} {\bibfnamefont {V.~G.}\ \bibnamefont
  {Menta}},\ }\bibfield  {title} {\bibinfo {title} {Comprehensive {{Parametric
  Study}} of {{Blockage Effect}} on the {{Performance}} of {{Horizontal Axis
  Hydrokinetic Turbines}}},\ }\href {https://doi.org/10.3390/en15072585}
  {\bibfield  {journal} {\bibinfo  {journal} {Energies}\ }\textbf {\bibinfo
  {volume} {15}},\ \bibinfo {pages} {2585} (\bibinfo {year}
  {2022})}\BibitemShut {NoStop}%
\bibitem [{\citenamefont {Chen}\ and\ \citenamefont
  {Liou}(2011)}]{chen_blockage_2011}%
  \BibitemOpen
  \bibfield  {author} {\bibinfo {author} {\bibfnamefont {T.~Y.}\ \bibnamefont
  {Chen}}\ and\ \bibinfo {author} {\bibfnamefont {L.~R.}\ \bibnamefont
  {Liou}},\ }\bibfield  {title} {\bibinfo {title} {Blockage corrections in wind
  tunnel tests of small horizontal-axis wind turbines},\ }\href
  {https://doi.org/10.1016/j.expthermflusci.2010.12.005} {\bibfield  {journal}
  {\bibinfo  {journal} {Experimental Thermal and Fluid Science}\ }\textbf
  {\bibinfo {volume} {35}},\ \bibinfo {pages} {565} (\bibinfo {year}
  {2011})}\BibitemShut {NoStop}%
\bibitem [{\citenamefont {Ross}\ and\ \citenamefont
  {Altman}(2011)}]{ross_wind_2011}%
  \BibitemOpen
  \bibfield  {author} {\bibinfo {author} {\bibfnamefont {I.}~\bibnamefont
  {Ross}}\ and\ \bibinfo {author} {\bibfnamefont {A.}~\bibnamefont {Altman}},\
  }\bibfield  {title} {\bibinfo {title} {Wind tunnel blockage corrections:
  {{Review}} and application to {{Savonius}} vertical-axis wind turbines},\
  }\href {https://doi.org/10.1016/j.jweia.2011.02.002} {\bibfield  {journal}
  {\bibinfo  {journal} {Journal of Wind Engineering and Industrial
  Aerodynamics}\ }\textbf {\bibinfo {volume} {99}},\ \bibinfo {pages} {523}
  (\bibinfo {year} {2011})}\BibitemShut {NoStop}%
\bibitem [{\citenamefont {Jeong}\ \emph {et~al.}(2018)\citenamefont {Jeong},
  \citenamefont {Lee},\ and\ \citenamefont {Kwon}}]{jeong_blockage_2018}%
  \BibitemOpen
  \bibfield  {author} {\bibinfo {author} {\bibfnamefont {H.}~\bibnamefont
  {Jeong}}, \bibinfo {author} {\bibfnamefont {S.}~\bibnamefont {Lee}},\ and\
  \bibinfo {author} {\bibfnamefont {S.-D.}\ \bibnamefont {Kwon}},\ }\bibfield
  {title} {\bibinfo {title} {Blockage corrections for wind tunnel tests
  conducted on a {{Darrieus}} wind turbine},\ }\href
  {https://doi.org/10.1016/j.jweia.2018.06.002} {\bibfield  {journal} {\bibinfo
   {journal} {Journal of Wind Engineering and Industrial Aerodynamics}\
  }\textbf {\bibinfo {volume} {179}},\ \bibinfo {pages} {229} (\bibinfo {year}
  {2018})}\BibitemShut {NoStop}%
\bibitem [{\citenamefont {Ross}\ and\ \citenamefont
  {Polagye}(2020{\natexlab{b}})}]{ross_experimental_2020}%
  \BibitemOpen
  \bibfield  {author} {\bibinfo {author} {\bibfnamefont {H.}~\bibnamefont
  {Ross}}\ and\ \bibinfo {author} {\bibfnamefont {B.}~\bibnamefont {Polagye}},\
  }\bibfield  {title} {\bibinfo {title} {An experimental assessment of
  analytical blockage corrections for turbines},\ }\href
  {https://doi.org/10.1016/j.renene.2020.01.135} {\bibfield  {journal}
  {\bibinfo  {journal} {Renewable Energy}\ }\textbf {\bibinfo {volume} {152}},\
  \bibinfo {pages} {1328} (\bibinfo {year} {2020}{\natexlab{b}})}\BibitemShut
  {NoStop}%
\bibitem [{\citenamefont {Steiros}\ \emph {et~al.}(2022)\citenamefont
  {Steiros}, \citenamefont {Bempedelis},\ and\ \citenamefont
  {Cicolin}}]{steiros_analytical_2022}%
  \BibitemOpen
  \bibfield  {author} {\bibinfo {author} {\bibfnamefont {K.}~\bibnamefont
  {Steiros}}, \bibinfo {author} {\bibfnamefont {N.}~\bibnamefont
  {Bempedelis}},\ and\ \bibinfo {author} {\bibfnamefont {M.~M.}\ \bibnamefont
  {Cicolin}},\ }\bibfield  {title} {\bibinfo {title} {An analytical blockage
  correction model for high-solidity turbines},\ }\href
  {https://doi.org/10.1017/jfm.2022.735} {\bibfield  {journal} {\bibinfo
  {journal} {Journal of Fluid Mechanics}\ }\textbf {\bibinfo {volume} {948}},\
  \bibinfo {pages} {A57} (\bibinfo {year} {2022})}\BibitemShut {NoStop}%
\bibitem [{\citenamefont {{Zilic de Arcos}}\ \emph {et~al.}(2020)\citenamefont
  {{Zilic de Arcos}}, \citenamefont {Tampier},\ and\ \citenamefont
  {Vogel}}]{zilic_de_arcos_numerical_2020}%
  \BibitemOpen
  \bibfield  {author} {\bibinfo {author} {\bibfnamefont {F.}~\bibnamefont
  {{Zilic de Arcos}}}, \bibinfo {author} {\bibfnamefont {G.}~\bibnamefont
  {Tampier}},\ and\ \bibinfo {author} {\bibfnamefont {C.~R.}\ \bibnamefont
  {Vogel}},\ }\bibfield  {title} {\bibinfo {title} {Numerical analysis of
  blockage correction methods for tidal turbines},\ }\href
  {https://doi.org/10.1007/s40722-020-00168-6} {\bibfield  {journal} {\bibinfo
  {journal} {Journal of Ocean Engineering and Marine Energy}\ }\textbf
  {\bibinfo {volume} {6}},\ \bibinfo {pages} {183} (\bibinfo {year}
  {2020})}\BibitemShut {NoStop}%
\bibitem [{\citenamefont {Zhang}\ \emph {et~al.}(2023)\citenamefont {Zhang},
  \citenamefont {Guo}, \citenamefont {Cheng}, \citenamefont {Hu},\ and\
  \citenamefont {Li}}]{zhang_analysis_2023}%
  \BibitemOpen
  \bibfield  {author} {\bibinfo {author} {\bibfnamefont {D.}~\bibnamefont
  {Zhang}}, \bibinfo {author} {\bibfnamefont {P.}~\bibnamefont {Guo}}, \bibinfo
  {author} {\bibfnamefont {Y.}~\bibnamefont {Cheng}}, \bibinfo {author}
  {\bibfnamefont {Q.}~\bibnamefont {Hu}},\ and\ \bibinfo {author}
  {\bibfnamefont {J.}~\bibnamefont {Li}},\ }\bibfield  {title} {\bibinfo
  {title} {Analysis of blockage correction methods for high-solidity
  hydrokinetic turbines: {{Experimental}} and numerical investigations},\
  }\href {https://doi.org/10.1016/j.oceaneng.2023.115185} {\bibfield  {journal}
  {\bibinfo  {journal} {Ocean Engineering}\ }\textbf {\bibinfo {volume}
  {283}},\ \bibinfo {pages} {115185} (\bibinfo {year} {2023})}\BibitemShut
  {NoStop}%
\bibitem [{\citenamefont {Glauert}(1935)}]{glauert_airplane_1935}%
  \BibitemOpen
  \bibfield  {author} {\bibinfo {author} {\bibfnamefont {H.}~\bibnamefont
  {Glauert}},\ }\bibfield  {title} {\bibinfo {title} {Airplane
  {{Propellers}}},\ }in\ \href {https://doi.org/10.1007/978-3-642-91487-4_3}
  {\emph {\bibinfo {booktitle} {Aerodynamic {{Theory}}: {{A General Review}} of
  {{Progress Under}} a {{Grant}} of the {{Guggenheim Fund}} for the
  {{Promotion}} of {{Aeronautics}}}}},\ \bibinfo {editor} {edited by\ \bibinfo
  {editor} {\bibfnamefont {W.~F.}\ \bibnamefont {Durand}}}\ (\bibinfo
  {publisher} {Springer},\ \bibinfo {address} {Berlin, Heidelberg},\ \bibinfo
  {year} {1935})\ pp.\ \bibinfo {pages} {169--360}\BibitemShut {NoStop}%
\bibitem [{\citenamefont {Barnsley}\ and\ \citenamefont
  {Wellicome}(1990)}]{barnsley_final_1990}%
  \BibitemOpen
  \bibfield  {author} {\bibinfo {author} {\bibfnamefont {M.~J.}\ \bibnamefont
  {Barnsley}}\ and\ \bibinfo {author} {\bibfnamefont {J.~F.}\ \bibnamefont
  {Wellicome}},\ }\href@noop {} {\emph {\bibinfo {title} {Final Report on the
  2nd Phase of Development and Testing of a Horizontal Axis Wind Turbine Test
  Rig for the Investigation of Stall Regulation Aerodynamics}}},\ \bibinfo
  {type} {Technical {{Report}}}\ \bibinfo {number} {E.5A/CON5103/ 1746}\
  (\bibinfo  {institution} {ETSU},\ \bibinfo {year} {1990})\BibitemShut
  {NoStop}%
\bibitem [{\citenamefont {Dehtyriov}\ \emph {et~al.}(2025)\citenamefont
  {Dehtyriov}, \citenamefont {Vogel},\ and\ \citenamefont
  {Willden}}]{dehtyriov_twoscale_2025}%
  \BibitemOpen
  \bibfield  {author} {\bibinfo {author} {\bibfnamefont {D.}~\bibnamefont
  {Dehtyriov}}, \bibinfo {author} {\bibfnamefont {C.}~\bibnamefont {Vogel}},\
  and\ \bibinfo {author} {\bibfnamefont {R.}~\bibnamefont {Willden}},\
  }\bibfield  {title} {\bibinfo {title} {A two-scale blockage correction for an
  array of tidal turbines},\ }\href {https://doi.org/10.36688/imej.8.101-108}
  {\bibfield  {journal} {\bibinfo  {journal} {International Marine Energy
  Journal}\ }\textbf {\bibinfo {volume} {8}},\ \bibinfo {pages} {101} (\bibinfo
  {year} {2025})}\BibitemShut {NoStop}%
\bibitem [{\citenamefont {Steiros}\ and\ \citenamefont
  {Hultmark}(2018)}]{steiros_drag_2018}%
  \BibitemOpen
  \bibfield  {author} {\bibinfo {author} {\bibfnamefont {K.}~\bibnamefont
  {Steiros}}\ and\ \bibinfo {author} {\bibfnamefont {M.}~\bibnamefont
  {Hultmark}},\ }\bibfield  {title} {\bibinfo {title} {Drag on flat plates of
  arbitrary porosity},\ }\href {https://doi.org/10.1017/jfm.2018.621}
  {\bibfield  {journal} {\bibinfo  {journal} {Journal of Fluid Mechanics}\
  }\textbf {\bibinfo {volume} {853}},\ \bibinfo {pages} {R3} (\bibinfo {year}
  {2018})}\BibitemShut {NoStop}%
\bibitem [{\citenamefont {Liew}\ \emph {et~al.}(2024)\citenamefont {Liew},
  \citenamefont {Heck},\ and\ \citenamefont {Howland}}]{liew_unified_2024a}%
  \BibitemOpen
  \bibfield  {author} {\bibinfo {author} {\bibfnamefont {J.}~\bibnamefont
  {Liew}}, \bibinfo {author} {\bibfnamefont {K.~S.}\ \bibnamefont {Heck}},\
  and\ \bibinfo {author} {\bibfnamefont {M.~F.}\ \bibnamefont {Howland}},\
  }\bibfield  {title} {\bibinfo {title} {Unified momentum model for rotor
  aerodynamics across operating regimes},\ }\href
  {https://doi.org/10.1038/s41467-024-50756-5} {\bibfield  {journal} {\bibinfo
  {journal} {Nature Communications}\ }\textbf {\bibinfo {volume} {15}},\
  \bibinfo {pages} {6658} (\bibinfo {year} {2024})}\BibitemShut {NoStop}%
\bibitem [{\citenamefont {Taylor}(1944)}]{taylor_air_1944}%
  \BibitemOpen
  \bibfield  {author} {\bibinfo {author} {\bibfnamefont {G.}~\bibnamefont
  {Taylor}},\ }\bibfield  {title} {\bibinfo {title} {Air resistance of a flat
  plate of very porous material},\ }\href@noop {} {\bibfield  {journal}
  {\bibinfo  {journal} {Aeronautical Research Council, Reports and Memoranda}\
  }\textbf {\bibinfo {volume} {2236}},\ \bibinfo {pages} {159} (\bibinfo {year}
  {1944})}\BibitemShut {NoStop}%
\bibitem [{\citenamefont {Koo}\ and\ \citenamefont
  {James}(1973)}]{koo_fluid_1973}%
  \BibitemOpen
  \bibfield  {author} {\bibinfo {author} {\bibfnamefont {J.-K.}\ \bibnamefont
  {Koo}}\ and\ \bibinfo {author} {\bibfnamefont {D.~F.}\ \bibnamefont
  {James}},\ }\bibfield  {title} {\bibinfo {title} {Fluid flow around and
  through a screen},\ }\href {https://doi.org/10.1017/S0022112073000327}
  {\bibfield  {journal} {\bibinfo  {journal} {Journal of Fluid Mechanics}\
  }\textbf {\bibinfo {volume} {60}},\ \bibinfo {pages} {513} (\bibinfo {year}
  {1973})}\BibitemShut {NoStop}%
\bibitem [{\citenamefont {Takamatsu}\ \emph {et~al.}(1985)\citenamefont
  {Takamatsu}, \citenamefont {Furukawa}, \citenamefont {Okuma},\ and\
  \citenamefont {Shimogawa}}]{takamatsu_study_1985}%
  \BibitemOpen
  \bibfield  {author} {\bibinfo {author} {\bibfnamefont {Y.}~\bibnamefont
  {Takamatsu}}, \bibinfo {author} {\bibfnamefont {A.}~\bibnamefont {Furukawa}},
  \bibinfo {author} {\bibfnamefont {K.}~\bibnamefont {Okuma}},\ and\ \bibinfo
  {author} {\bibfnamefont {Y.}~\bibnamefont {Shimogawa}},\ }\bibfield  {title}
  {\bibinfo {title} {Study on {{Hydrodynamic Performance}} of {{Darrieus-type
  Cross-flow Water Turbine}}},\ }\href
  {https://doi.org/10.1299/jsme1958.28.1119} {\bibfield  {journal} {\bibinfo
  {journal} {Bulletin of JSME}\ }\textbf {\bibinfo {volume} {28}},\ \bibinfo
  {pages} {1119} (\bibinfo {year} {1985})}\BibitemShut {NoStop}%
\bibitem [{\citenamefont {Consul}\ \emph {et~al.}(2013)\citenamefont {Consul},
  \citenamefont {Willden},\ and\ \citenamefont
  {McIntosh}}]{consul_blockage_2013}%
  \BibitemOpen
  \bibfield  {author} {\bibinfo {author} {\bibfnamefont {C.~A.}\ \bibnamefont
  {Consul}}, \bibinfo {author} {\bibfnamefont {R.~H.}\ \bibnamefont
  {Willden}},\ and\ \bibinfo {author} {\bibfnamefont {S.~C.}\ \bibnamefont
  {McIntosh}},\ }\bibfield  {title} {\bibinfo {title} {Blockage effects on the
  hydrodynamic performance of a marine cross-flow turbine},\ }\href
  {https://doi.org/10.1098/rsta.2012.0299} {\bibfield  {journal} {\bibinfo
  {journal} {Philosophical Transactions of the Royal Society A: Mathematical,
  Physical and Engineering Sciences}\ }\textbf {\bibinfo {volume} {371}},\
  \bibinfo {pages} {20120299} (\bibinfo {year} {2013})}\BibitemShut {NoStop}%
\bibitem [{\citenamefont {McAdam}\ \emph
  {et~al.}(2013{\natexlab{a}})\citenamefont {McAdam}, \citenamefont {Houlsby},\
  and\ \citenamefont {Oldfield}}]{mcadam_experimental_2013}%
  \BibitemOpen
  \bibfield  {author} {\bibinfo {author} {\bibfnamefont {R.~A.}\ \bibnamefont
  {McAdam}}, \bibinfo {author} {\bibfnamefont {G.~T.}\ \bibnamefont
  {Houlsby}},\ and\ \bibinfo {author} {\bibfnamefont {M.~L.~G.}\ \bibnamefont
  {Oldfield}},\ }\bibfield  {title} {\bibinfo {title} {Experimental
  measurements of the hydrodynamic performance and structural loading of the
  {{Transverse Horizontal Axis Water Turbine}}: {{Part}} 1},\ }\href
  {https://doi.org/10.1016/j.renene.2013.03.016} {\bibfield  {journal}
  {\bibinfo  {journal} {Renewable Energy}\ }\textbf {\bibinfo {volume} {59}},\
  \bibinfo {pages} {105} (\bibinfo {year} {2013}{\natexlab{a}})}\BibitemShut
  {NoStop}%
\bibitem [{\citenamefont {McAdam}\ \emph
  {et~al.}(2013{\natexlab{b}})\citenamefont {McAdam}, \citenamefont {Houlsby},\
  and\ \citenamefont {Oldfield}}]{mcadam_experimental_2013a}%
  \BibitemOpen
  \bibfield  {author} {\bibinfo {author} {\bibfnamefont {R.~A.}\ \bibnamefont
  {McAdam}}, \bibinfo {author} {\bibfnamefont {G.~T.}\ \bibnamefont
  {Houlsby}},\ and\ \bibinfo {author} {\bibfnamefont {M.~L.~G.}\ \bibnamefont
  {Oldfield}},\ }\bibfield  {title} {\bibinfo {title} {Experimental
  measurements of the hydrodynamic performance and structural loading of the
  transverse horizontal axis water turbine: {{Part}} 2},\ }\href
  {https://doi.org/10.1016/j.renene.2013.03.015} {\bibfield  {journal}
  {\bibinfo  {journal} {Renewable Energy}\ }\textbf {\bibinfo {volume} {59}},\
  \bibinfo {pages} {141} (\bibinfo {year} {2013}{\natexlab{b}})}\BibitemShut
  {NoStop}%
\bibitem [{\citenamefont {Kolekar}\ and\ \citenamefont
  {Banerjee}(2015)}]{kolekar_performance_2015}%
  \BibitemOpen
  \bibfield  {author} {\bibinfo {author} {\bibfnamefont {N.}~\bibnamefont
  {Kolekar}}\ and\ \bibinfo {author} {\bibfnamefont {A.}~\bibnamefont
  {Banerjee}},\ }\bibfield  {title} {\bibinfo {title} {Performance
  characterization and placement of a marine hydrokinetic turbine in a tidal
  channel under boundary proximity and blockage effects},\ }\href
  {https://doi.org/10.1016/j.apenergy.2015.03.052} {\bibfield  {journal}
  {\bibinfo  {journal} {Applied Energy}\ }\textbf {\bibinfo {volume} {148}},\
  \bibinfo {pages} {121} (\bibinfo {year} {2015})}\BibitemShut {NoStop}%
\bibitem [{\citenamefont {Birjandi}\ \emph {et~al.}(2013)\citenamefont
  {Birjandi}, \citenamefont {Bibeau}, \citenamefont {Chatoorgoon},\ and\
  \citenamefont {Kumar}}]{birjandi_power_2013}%
  \BibitemOpen
  \bibfield  {author} {\bibinfo {author} {\bibfnamefont {A.~H.}\ \bibnamefont
  {Birjandi}}, \bibinfo {author} {\bibfnamefont {E.~L.}\ \bibnamefont
  {Bibeau}}, \bibinfo {author} {\bibfnamefont {V.}~\bibnamefont
  {Chatoorgoon}},\ and\ \bibinfo {author} {\bibfnamefont {A.}~\bibnamefont
  {Kumar}},\ }\bibfield  {title} {\bibinfo {title} {Power measurement of
  hydrokinetic turbines with free-surface and blockage effect},\ }\href
  {https://doi.org/10.1016/j.oceaneng.2013.05.023} {\bibfield  {journal}
  {\bibinfo  {journal} {Ocean Engineering}\ }\textbf {\bibinfo {volume} {69}},\
  \bibinfo {pages} {9} (\bibinfo {year} {2013})}\BibitemShut {NoStop}%
\bibitem [{\citenamefont {Gaurier}\ \emph {et~al.}(2015)\citenamefont
  {Gaurier}, \citenamefont {Germain}, \citenamefont {Facq}, \citenamefont
  {Johnstone}, \citenamefont {Grant}, \citenamefont {Day}, \citenamefont
  {Nixon}, \citenamefont {Di~Felice},\ and\ \citenamefont
  {Costanzo}}]{gaurier_tidal_2015}%
  \BibitemOpen
  \bibfield  {author} {\bibinfo {author} {\bibfnamefont {B.}~\bibnamefont
  {Gaurier}}, \bibinfo {author} {\bibfnamefont {G.}~\bibnamefont {Germain}},
  \bibinfo {author} {\bibfnamefont {J.~V.}\ \bibnamefont {Facq}}, \bibinfo
  {author} {\bibfnamefont {C.~M.}\ \bibnamefont {Johnstone}}, \bibinfo {author}
  {\bibfnamefont {A.~D.}\ \bibnamefont {Grant}}, \bibinfo {author}
  {\bibfnamefont {A.~H.}\ \bibnamefont {Day}}, \bibinfo {author} {\bibfnamefont
  {E.}~\bibnamefont {Nixon}}, \bibinfo {author} {\bibfnamefont
  {F.}~\bibnamefont {Di~Felice}},\ and\ \bibinfo {author} {\bibfnamefont
  {M.}~\bibnamefont {Costanzo}},\ }\bibfield  {title} {\bibinfo {title} {Tidal
  energy ``{{Round Robin}}'' tests comparisons between towing tank and
  circulating tank results},\ }\href
  {https://doi.org/10.1016/j.ijome.2015.05.005} {\bibfield  {journal} {\bibinfo
   {journal} {International Journal of Marine Energy}\ }\bibinfo {series}
  {Special {{Issue}} on {{Marine Renewables Infrastructure Network}}},\ \textbf
  {\bibinfo {volume} {12}},\ \bibinfo {pages} {87} (\bibinfo {year}
  {2015})}\BibitemShut {NoStop}%
\bibitem [{\citenamefont {Ryi}\ \emph {et~al.}(2015)\citenamefont {Ryi},
  \citenamefont {Rhee}, \citenamefont {Chang~Hwang},\ and\ \citenamefont
  {Choi}}]{ryi_blockage_2015}%
  \BibitemOpen
  \bibfield  {author} {\bibinfo {author} {\bibfnamefont {J.}~\bibnamefont
  {Ryi}}, \bibinfo {author} {\bibfnamefont {W.}~\bibnamefont {Rhee}}, \bibinfo
  {author} {\bibfnamefont {U.}~\bibnamefont {Chang~Hwang}},\ and\ \bibinfo
  {author} {\bibfnamefont {J.-S.}\ \bibnamefont {Choi}},\ }\bibfield  {title}
  {\bibinfo {title} {Blockage effect correction for a scaled wind turbine rotor
  by using wind tunnel test data},\ }\href
  {https://doi.org/10.1016/j.renene.2014.11.057} {\bibfield  {journal}
  {\bibinfo  {journal} {Renewable Energy}\ }\bibinfo {series} {Selected
  {{Papers}} on {{Renewable Energy}}: {{AFORE}} 2013},\ \textbf {\bibinfo
  {volume} {79}},\ \bibinfo {pages} {227} (\bibinfo {year} {2015})}\BibitemShut
  {NoStop}%
\bibitem [{\citenamefont {Badshah}\ \emph {et~al.}(2019)\citenamefont
  {Badshah}, \citenamefont {VanZwieten}, \citenamefont {Badshah},\ and\
  \citenamefont {Jan}}]{badshah_cfd_2019}%
  \BibitemOpen
  \bibfield  {author} {\bibinfo {author} {\bibfnamefont {M.}~\bibnamefont
  {Badshah}}, \bibinfo {author} {\bibfnamefont {J.}~\bibnamefont {VanZwieten}},
  \bibinfo {author} {\bibfnamefont {S.}~\bibnamefont {Badshah}},\ and\ \bibinfo
  {author} {\bibfnamefont {S.}~\bibnamefont {Jan}},\ }\bibfield  {title}
  {\bibinfo {title} {{{CFD}} study of blockage ratio and boundary proximity
  effects on the performance of a tidal turbine},\ }\href
  {https://doi.org/10.1049/iet-rpg.2018.5134} {\bibfield  {journal} {\bibinfo
  {journal} {IET Renewable Power Generation}\ }\textbf {\bibinfo {volume}
  {13}},\ \bibinfo {pages} {744} (\bibinfo {year} {2019})}\BibitemShut
  {NoStop}%
\bibitem [{\citenamefont {{Gauvin-Tremblay}}\ and\ \citenamefont
  {Dumas}(2020)}]{gauvin-tremblay_twoway_2020}%
  \BibitemOpen
  \bibfield  {author} {\bibinfo {author} {\bibfnamefont {O.}~\bibnamefont
  {{Gauvin-Tremblay}}}\ and\ \bibinfo {author} {\bibfnamefont {G.}~\bibnamefont
  {Dumas}},\ }\bibfield  {title} {\bibinfo {title} {Two-way interaction between
  river and deployed cross-flow hydrokinetic turbines},\ }\href
  {https://doi.org/10.1063/5.0004492} {\bibfield  {journal} {\bibinfo
  {journal} {Journal of Renewable and Sustainable Energy}\ }\textbf {\bibinfo
  {volume} {12}},\ \bibinfo {pages} {034501} (\bibinfo {year}
  {2020})}\BibitemShut {NoStop}%
\bibitem [{\citenamefont {Hunt}\ and\ \citenamefont
  {Polagye}(2023)}]{hunt_experimental_2023}%
  \BibitemOpen
  \bibfield  {author} {\bibinfo {author} {\bibfnamefont {A.}~\bibnamefont
  {Hunt}}\ and\ \bibinfo {author} {\bibfnamefont {B.}~\bibnamefont {Polagye}},\
  }\bibfield  {title} {\bibinfo {title} {Experimental techniques for evaluating
  the performance of high-blockage cross-flow turbine arrays},\ }in\ \href
  {https://doi.org/10.36688/ewtec-2023-203} {\emph {\bibinfo {booktitle}
  {Proceedings of the 15th {{European Wave}} and {{Tidal Energy
  Conference}}}}},\ Vol.~\bibinfo {volume} {15}\ (\bibinfo {year}
  {2023})\BibitemShut {NoStop}%
\bibitem [{\citenamefont {Ross}\ and\ \citenamefont
  {Polagye}(2022)}]{ross_effects_2022}%
  \BibitemOpen
  \bibfield  {author} {\bibinfo {author} {\bibfnamefont {H.}~\bibnamefont
  {Ross}}\ and\ \bibinfo {author} {\bibfnamefont {B.}~\bibnamefont {Polagye}},\
  }\bibfield  {title} {\bibinfo {title} {Effects of dimensionless parameters on
  the performance of a cross-flow current turbine},\ }\href
  {https://doi.org/10.1016/j.jfluidstructs.2022.103726} {\bibfield  {journal}
  {\bibinfo  {journal} {Journal of Fluids and Structures}\ }\textbf {\bibinfo
  {volume} {114}},\ \bibinfo {pages} {103726} (\bibinfo {year}
  {2022})}\BibitemShut {NoStop}%
\bibitem [{\citenamefont {Polagye}\ \emph {et~al.}(2019)\citenamefont
  {Polagye}, \citenamefont {Strom}, \citenamefont {Ross}, \citenamefont
  {Forbush},\ and\ \citenamefont {Cavagnaro}}]{polagye_comparison_2019}%
  \BibitemOpen
  \bibfield  {author} {\bibinfo {author} {\bibfnamefont {B.}~\bibnamefont
  {Polagye}}, \bibinfo {author} {\bibfnamefont {B.}~\bibnamefont {Strom}},
  \bibinfo {author} {\bibfnamefont {H.}~\bibnamefont {Ross}}, \bibinfo {author}
  {\bibfnamefont {D.}~\bibnamefont {Forbush}},\ and\ \bibinfo {author}
  {\bibfnamefont {R.~J.}\ \bibnamefont {Cavagnaro}},\ }\bibfield  {title}
  {\bibinfo {title} {Comparison of cross-flow turbine performance under
  torque-regulated and speed-regulated control},\ }\href
  {https://doi.org/10.1063/1.5087476} {\bibfield  {journal} {\bibinfo
  {journal} {Journal of Renewable and Sustainable Energy}\ }\textbf {\bibinfo
  {volume} {11}},\ \bibinfo {pages} {044501} (\bibinfo {year}
  {2019})}\BibitemShut {NoStop}%
\bibitem [{\citenamefont {Araya}\ and\ \citenamefont
  {Dabiri}(2015)}]{araya_comparison_2015}%
  \BibitemOpen
  \bibfield  {author} {\bibinfo {author} {\bibfnamefont {D.~B.}\ \bibnamefont
  {Araya}}\ and\ \bibinfo {author} {\bibfnamefont {J.~O.}\ \bibnamefont
  {Dabiri}},\ }\bibfield  {title} {\bibinfo {title} {A comparison of wake
  measurements in motor-driven and flow-driven turbine experiments},\ }\href
  {https://doi.org/10.1007/s00348-015-2022-7} {\bibfield  {journal} {\bibinfo
  {journal} {Experiments in Fluids}\ }\textbf {\bibinfo {volume} {56}},\
  \bibinfo {pages} {150} (\bibinfo {year} {2015})}\BibitemShut {NoStop}%
\bibitem [{\citenamefont {Goring}\ and\ \citenamefont
  {Nikora}(2002)}]{goring_despiking_2002}%
  \BibitemOpen
  \bibfield  {author} {\bibinfo {author} {\bibfnamefont {D.~G.}\ \bibnamefont
  {Goring}}\ and\ \bibinfo {author} {\bibfnamefont {V.~I.}\ \bibnamefont
  {Nikora}},\ }\bibfield  {title} {\bibinfo {title} {Despiking {{Acoustic
  Doppler Velocimeter Data}}},\ }\href
  {https://doi.org/10.1061/(ASCE)0733-9429(2002)128:1(117)} {\bibfield
  {journal} {\bibinfo  {journal} {Journal of Hydraulic Engineering}\ }\textbf
  {\bibinfo {volume} {128}},\ \bibinfo {pages} {117} (\bibinfo {year}
  {2002})}\BibitemShut {NoStop}%
\bibitem [{\citenamefont {Hunt}\ \emph {et~al.}(2020)\citenamefont {Hunt},
  \citenamefont {Stringer},\ and\ \citenamefont {Polagye}}]{hunt_effect_2020}%
  \BibitemOpen
  \bibfield  {author} {\bibinfo {author} {\bibfnamefont {A.}~\bibnamefont
  {Hunt}}, \bibinfo {author} {\bibfnamefont {C.}~\bibnamefont {Stringer}},\
  and\ \bibinfo {author} {\bibfnamefont {B.}~\bibnamefont {Polagye}},\
  }\bibfield  {title} {\bibinfo {title} {Effect of aspect ratio on cross-flow
  turbine performance},\ }\href {https://doi.org/10.1063/5.0016753} {\bibfield
  {journal} {\bibinfo  {journal} {Journal of Renewable and Sustainable Energy}\
  }\textbf {\bibinfo {volume} {12}},\ \bibinfo {pages} {054501} (\bibinfo
  {year} {2020})}\BibitemShut {NoStop}%
\bibitem [{\citenamefont {Kolekar}\ \emph {et~al.}(2019)\citenamefont
  {Kolekar}, \citenamefont {Vinod},\ and\ \citenamefont
  {Banerjee}}]{kolekar_blockage_2019}%
  \BibitemOpen
  \bibfield  {author} {\bibinfo {author} {\bibfnamefont {N.}~\bibnamefont
  {Kolekar}}, \bibinfo {author} {\bibfnamefont {A.}~\bibnamefont {Vinod}},\
  and\ \bibinfo {author} {\bibfnamefont {A.}~\bibnamefont {Banerjee}},\
  }\bibfield  {title} {\bibinfo {title} {On {{Blockage Effects}} for a {{Tidal
  Turbine}} in {{Free Surface Proximity}}},\ }\href
  {https://doi.org/10.3390/en12173325} {\bibfield  {journal} {\bibinfo
  {journal} {Energies}\ }\textbf {\bibinfo {volume} {12}},\ \bibinfo {pages}
  {3325} (\bibinfo {year} {2019})}\BibitemShut {NoStop}%
\bibitem [{\citenamefont {Miller}\ \emph {et~al.}(2018)\citenamefont {Miller},
  \citenamefont {Duvvuri}, \citenamefont {Brownstein}, \citenamefont {Lee},
  \citenamefont {Dabiri},\ and\ \citenamefont
  {Hultmark}}]{miller_verticalaxis_2018}%
  \BibitemOpen
  \bibfield  {author} {\bibinfo {author} {\bibfnamefont {M.~A.}\ \bibnamefont
  {Miller}}, \bibinfo {author} {\bibfnamefont {S.}~\bibnamefont {Duvvuri}},
  \bibinfo {author} {\bibfnamefont {I.}~\bibnamefont {Brownstein}}, \bibinfo
  {author} {\bibfnamefont {M.}~\bibnamefont {Lee}}, \bibinfo {author}
  {\bibfnamefont {J.~O.}\ \bibnamefont {Dabiri}},\ and\ \bibinfo {author}
  {\bibfnamefont {M.}~\bibnamefont {Hultmark}},\ }\bibfield  {title} {\bibinfo
  {title} {Vertical-{{Axis Wind Turbine Experiments}} at {{Full Dynamic
  Similarity}}},\ }\href {https://doi.org/10.1017/jfm.2018.197} {\bibfield
  {journal} {\bibinfo  {journal} {Journal of Fluid Mechanics}\ }\textbf
  {\bibinfo {volume} {844}},\ \bibinfo {pages} {707} (\bibinfo {year}
  {2018})}\BibitemShut {NoStop}%
\bibitem [{\citenamefont {Miller}\ \emph {et~al.}(2021)\citenamefont {Miller},
  \citenamefont {Duvvuri},\ and\ \citenamefont
  {Hultmark}}]{miller_solidity_2021}%
  \BibitemOpen
  \bibfield  {author} {\bibinfo {author} {\bibfnamefont {M.~A.}\ \bibnamefont
  {Miller}}, \bibinfo {author} {\bibfnamefont {S.}~\bibnamefont {Duvvuri}},\
  and\ \bibinfo {author} {\bibfnamefont {M.}~\bibnamefont {Hultmark}},\
  }\bibfield  {title} {\bibinfo {title} {Solidity effects on the performance of
  vertical-axis wind turbines},\ }\bibfield  {journal} {\bibinfo  {journal}
  {Flow}\ }\textbf {\bibinfo {volume} {1}},\ \href
  {https://doi.org/10.1017/flo.2021.9} {10.1017/flo.2021.9} (\bibinfo {year}
  {2021})\BibitemShut {NoStop}%
\bibitem [{\citenamefont {Bachant}\ and\ \citenamefont
  {Wosnik}(2016)}]{bachant_effects_2016}%
  \BibitemOpen
  \bibfield  {author} {\bibinfo {author} {\bibfnamefont {P.}~\bibnamefont
  {Bachant}}\ and\ \bibinfo {author} {\bibfnamefont {M.}~\bibnamefont
  {Wosnik}},\ }\bibfield  {title} {\bibinfo {title} {Effects of {{Reynolds
  Number}} on the {{Energy Conversion}} and {{Near-Wake Dynamics}} of a {{High
  Solidity Vertical-Axis Cross-Flow Turbine}}},\ }\href
  {https://doi.org/10.3390/en9020073} {\bibfield  {journal} {\bibinfo
  {journal} {Energies}\ }\textbf {\bibinfo {volume} {9}},\ \bibinfo {pages}
  {73} (\bibinfo {year} {2016})}\BibitemShut {NoStop}%
\bibitem [{\citenamefont {Hunt}\ \emph
  {et~al.}(2024{\natexlab{a}})\citenamefont {Hunt}, \citenamefont {Strom},
  \citenamefont {Talpey}, \citenamefont {Ross}, \citenamefont {Scherl},
  \citenamefont {Brunton}, \citenamefont {Wosnik},\ and\ \citenamefont
  {Polagye}}]{hunt_experimental_2024}%
  \BibitemOpen
  \bibfield  {author} {\bibinfo {author} {\bibfnamefont {A.}~\bibnamefont
  {Hunt}}, \bibinfo {author} {\bibfnamefont {B.}~\bibnamefont {Strom}},
  \bibinfo {author} {\bibfnamefont {G.}~\bibnamefont {Talpey}}, \bibinfo
  {author} {\bibfnamefont {H.}~\bibnamefont {Ross}}, \bibinfo {author}
  {\bibfnamefont {I.}~\bibnamefont {Scherl}}, \bibinfo {author} {\bibfnamefont
  {S.}~\bibnamefont {Brunton}}, \bibinfo {author} {\bibfnamefont
  {M.}~\bibnamefont {Wosnik}},\ and\ \bibinfo {author} {\bibfnamefont
  {B.}~\bibnamefont {Polagye}},\ }\bibfield  {title} {\bibinfo {title} {An
  experimental evaluation of the interplay between geometry and scale on
  cross-flow turbine performance},\ }\href
  {https://doi.org/10.1016/j.rser.2024.114848} {\bibfield  {journal} {\bibinfo
  {journal} {Renewable and Sustainable Energy Reviews}\ }\textbf {\bibinfo
  {volume} {206}},\ \bibinfo {pages} {114848} (\bibinfo {year}
  {2024}{\natexlab{a}})}\BibitemShut {NoStop}%
\bibitem [{\citenamefont {Young}\ \emph {et~al.}(2017)\citenamefont {Young},
  \citenamefont {Harwood}, \citenamefont {Montero}, \citenamefont {Ward},\ and\
  \citenamefont {Ceccio}}]{young_ventilation_2017}%
  \BibitemOpen
  \bibfield  {author} {\bibinfo {author} {\bibfnamefont {Y.~L.}\ \bibnamefont
  {Young}}, \bibinfo {author} {\bibfnamefont {C.~M.}\ \bibnamefont {Harwood}},
  \bibinfo {author} {\bibfnamefont {F.~M.}\ \bibnamefont {Montero}}, \bibinfo
  {author} {\bibfnamefont {J.~C.}\ \bibnamefont {Ward}},\ and\ \bibinfo
  {author} {\bibfnamefont {S.~L.}\ \bibnamefont {Ceccio}},\ }\bibfield  {title}
  {\bibinfo {title} {Ventilation of {{Lifting Bodies}}: {{Review}} of the
  {{Physics}} and {{Discussion}} of {{Scaling Effects}}},\ }\href
  {https://doi.org/10.1115/1.4035360} {\bibfield  {journal} {\bibinfo
  {journal} {Applied Mechanics Reviews}\ }\textbf {\bibinfo {volume} {69}},\
  \bibinfo {pages} {010801} (\bibinfo {year} {2017})}\BibitemShut {NoStop}%
\bibitem [{\citenamefont {Zanforlin}\ and\ \citenamefont
  {Nishino}(2016)}]{zanforlin_fluid_2016}%
  \BibitemOpen
  \bibfield  {author} {\bibinfo {author} {\bibfnamefont {S.}~\bibnamefont
  {Zanforlin}}\ and\ \bibinfo {author} {\bibfnamefont {T.}~\bibnamefont
  {Nishino}},\ }\bibfield  {title} {\bibinfo {title} {Fluid dynamic mechanisms
  of enhanced power generation by closely spaced vertical axis wind turbines},\
  }\href {https://doi.org/10.1016/j.renene.2016.08.015} {\bibfield  {journal}
  {\bibinfo  {journal} {Renewable Energy}\ }\textbf {\bibinfo {volume} {99}},\
  \bibinfo {pages} {1213} (\bibinfo {year} {2016})}\BibitemShut {NoStop}%
\bibitem [{\citenamefont {Scherl}(2022)}]{scherl_optimization_2022}%
  \BibitemOpen
  \bibfield  {author} {\bibinfo {author} {\bibfnamefont {I.}~\bibnamefont
  {Scherl}},\ }\emph {\bibinfo {title} {Optimization, {{Modeling}}, and
  {{Control}} of {{Cross-Flow Turbine Arrays}}}},\ \href@noop {} {Ph.D.
  thesis},\ \bibinfo  {school} {University of Washington}, \bibinfo {address}
  {United States -- Washington} (\bibinfo {year} {2022})\BibitemShut {NoStop}%
\bibitem [{\citenamefont {{Gauvin-Tremblay}}\ and\ \citenamefont
  {Dumas}(2022)}]{gauvin-tremblay_hydrokinetic_2022}%
  \BibitemOpen
  \bibfield  {author} {\bibinfo {author} {\bibfnamefont {O.}~\bibnamefont
  {{Gauvin-Tremblay}}}\ and\ \bibinfo {author} {\bibfnamefont {G.}~\bibnamefont
  {Dumas}},\ }\bibfield  {title} {\bibinfo {title} {Hydrokinetic turbine array
  analysis and optimization integrating blockage effects and turbine-wake
  interactions},\ }\href {https://doi.org/10.1016/j.renene.2021.09.003}
  {\bibfield  {journal} {\bibinfo  {journal} {Renewable Energy}\ }\textbf
  {\bibinfo {volume} {181}},\ \bibinfo {pages} {851} (\bibinfo {year}
  {2022})}\BibitemShut {NoStop}%
\bibitem [{hun()}]{hunt_supplemental}%
  \BibitemOpen
  \href@noop {} {\bibinfo {title} {See supplemental material at
  \textcolor{red}{[URL will be inserted by publisher]} for lateral force
  measurements, estimates of hydraulic efficiency, application of steiros
  \textit{et al.}'s model to the experimental data, a visualization of the
  solution space for the generalized blockage adjustment method, sensitivity
  analysis of the phase-averaged and time-averaged piv flow fields, and
  additional figures of piv flow fields at $0.6d$ and $1.5d$ downstream of the
  array and $1.0d$ upstream of the array.}}\BibitemShut {Stop}%
\bibitem [{\citenamefont {Westerweel}\ and\ \citenamefont
  {Scarano}(2005)}]{westerweel_universal_2005}%
  \BibitemOpen
  \bibfield  {author} {\bibinfo {author} {\bibfnamefont {J.}~\bibnamefont
  {Westerweel}}\ and\ \bibinfo {author} {\bibfnamefont {F.}~\bibnamefont
  {Scarano}},\ }\bibfield  {title} {\bibinfo {title} {Universal outlier
  detection for {{PIV}} data},\ }\href
  {https://doi.org/10.1007/s00348-005-0016-6} {\bibfield  {journal} {\bibinfo
  {journal} {Experiments in Fluids}\ }\textbf {\bibinfo {volume} {39}},\
  \bibinfo {pages} {1096} (\bibinfo {year} {2005})}\BibitemShut {NoStop}%
\bibitem [{\citenamefont {Pao}\ and\ \citenamefont
  {Johnson}(2009)}]{pao_tutorial_2009}%
  \BibitemOpen
  \bibfield  {author} {\bibinfo {author} {\bibfnamefont {L.~Y.}\ \bibnamefont
  {Pao}}\ and\ \bibinfo {author} {\bibfnamefont {K.~E.}\ \bibnamefont
  {Johnson}},\ }\bibfield  {title} {\bibinfo {title} {A tutorial on the
  dynamics and control of wind turbines and wind farms},\ }in\ \href
  {https://doi.org/10.1109/ACC.2009.5160195} {\emph {\bibinfo {booktitle} {2009
  {{American Control Conference}}}}}\ (\bibinfo {year} {2009})\ pp.\ \bibinfo
  {pages} {2076--2089}\BibitemShut {NoStop}%
\bibitem [{\citenamefont {Long}(2024)}]{long_threedimensionality_2024}%
  \BibitemOpen
  \bibfield  {author} {\bibinfo {author} {\bibfnamefont {B.}~\bibnamefont
  {Long}},\ }\emph {\bibinfo {title} {Three-Dimensionality of {{Vertical Axis
  Cross-Flow Turbine Flow In High Confinement}}}},\ \href@noop {} {Master's
  thesis},\ \bibinfo  {school} {University of Washington} (\bibinfo {year}
  {2024})\BibitemShut {NoStop}%
\bibitem [{\citenamefont {Bachant}\ and\ \citenamefont
  {Wosnik}(2015)}]{bachant_characterising_2015}%
  \BibitemOpen
  \bibfield  {author} {\bibinfo {author} {\bibfnamefont {P.}~\bibnamefont
  {Bachant}}\ and\ \bibinfo {author} {\bibfnamefont {M.}~\bibnamefont
  {Wosnik}},\ }\bibfield  {title} {\bibinfo {title} {Characterising the
  near-wake of a cross-flow turbine},\ }\href
  {https://doi.org/10.1080/14685248.2014.1001852} {\bibfield  {journal}
  {\bibinfo  {journal} {Journal of Turbulence}\ }\textbf {\bibinfo {volume}
  {16}},\ \bibinfo {pages} {392} (\bibinfo {year} {2015})}\BibitemShut
  {NoStop}%
\bibitem [{\citenamefont {Posa}(2022)}]{posa_wake_2022}%
  \BibitemOpen
  \bibfield  {author} {\bibinfo {author} {\bibfnamefont {A.}~\bibnamefont
  {Posa}},\ }\bibfield  {title} {\bibinfo {title} {Wake characterization of
  paired cross-flow turbines},\ }\href
  {https://doi.org/10.1016/j.renene.2022.07.002} {\bibfield  {journal}
  {\bibinfo  {journal} {Renewable Energy}\ }\textbf {\bibinfo {volume} {196}},\
  \bibinfo {pages} {1064} (\bibinfo {year} {2022})}\BibitemShut {NoStop}%
\bibitem [{\citenamefont {Paraschivoiu}(2002)}]{paraschivoiu_wind_2002}%
  \BibitemOpen
  \bibfield  {author} {\bibinfo {author} {\bibfnamefont {I.}~\bibnamefont
  {Paraschivoiu}},\ }\href@noop {} {\emph {\bibinfo {title} {Wind {{Turbine
  Design}} with {{Emphasis}} on {{Darrieus Concept}}}}}\ (\bibinfo  {publisher}
  {Polytechnic International Press},\ \bibinfo {year} {2002})\BibitemShut
  {NoStop}%
\bibitem [{\citenamefont {Ayati}\ \emph {et~al.}(2019)\citenamefont {Ayati},
  \citenamefont {Steiros}, \citenamefont {Miller}, \citenamefont {Duvvuri},\
  and\ \citenamefont {Hultmark}}]{ayati_doublemultiple_2019}%
  \BibitemOpen
  \bibfield  {author} {\bibinfo {author} {\bibfnamefont {A.~A.}\ \bibnamefont
  {Ayati}}, \bibinfo {author} {\bibfnamefont {K.}~\bibnamefont {Steiros}},
  \bibinfo {author} {\bibfnamefont {M.~A.}\ \bibnamefont {Miller}}, \bibinfo
  {author} {\bibfnamefont {S.}~\bibnamefont {Duvvuri}},\ and\ \bibinfo {author}
  {\bibfnamefont {M.}~\bibnamefont {Hultmark}},\ }\bibfield  {title} {\bibinfo
  {title} {A double-multiple streamtube model for vertical axis wind turbines
  of arbitrary rotor loading},\ }\href {https://doi.org/10.5194/wes-4-653-2019}
  {\bibfield  {journal} {\bibinfo  {journal} {Wind Energy Science}\ }\textbf
  {\bibinfo {volume} {4}},\ \bibinfo {pages} {653} (\bibinfo {year}
  {2019})}\BibitemShut {NoStop}%
\bibitem [{\citenamefont {Houlsby}\ and\ \citenamefont
  {Vogel}(2017)}]{houlsby_power_2017}%
  \BibitemOpen
  \bibfield  {author} {\bibinfo {author} {\bibfnamefont {G.}~\bibnamefont
  {Houlsby}}\ and\ \bibinfo {author} {\bibfnamefont {C.}~\bibnamefont
  {Vogel}},\ }\bibfield  {title} {\bibinfo {title} {The power available to
  tidal turbines in an open channel flow},\ }\href
  {https://doi.org/10.1680/jener.15.00035} {\bibfield  {journal} {\bibinfo
  {journal} {Proceedings of Institution of Civil Engineers: Energy}\ }\textbf
  {\bibinfo {volume} {170}},\ \bibinfo {pages} {12} (\bibinfo {year}
  {2017})}\BibitemShut {NoStop}%
\bibitem [{\citenamefont {Ryan}\ \emph {et~al.}(2016)\citenamefont {Ryan},
  \citenamefont {Coletti}, \citenamefont {Elkins}, \citenamefont {Dabiri},\
  and\ \citenamefont {Eaton}}]{ryan_threedimensional_2016}%
  \BibitemOpen
  \bibfield  {author} {\bibinfo {author} {\bibfnamefont {K.~J.}\ \bibnamefont
  {Ryan}}, \bibinfo {author} {\bibfnamefont {F.}~\bibnamefont {Coletti}},
  \bibinfo {author} {\bibfnamefont {C.~J.}\ \bibnamefont {Elkins}}, \bibinfo
  {author} {\bibfnamefont {J.~O.}\ \bibnamefont {Dabiri}},\ and\ \bibinfo
  {author} {\bibfnamefont {J.~K.}\ \bibnamefont {Eaton}},\ }\bibfield  {title}
  {\bibinfo {title} {Three-dimensional flow field around and downstream of a
  subscale model rotating vertical axis wind turbine},\ }\href
  {https://doi.org/10.1007/s00348-016-2122-z} {\bibfield  {journal} {\bibinfo
  {journal} {Experiments in Fluids}\ }\textbf {\bibinfo {volume} {57}},\
  \bibinfo {pages} {1} (\bibinfo {year} {2016})}\BibitemShut {NoStop}%
\bibitem [{\citenamefont {Bahaj}\ \emph {et~al.}(2007)\citenamefont {Bahaj},
  \citenamefont {Molland}, \citenamefont {Chaplin},\ and\ \citenamefont
  {Batten}}]{bahaj_power_2007}%
  \BibitemOpen
  \bibfield  {author} {\bibinfo {author} {\bibfnamefont {A.~S.}\ \bibnamefont
  {Bahaj}}, \bibinfo {author} {\bibfnamefont {A.~F.}\ \bibnamefont {Molland}},
  \bibinfo {author} {\bibfnamefont {J.~R.}\ \bibnamefont {Chaplin}},\ and\
  \bibinfo {author} {\bibfnamefont {W.~M.~J.}\ \bibnamefont {Batten}},\
  }\bibfield  {title} {\bibinfo {title} {Power and thrust measurements of
  marine current turbines under various hydrodynamic flow conditions in a
  cavitation tunnel and a towing tank},\ }\href
  {https://doi.org/10.1016/j.renene.2006.01.012} {\bibfield  {journal}
  {\bibinfo  {journal} {Renewable Energy}\ }\textbf {\bibinfo {volume} {32}},\
  \bibinfo {pages} {407} (\bibinfo {year} {2007})}\BibitemShut {NoStop}%
\bibitem [{\citenamefont {Medici}\ and\ \citenamefont
  {Alfredsson}(2006)}]{medici_measurements_2006}%
  \BibitemOpen
  \bibfield  {author} {\bibinfo {author} {\bibfnamefont {D.}~\bibnamefont
  {Medici}}\ and\ \bibinfo {author} {\bibfnamefont {P.~H.}\ \bibnamefont
  {Alfredsson}},\ }\bibfield  {title} {\bibinfo {title} {Measurements on a wind
  turbine wake: {{3D}} effects and bluff body vortex shedding},\ }\href
  {https://doi.org/10.1002/we.156} {\bibfield  {journal} {\bibinfo  {journal}
  {Wind Energy}\ }\textbf {\bibinfo {volume} {9}},\ \bibinfo {pages} {219}
  (\bibinfo {year} {2006})}\BibitemShut {NoStop}%
\bibitem [{\citenamefont {Chamorro}\ \emph {et~al.}(2012)\citenamefont
  {Chamorro}, \citenamefont {Arndt},\ and\ \citenamefont
  {Sotiropoulos}}]{chamorro_reynolds_2012}%
  \BibitemOpen
  \bibfield  {author} {\bibinfo {author} {\bibfnamefont {L.~P.}\ \bibnamefont
  {Chamorro}}, \bibinfo {author} {\bibfnamefont {R.}~\bibnamefont {Arndt}},\
  and\ \bibinfo {author} {\bibfnamefont {F.}~\bibnamefont {Sotiropoulos}},\
  }\bibfield  {title} {\bibinfo {title} {Reynolds number dependence of
  turbulence statistics in the wake of wind turbines},\ }\href
  {https://doi.org/10.1002/we.501} {\bibfield  {journal} {\bibinfo  {journal}
  {Wind Energy}\ }\textbf {\bibinfo {volume} {15}},\ \bibinfo {pages} {733}
  (\bibinfo {year} {2012})}\BibitemShut {NoStop}%
\bibitem [{\citenamefont {Araya}\ \emph {et~al.}(2017)\citenamefont {Araya},
  \citenamefont {Colonius},\ and\ \citenamefont
  {Dabiri}}]{araya_transition_2017}%
  \BibitemOpen
  \bibfield  {author} {\bibinfo {author} {\bibfnamefont {D.~B.}\ \bibnamefont
  {Araya}}, \bibinfo {author} {\bibfnamefont {T.}~\bibnamefont {Colonius}},\
  and\ \bibinfo {author} {\bibfnamefont {J.~O.}\ \bibnamefont {Dabiri}},\
  }\bibfield  {title} {\bibinfo {title} {Transition to bluff-body dynamics in
  the wake of vertical-axis wind turbines},\ }\href
  {https://doi.org/10.1017/jfm.2016.862} {\bibfield  {journal} {\bibinfo
  {journal} {Journal of Fluid Mechanics}\ }\textbf {\bibinfo {volume} {813}},\
  \bibinfo {pages} {346} (\bibinfo {year} {2017})}\BibitemShut {NoStop}%
\bibitem [{\citenamefont {Alexander}\ and\ \citenamefont
  {Holownia}(1978)}]{alexander_wind_1978}%
  \BibitemOpen
  \bibfield  {author} {\bibinfo {author} {\bibfnamefont {A.~J.}\ \bibnamefont
  {Alexander}}\ and\ \bibinfo {author} {\bibfnamefont {B.~P.}\ \bibnamefont
  {Holownia}},\ }\bibfield  {title} {\bibinfo {title} {Wind tunnel tests on a
  savonius rotor},\ }\href {https://doi.org/10.1016/0167-6105(78)90037-5}
  {\bibfield  {journal} {\bibinfo  {journal} {Journal of Wind Engineering and
  Industrial Aerodynamics}\ }\textbf {\bibinfo {volume} {3}},\ \bibinfo {pages}
  {343} (\bibinfo {year} {1978})}\BibitemShut {NoStop}%
\bibitem [{\citenamefont {Hunt}\ \emph
  {et~al.}(2024{\natexlab{b}})\citenamefont {Hunt}, \citenamefont {Athair},
  \citenamefont {Williams},\ and\ \citenamefont {Polagye}}]{hunt_data_2024a}%
  \BibitemOpen
  \bibfield  {author} {\bibinfo {author} {\bibfnamefont {A.}~\bibnamefont
  {Hunt}}, \bibinfo {author} {\bibfnamefont {A.}~\bibnamefont {Athair}},
  \bibinfo {author} {\bibfnamefont {O.}~\bibnamefont {Williams}},\ and\
  \bibinfo {author} {\bibfnamefont {B.}~\bibnamefont {Polagye}},\ }\href
  {https://doi.org/10.5061/dryad.dfn2z35b5} {\bibinfo {title} {Data from:
  {{Experimental}} validation of a linear momentum and bluff-body model for
  high-blockage cross-flow turbine arrays. [{{Dataset}}]. {{Dryad}}.}}
  (\bibinfo {year} {2024}{\natexlab{b}})\BibitemShut {NoStop}%
\end{thebibliography}%


\newpage
\nolinenumbers

\clearpage
\pagenumbering{arabic}
\renewcommand*{\thepage}{S\arabic{page}}

\renewcommand{\thefigure}{S.\arabic{figure}}
\setcounter{figure}{0}
\renewcommand{\thetable}{S.\arabic{table}}
\setcounter{table}{0}
\renewcommand{\theequation}{S.\arabic{equation}}
\setcounter{equation}{0}
\renewcommand{\thesection}{S.\arabic{section}}
\setcounter{section}{0}

\begin{centering}
    \Large
    \textbf{Supplemental material for ``Experimental validation of a linear momentum and bluff-body model for high-blockage cross-flow turbine arrays''}

    \vspace{1cm}

    \large
    Aidan Hunt\textsuperscript{1,$\ast$},
    Ari Athair\textsuperscript{2},
    Owen Williams\textsuperscript{2},
    and Brian Polagye\textsuperscript{1}

    \normalsize
    \textsuperscript{1} Department of Mechanical Engineering, University of Washington.
    
    \textsuperscript{2} Department of Aeronautics and Astronautics, University of Washington.
    
    \textsuperscript{$\ast$:} To whom correspondence should be addressed: ahunt94@uw.edu.

    \vspace{2cm}

\end{centering}

This PDF file contains the following supplementary material:

\begin{itemize}[align=right,
                  leftmargin=2em,
                  topsep=0.1em,
                  itemsep=0.1em]
    \item Lateral force coefficients for the cross-flow turbine array,
    \item Estimations of hydraulic efficiency for the cross-flow turbine array,
    \item Evaluation of the analytical model of \citet{steiros_analytical_2022},
    \item Solution space for the generalized blockage adjustment method presented in \cref{sec:lmadForecasting},
    \item Sensitivity analysis for the phase-averaged and time-averaged PIV flow fields.
    \item Additional figures of PIV flow fields $0.6D$ and $1.5D$ downstream of the array and $1.0D$ upstream of the array.
\end{itemize}

\newpage
\section*{Lateral force coefficients}

Since the turbines in the array are counter-rotated with zero phase offset between them, the net lateral force on the array is approximately zero since the array is symmetric about its centerline.
However, by defining the directions of positive $L_{A}$ and $L_{B}$ with this counter-rotation in mind as in \cref{fig:arrayOverhead}, the array-average lateral force coefficient is nonzero and represents the average lateral force coefficient experienced by an individual rotor in the array as

\begin{equation}
    C_{L,\mathrm{rotor}} = \frac{L_A + L_B}{\frac{1}{2}\rho U_{\infty}^2 A_{\mathrm{turbines}}} \ \ .
\end{equation}

\begin{figure}[hbt!]
     \centering
     \includegraphics[width=0.6\textwidth]{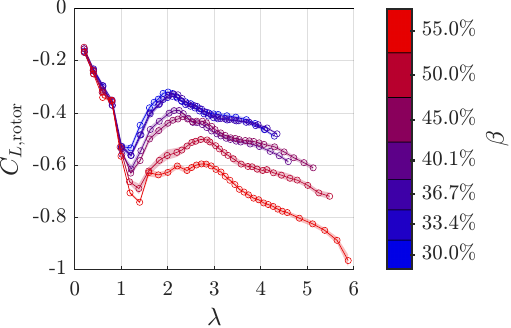}
     \caption{Time-averaged rotor-average lateral force coefficient as a function of $\beta$ and $\lambda$. The shaded regions indicate the interquartile range of the array- and cycle-averaged performance at each $\beta$ and $\lambda$ (the vertical span of the shaded region at each point is similar to the size of the markers for most operating conditions).}
     \label{fig:arrayLat}
 \end{figure}

 \Cref{fig:arrayLat} shows the effect of blockage on $C_{L,\mathrm{rotor}} - \lambda$ curves across the tested $\beta$ and $\lambda$.
 As for $C_{P}$ and $C_{T}$, the magnitude of $C_{L,\mathrm{rotor}}$ at a particular $\lambda$ tends to increase with the blockage ratio.
 Additionally in a similar manner to the array power and thrust coefficients, $C_{L,\mathrm{rotor}}$ also exhibits invariance to $\beta$ at lower tip-speed ratios ($\lambda \leq 1$).

\newpage
\section*{Estimating Hydraulic Efficiency}

As noted in \Cref{sec:results}, the conventional definition of the array power coefficient (\cref{eq:cp}) neglects the potential energy that may be harnessed by the array in a confined flow.
Consequently, at higher blockage, array efficiency may be better described using a hydraulic efficiency ($\eta$) that considers both the static and dynamic head in the flow:
\begin{equation}
    \eta = \frac{Q_{A}\omega_{A} + Q_{B}\omega_{B}}{\dot{m} g \Delta h} \ \ ,
    \label{eq:hEff}
\end{equation}
where $\dot{m}$ is the mass flow rate through the channel and $\Delta h$ is the difference in the water depth between far upstream and far downstream.
Similar definitions of efficiency were employed by \citet{takamatsu_study_1985} and \citet{mcadam_experimental_2013} in their studies of high-blockage cross-flow turbines.
However, both $\dot{m}$ and $\Delta h$ can be difficult to measure in practice.
In this study, instrumentation to measure the volumetric flow rate through the flume was unavailable, and the volumetric flow rate is expected to vary with $\lambda$ at each $\beta$ due to interactions between the array and the channel \citep{hunt_experimental_2023}.
Additionally, accurate measurements of $\Delta h$ require a channel of sufficient length to allow the flow to fully mix downstream of the array and sufficient measurement resolution to identify this condition.
For the experimental set-up in this study, the finite length of the flume downstream of the cross-flow turbine array is likely insufficient to allow the flow to fully mix, and the furthest downstream measurement of the water depth was taken only $2D$ downstream of the array.

\begin{figure}[t]
     \centering
     \includegraphics[width=\textwidth]{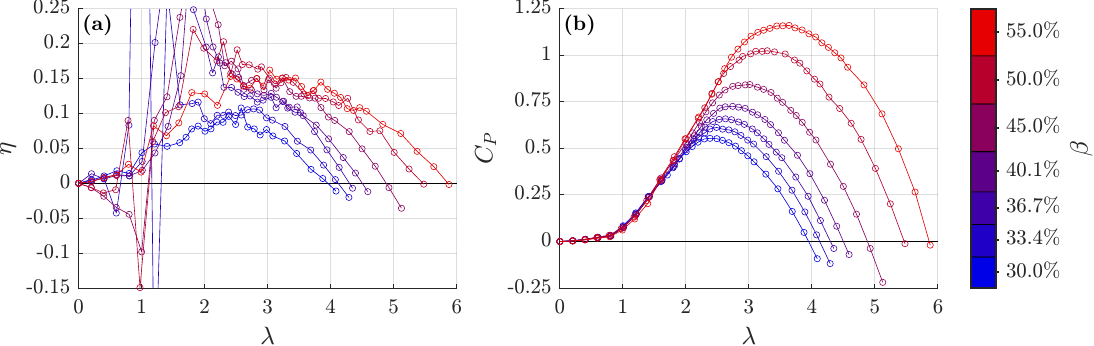}
      {\phantomsubcaption\label{fig:hEff_eta}
       \phantomsubcaption\label{fig:hEff_cp}
      }
     \caption{\subref{fig:hEff_eta} Time-average hydraulic efficiency estimated for the array, compared to \subref{fig:hEff_cp} the time-average array-average $C_P$ (reproduced from \cref{fig:arrayPerf_cp}). The axes limits in \cref{fig:hEff_eta} are cropped to focus on non-spurious values of $\eta$.}
     \label{fig:hEff}
 \end{figure}

Although precise measurements of $\dot{m}$ and $\Delta h$ are not available in this study, we present a rough estimate of $\eta$ for the experimental array across the tested range of $\beta$ and $\lambda$ using the available velocity and depth measurements.
Assuming a uniform velocity profile in the Tyler flume, we estimate $\dot{m}\approx \rho U_{\infty}hw$.
$\Delta h$ is estimated as the difference between the furthest upstream measured water depth (at $x/D = -5.8$) and furthest downstream measured water depth (at $x/D = 2$).
The resulting hydraulic efficiencies are expressed as a function of $\beta$ and $\lambda$ in \cref{fig:hEff_eta}.
Unlike $C_P$, the values of $\eta$ are $\leq 1$, with the exception of several spikes in the estimated $\eta$ at lower $\lambda$.
These spikes occur when $\Delta h \approx 0$ between the two sample locations resulting from an irregular and unsteady free surface, and emphasizes the spatial and temporal resolution required to characterize the free surface deformation for the purposes of computing hydraulic efficiency.
As $\lambda$ increases, the thrust on the array results in a more defined free surface profile and less noise in the estimates of hydraulic efficiency, with trends in $\eta$ with blockage largely following those of $C_P$ (\cref{fig:hEff_cp}).
Nonetheless, the $\eta$ shown are likely lower than the true hydraulic efficiencies, due to overestimation of the mass flow rate and total head drop from the available velocity and depth measurements.

\newpage
\section*{Application of Steiros et al.'s model to the experimental data}

\subsection{Prediction of Near-Wake Velocities}

Here, we apply a confinement model based on two-dimensional potential flow theory developed by \citet{steiros_analytical_2022} to the experimental data to predict the bypass velocities ($u_b$) and wake velocities ($u_w$) from measurements of $C_T$ and $U_{\infty}$ at each $\beta$.
In this model, free surface deformation is assumed to be negligible (i.e., $Fr_h \rightarrow 0$).
Key details of the modeling approach and resulting equation set are summarized here; a detailed derivation is provided in \citet{steiros_analytical_2022}, with the underlying porous plate theory described by \citet{steiros_drag_2018}.
\citeauthor{steiros_analytical_2022} follow the theory of \citet{taylor_air_1944} for flow through and around a porous plate by representing the turbine as a line of evenly distributed potential sources in a freestream flow.
The induced flow field through and around the turbine is then determined through superposition, with adjustments applied to ensure continuity through the turbine \citep{koo_fluid_1973} and to allow the wake pressure to fall below the ambient pressure \citep{steiros_analytical_2022}.
Flow confinement is imposed through the method of mirror images by considering an infinite number of turbines arranged side-by-side with a distance $w$ between their centers.
The thrust on the turbine is then computed by applying conservation of mass, momentum, and energy in a control volume encompassing the modeled flow field around the turbine, yielding
\begin{equation}
    C_T = \frac{4(u^{\ast} -1)(1 - u^{\ast})}{(1-\beta)(2 - u^{\ast} - u^{\ast}\beta)} \left( \frac{1 - u{\ast}}{3} - \frac{1 - 2u^{\ast}\beta + \beta}{1 - \beta} \right) ,
    \label{eq:CT_steiros}
\end{equation}
\noindent where $u^{\ast} = u_t/U_{\infty}$ is the normalized velocity through the turbine rotor.

As for the LMADT model of \citet{houlsby_application_2008}, in applying the model of \citet{steiros_analytical_2022} we treat the array as a single turbine with the same blockage ratio as the array.
Using the measured $C_T$ and $\beta$, \cref{eq:CT_steiros} may be solved numerically for $u^{\ast}$.
Then, the bypass velocity and wake velocity may be calculated respectively as
\begin{equation}
    u_b = U_{\infty} \left( \frac{1 - 2u^{\ast}\beta + \beta}{1 - \beta} \right) \ ,
    \label{eq:ub_steiros}
\end{equation}
and
\begin{equation}
    u_w = U_{\infty} \left( \frac{u^{\ast}(1 - 2u^{\ast}\beta + \beta)}{2 - u^{\ast} - u^{\ast}\beta} \right) \ .
    \label{eq:uw_steiros}
\end{equation}

\begin{figure}
    \centering
    \includegraphics[width=\textwidth]{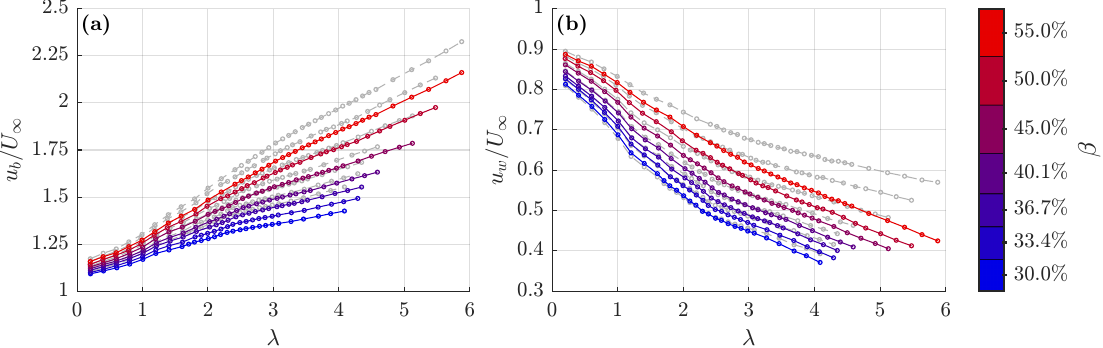}
    {\phantomsubcaption\label{fig:steiros_ub}
    \phantomsubcaption\label{fig:steiros_uw}
    }
    \caption{The \subref{fig:steiros_ub} bypass velocities, and \subref{fig:steiros_uw} wake velocities (all normalized by the freestream velocity) predicted by the model of \citet{steiros_analytical_2022} from measurements of array thrust and flow conditions. The near-wake velocities predicted by the LMADT model of \citet{houlsby_application_2008} from \Cref{fig:lmadVels} are shown in gray for comparison.}
    \label{fig:steiros_vels}
\end{figure}

\begin{figure}
    \centering
    \includegraphics[width=\textwidth]{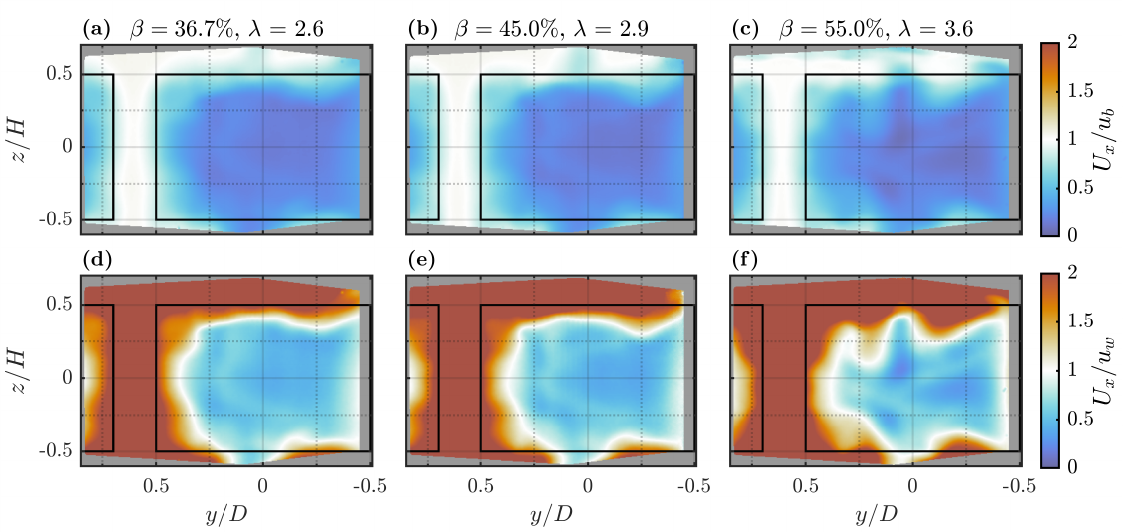}
    {\phantomsubcaption\label{fig:pivSteiros_bypass_36}
    \phantomsubcaption\label{fig:pivSteiros_bypass_45}
    \phantomsubcaption\label{fig:pivSteiros_bypass_55}
    \phantomsubcaption\label{fig:pivSteiros_wake_36}
    \phantomsubcaption\label{fig:pivSteiros_wake_45}
    \phantomsubcaption\label{fig:pivSteiros_wake_55}
    }
    \caption{Time-averaged streamwise velocities measured $0.6D$ downstream of the array at $\beta = 36.7\%$, $45.0\%$, and $55.0\%$ and optimal $\lambda$, normalized by the corresponding \subref{fig:pivScaled_bypass_36}-\subref{fig:pivScaled_bypass_55} bypass velocity predicted by the model of \citet{steiros_analytical_2022}, and \subref{fig:pivScaled_wake_36}-\subref{fig:pivScaled_wake_55} wake velocity predicted by the model of \citet{steiros_analytical_2022}. The black rectangles indicate the projected area of the turbines in the field of view.}
    \label{fig:pivScaled_steiros}
\end{figure}

\begin{figure*}
     \centering
     \includegraphics[width=\textwidth]{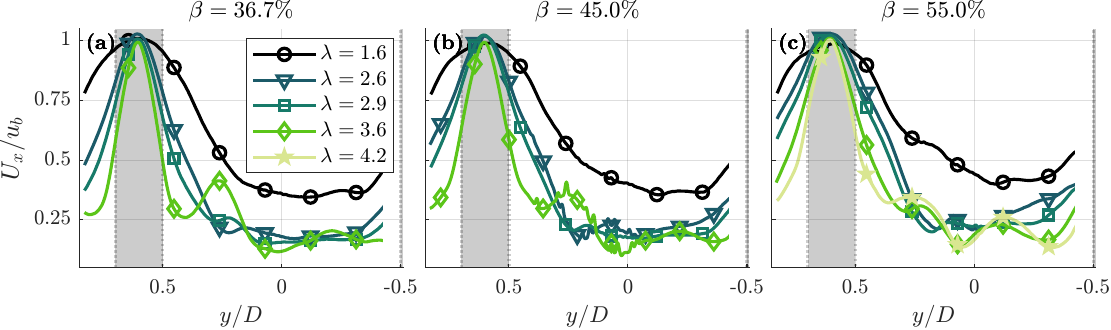}
      {\phantomsubcaption\label{fig:pivlineSteiros_bypass_36}
      \phantomsubcaption\label{fig:pivlineSteiros_bypass_45}
      \phantomsubcaption\label{fig:pivlineSteiros_bypass_55}
      }
     \caption{Time-averaged streamwise velocity profiles measured $0.6D$ downstream of the array normalized by the bypass velocity predicted from the model of \citet{steiros_analytical_2022} at \subref{fig:pivlineScaled_bypass_36} $\beta = 36.7\%$, \subref{fig:pivlineScaled_bypass_45} $45.0\%$, and \subref{fig:pivlineScaled_bypass_55} $55.0\%$ and various $\lambda$.
     The profiles shown correspond to the average velocity along the middle $70\%$ of the turbine blade span. Vertical gray rectangles correspond to the bypass region.}
     \label{fig:pivlineSteiros}
 \end{figure*}

\begin{figure*}
      \centering
      \includegraphics[width=\textwidth]{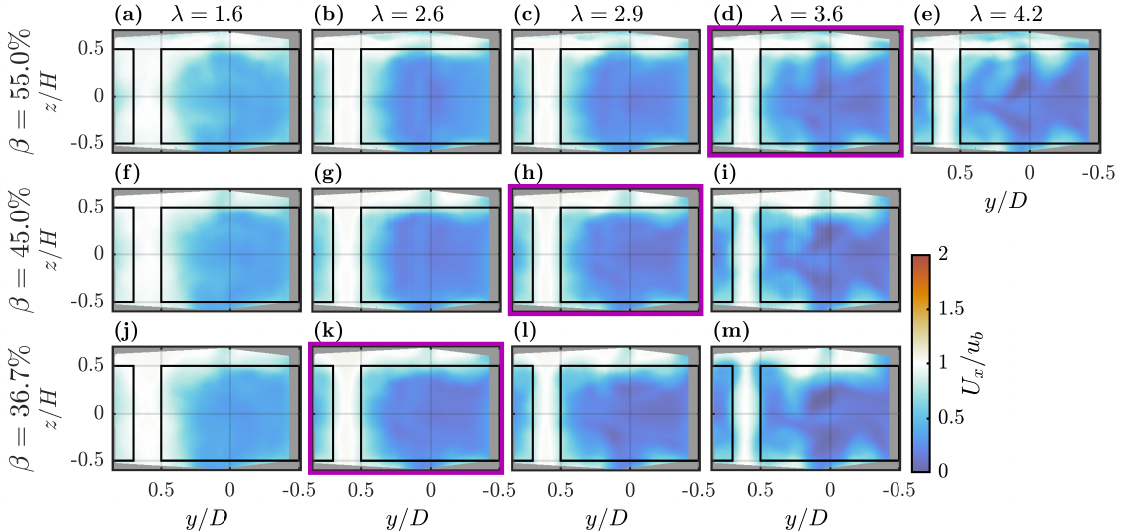}
      \caption{Time-averaged streamwise velocity $0.6D$ downstream of the array as function of $\beta$ and $\lambda$, normalized by the corresponding bypass velocity, $u_b$, predicted by \ul{the model of Steiros \textit{et al.}} \citep{steiros_analytical_2022}. The black rectangles in each tile indicate the projected area of the turbines in the field of view. Tiles corresponding to the optimal $\lambda$ for each blockage are outlined in purple (tiles (d), (h), and (k)), and correspond to \cref{fig:pivSteiros_bypass_36,fig:pivSteiros_bypass_45,fig:pivSteiros_bypass_55}.}
      \label{fig:pivD06Ub_steiros}
\end{figure*}

\begin{figure*}
      \centering
      \includegraphics[width=\textwidth]{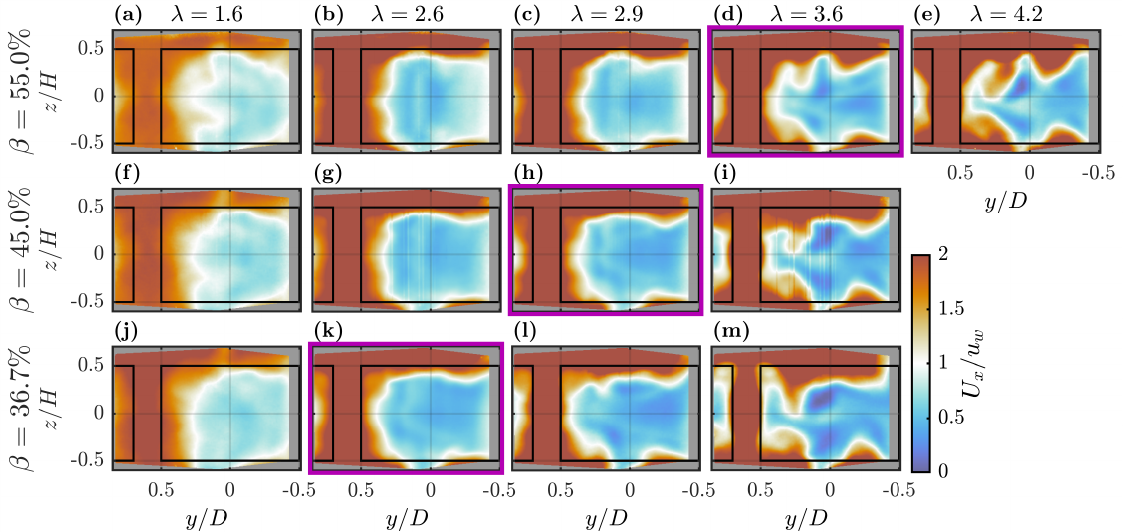}
      \caption{Time-averaged streamwise velocity $0.6D$ downstream of the array as function of $\beta$ and $\lambda$, normalized by the corresponding wake velocity, $u_w$, predicted by \ul{the model of Steiros \textit{et al.}} \citep{steiros_analytical_2022}. The black rectangles in each tile indicate the projected area of the turbines in the field of view. Flow fields corresponding to the optimal $\lambda$ for each blockage are outlined in purple (tiles (d), (h), and (k)), and correspond to \cref{fig:pivSteiros_wake_36,fig:pivSteiros_wake_45,fig:pivSteiros_wake_55}.}
      \label{fig:pivD06Uw_steiros}
\end{figure*}

\begin{figure*}
      \centering
      \includegraphics[width=\textwidth]{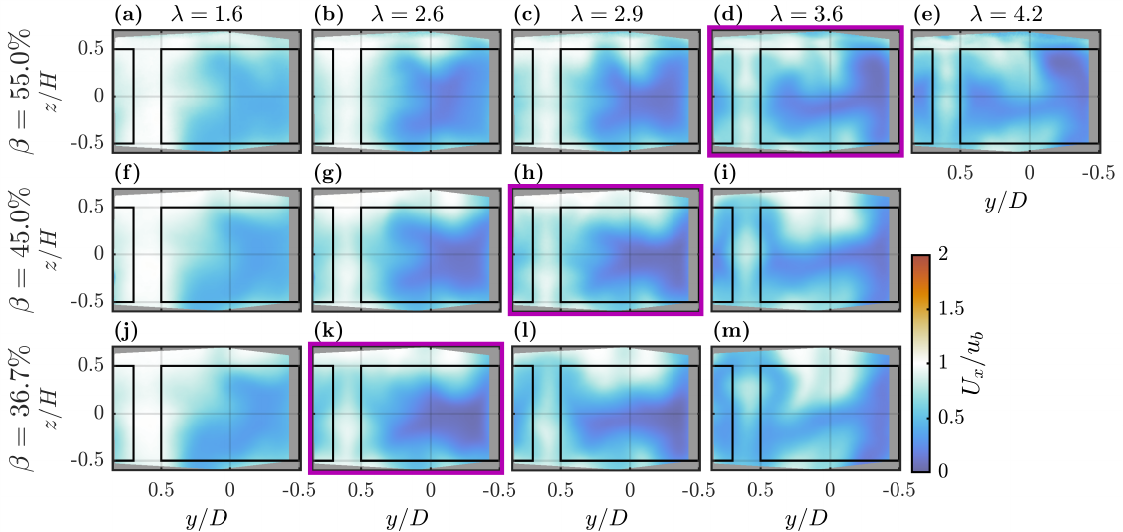}
      \caption{Time-averaged streamwise velocity $1.5D$ downstream of the array as function of $\beta$ and $\lambda$, normalized by the corresponding bypass velocity, $u_b$, predicted by \ul{the model of Steiros \textit{et al.}} \citep{steiros_analytical_2022}. The black rectangles in each tile indicate the projected area of the turbines in the field of view. Tiles corresponding to the optimal $\lambda$ for each blockage are outlined in purple (tiles (d), (h), and (k)).}
      \label{fig:pivD15Ub_steiros}
\end{figure*}

\begin{figure*}
      \centering
      \includegraphics[width=\textwidth]{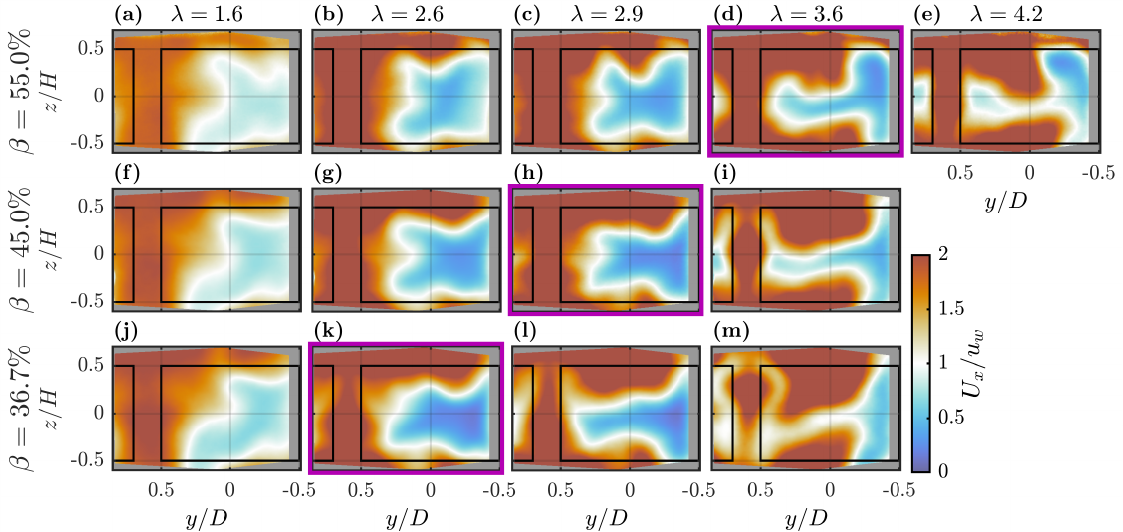}
      \caption{Time-averaged streamwise velocity $1.5D$ downstream of the array as function of $\beta$ and $\lambda$, normalized by the corresponding wake velocity, $u_w$, predicted by \ul{the model of Steiros \textit{et al.}} \citet{steiros_analytical_2022}. The black rectangles in each tile indicate the projected area of the turbines in the field of view. Flow fields corresponding to the optimal $\lambda$ for each blockage are outlined in purple (tiles (d), (h), and (k)).}
      \label{fig:pivD15Uw_steiros}
\end{figure*}

The resulting $u_b$ and $u_w$ are normalized by the freestream velocity and plotted in \Cref{fig:steiros_vels}.
Trends in the near-wake velocities predicted by the model of \citet{steiros_analytical_2022} are the same as those for the near-wake velocities predicted by \citet{houlsby_application_2008}: $u_b/U_{\infty}$ increases with $\lambda$ whereas $u_w/U_{\infty}$ decreases with $\lambda$, and both $u_b/U_{\infty}$ and $u_w/U_{\infty}$ increase with $\beta$.
While both models yield similar predictions for the wake and bypass velocities at low $\lambda$, \citeauthor{steiros_analytical_2022}'s model predicts lower $u_b$ and $u_w$ than \citeauthor{houlsby_application_2008}'s model as $\lambda$ increases.
Despite these differences, the $u_b$ obtained from \citeauthor{steiros_analytical_2022}'s model shows good agreement with the streamwise velocity measured in the bypass region $0.6D$ downstream of the array (\Cref{fig:pivSteiros_bypass_36,fig:pivSteiros_bypass_45,fig:pivSteiros_bypass_55}, \Cref{fig:pivlineSteiros}).
In contrast, the experimentally measured wake velocity is still poorly predicted by \citeauthor{steiros_analytical_2022}'s model at this location \Cref{fig:pivSteiros_bypass_36,fig:pivSteiros_bypass_55}.
Flow fields at other $\beta$, $\lambda$, and streamwise locations normalized by $u_b$ and $u_w$ obtained from \citeauthor{steiros_analytical_2022}'s model are subsequently provided in \Cref{fig:pivD06Ub_steiros,fig:pivD06Uw_steiros,fig:pivD15Ub_steiros,fig:pivD15Uw_steiros}.

\subsection{Blockage Correction}

Here, we summarize the blockage correction method proposed by \citet{steiros_analytical_2022}, which is used to obtain the ${C_T}'-\lambda'$ and ${C_P}'-\lambda'$ curves in \Cref{fig:compare_ct_UPrime,fig:compare_cp_UPrime}.
The blockage correction is performed by determining the unconfined freestream velocity, ${U_{\infty}}'$, that results in the same $u_t$ and dimensional thrust at $\beta=0\%$ as for the confined condition, i.e.
\begin{equation}
    u_t = {u_t}' ,
\end{equation}
\begin{equation}
    T = {T}' .
\end{equation}
Therefore, it follows that
\begin{equation}
    {C_T}'({U_{\infty}}')^2 = {C_T}(U_{\infty})^2 \ , 
\end{equation}
and consequently,
\begin{equation}
    {C_T}' 
    = {C_T}\left( \frac{U_{\infty}}{{U_{\infty}}'} \right)^2
    = {C_T}\left( \frac{u^{\ast}}{{u^{\ast}}'} \right)^2 
    \label{eq:ctPrime_steiros}
\end{equation}
since $u_t$ = ${u_t}'$.
Following the procedure outlined by \citet{steiros_analytical_2022}, the blockage correction is performed by first calculating $u^{\ast}$ from the known $C_T$ and $\beta$ at a given confined condition using \cref{eq:CT_steiros}.
Then, ${u^{\ast}}'$ is calculated by simultaneously solving \cref{eq:CT_steiros,eq:ctPrime_steiros} for the value of ${u^{\ast}}'$ that yields equivalent values of ${C_T}'$, with $\beta = 0\%$ used in \cref{eq:CT_steiros}.
Once ${u^{\ast}}'$ is known, ${C_T}'$ is computed from \cref{eq:ctPrime_steiros} and ${C_P}'$ and $\lambda'$ are computed as
\begin{equation}
    {C_P}' 
    = {C_P}\left( \frac{u^{\ast}}{{u^{\ast}}'} \right)^3 \,  
    \label{eq:cpPrime_steiros}
\end{equation}
\begin{equation}
    {\lambda}' 
    = {\lambda}\left( \frac{u^{\ast}}{{u^{\ast}}'} \right)
    \label{eq:lambdaPrime_steiros}
\end{equation}

\subsection{Generalized Blockage Adjustment using Steiros' et Al's model}

\begin{figure*}
     \centering
     \includegraphics[width=\textwidth]{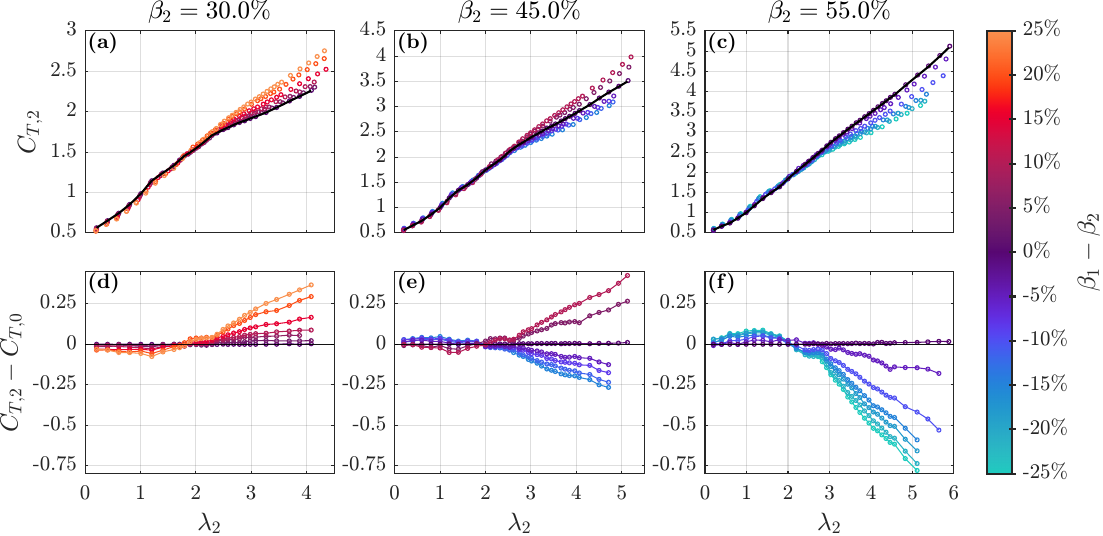}
     {\phantomsubcaption\label{fig:steirosCT_30}
      \phantomsubcaption\label{fig:steirosCT_45}
      \phantomsubcaption\label{fig:steirosCT_55}
      \phantomsubcaption\label{fig:steirosCT_30_error}
      \phantomsubcaption\label{fig:steirosCT_45_error}
      \phantomsubcaption\label{fig:steirosCT_55_error}
     }
     \caption{Confined $C_T$ predicted from the experimental data using the Glauert-derived blockage correction method proposed by \citet{steiros_analytical_2022}. \subref{fig:steirosCT_30}-\subref{fig:steirosCT_55} Predicted thrust coefficient and \subref{fig:steirosCT_30_error}-\subref{fig:steirosCT_55_error} error between the predicted thrust coefficient and measured thrust coefficient ($C_{T,0}$, shown as the gray lines in \subref{fig:steirosCT_30}-\subref{fig:steirosCT_55}) for $\beta_2 = 30.0\%$, $45.0\%$, and $55.0\%$. Color indicates the difference between the blockage ratio of the performance data that was used to make each prediction ($\beta_1$) and the target blockage ratio ($\beta_2$). Note the different axes limits in \subref{fig:steirosCT_30}-\subref{fig:steirosCT_55}.}
     \label{fig:forecastCT_steiros}
 \end{figure*}

\begin{figure*}
    \centering
    \includegraphics[width=\textwidth]{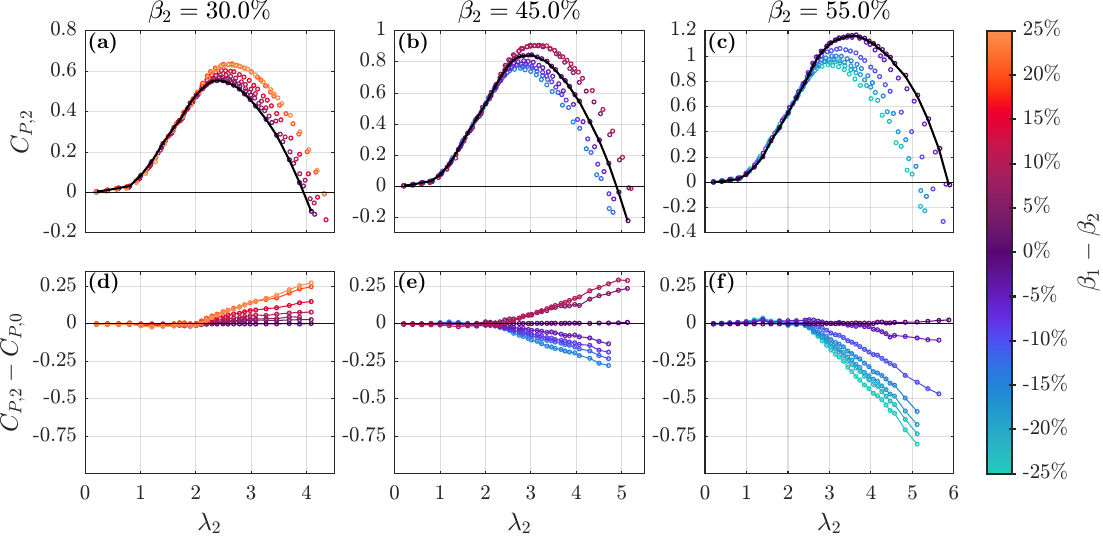}
    {\phantomsubcaption\label{fig:steirosCP_30}
    \phantomsubcaption\label{fig:steirosCP_45}
    \phantomsubcaption\label{fig:steirosCP_55}
    \phantomsubcaption\label{fig:steirosCP_30_error}
    \phantomsubcaption\label{fig:steirosCP_45_error}
    \phantomsubcaption\label{fig:steirosCP_55_error}
    }
    \caption{Confined $C_P$ predicted from the experimental data using Glauert-derived the blockage correction method proposed by \citet{steiros_analytical_2022}. \subref{fig:steirosCP_30}-\subref{fig:steirosCP_55} Predicted power coefficient and \subref{fig:steirosCP_30_error}-\subref{fig:steirosCP_55_error} error between the predicted power coefficient and measured power coefficient ($C_{T,0}$, shown as the gray lines in \subref{fig:steirosCP_30}-\subref{fig:steirosCP_55}) for $\beta_2 = 30.0\%$, $45.0\%$, and $55.0\%$. Color indicates the difference between the blockage ratio of the performance data that was used to make each prediction ($\beta_1$) and the target blockage ratio ($\beta_2$). Note the different axes limits in \subref{fig:steirosCP_30}-\subref{fig:steirosCP_55}.}
    \label{fig:forecastCP_steiros}
 \end{figure*}

The Glauert-derived blockage correction method of \citet{steiros_analytical_2022} may also used to ``correct'' confined performance performance at one blockage ratio ($\beta_1$) to a different blockage ratio ($\beta_2$) by simply specifying a non-zero target value of $\beta$ when solving \cref{eq:CT_steiros} for ${u^{\ast}}'$.
While \citeauthor{steiros_analytical_2022} note this generalized blockage adjustment capability, the evaluation of their model is limited to prediction of low-blockage ($\beta = 3\%$) performance.
Here, we apply this blockage adjustment approach to the experimental data and present the results in a similar manner to \Cref{sec:lmadForecasting}.
\Cref{fig:forecastCT_steiros} shows the predicted thrust coefficients at $\beta_2 = 30.0\%$, $45.0\%$, and $55.0\%$ when \citeauthor{steiros_analytical_2022}'s generalized blockage correction is applied to the experimental data.
$C_T$ is accurately predicted at lower $\lambda$, but error between the predicted and measured $C_T$ becomes large at higher $\lambda$.
Similar trends are observed for the predicted $C_P$ using this method (\Cref{fig:forecastCP_steiros}).
For both $C_T$ and $C_P$, prediction accuracy becomes worse for a larger difference between $\beta_1$ and $\beta_2$.
This result is consistent with \citeauthor{steiros_analytical_2022}'s observations of the limitations of their model's accuracy at high $\lambda$.
Trends in prediction accuracy resemble trends in blockage-corrected ${C_T}'-\lambda'$ and ${C_P}'-\lambda'$ curves \Cref{fig:compare_ct_UPrime,fig:compare_cp_UPrime}, which form the basis for a generalized blockage adjustment based in Glauert's theory.

While \Cref{fig:forecastCT_steiros,fig:forecastCP_steiros} demonstrate a Glauert-derived performance prediction method based upon the model of \citet{steiros_analytical_2022}, the collapse in the ${C_T}^b-\lambda^b$ curves in \Cref{fig:compare_ct_bypass} suggests that a bluff-body blockage adjustment based on this model may also be suitable.
Here, we present a blockage adjustment method based on \citeauthor{steiros_analytical_2022}'s model that leverages the relationship between $u_b$ and $C_T$ by following a similar procedure as in \Cref{sec:lmadForecasting}.
We assume that there is some freestream velocity $U_{\infty,2}$ at $\beta_2$ that yields the same dimensional thrust and bypass velocity at $\beta_2$ as at $\beta_1$:
\begin{equation}
    u_{b,1} = u_{b,2} = u_b \ ,
\end{equation}
\begin{equation}
    T_1 = T_2 = T\ .
\end{equation}
Therefore, as in \Cref{eq:ct2From1},
 \begin{equation}
    C_{T,2} = C_{T,1}{\left( \frac{U_{\infty,1}}{U_{\infty,2}} \right)}^2 \ .
    \label{eq:ct2From1_steiros}
\end{equation}
To predict performance at $\beta_2$ from known performance at $\beta_1$, a reasonable value of $U_{\infty, 2}$ is guessed.
Then, using this guess and the known $u_b$ (which is constant between $\beta_1$ and $\beta_2$), \Cref{eq:ub_steiros} is solved for $u^{\ast}_2$ at $\beta_2$.
$C_{T,2}$ is then solved from \Cref{eq:CT_steiros} using $\beta_2$ and $u^{\ast}_2$, and this value is compared to the $C_{T,2}$ obtained from \Cref{eq:ct2From1_steiros}.
This process is iterated until the difference between the values of $C_{T,2}$ obtained from \Cref{eq:CT_steiros,eq:ct2From1_steiros} is minimized.
Finally, with $U_{\infty,2}$ known, $C_{T,2}$ is obtained from \Cref{eq:ct2From1_steiros}, and $C_{P,2}$ and $\lambda_2$ are respectively calculated as
\begin{equation}
    C_{P,2} = C_{P,1}{\left( \frac{U_{\infty,1}}{U_{\infty,2}} \right)}^3 \ ,
    \label{eq:cp2From1_steiros}
\end{equation}
 \begin{equation}
    \lambda_{2} = \lambda_{1}{\left( \frac{U_{\infty,1}}{U_{\infty,2}} \right)} \ .
    \label{eq:TSR2From1_steiros}
\end{equation}

 \begin{figure*}
     \centering
     \includegraphics[width=\textwidth]{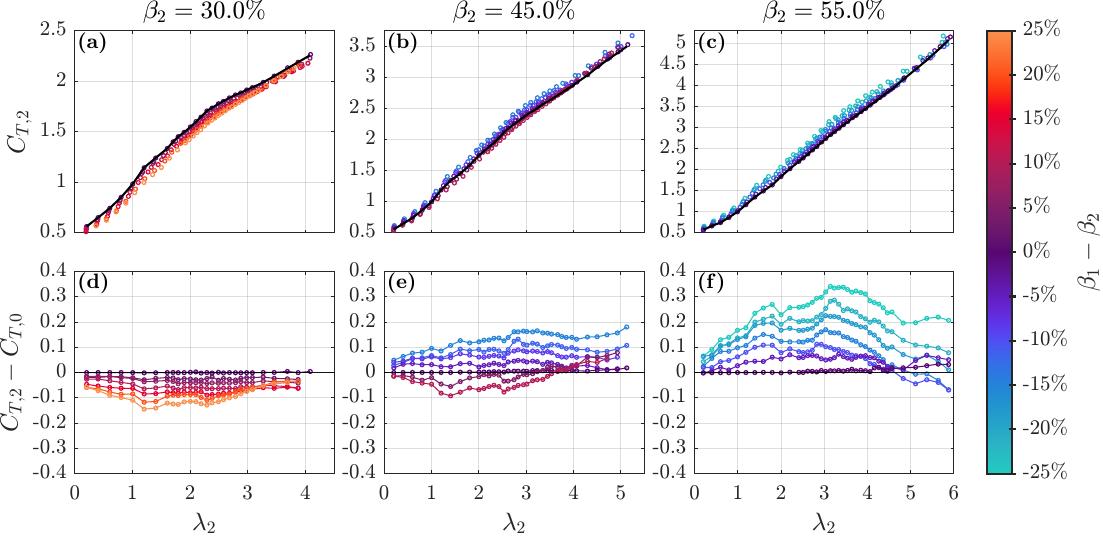}
     {\phantomsubcaption\label{fig:steirosBluffCT_30}
      \phantomsubcaption\label{fig:steirosBluffCT_45}
      \phantomsubcaption\label{fig:steirosBluffCT_55}
      \phantomsubcaption\label{fig:steirosBluffCT_30_error}
      \phantomsubcaption\label{fig:steirosBluffCT_45_error}
      \phantomsubcaption\label{fig:steirosBluffCT_55_error}
     }
     \caption{Confined $C_T$ predicted from the experimental data using a bluff-body method and the model of \citet{steiros_analytical_2022}. \subref{fig:steirosBluffCT_30}-\subref{fig:steirosBluffCT_55} Predicted thrust coefficient and \subref{fig:steirosBluffCT_30_error}-\subref{fig:steirosBluffCT_55_error} error between the predicted thrust coefficient and measured thrust coefficient ($C_{T,0}$, shown as the gray lines in \subref{fig:steirosBluffCT_30}-\subref{fig:steirosBluffCT_55}) for $\beta_2 = 30.0\%$, $45.0\%$, and $55.0\%$. Color indicates the difference between the blockage ratio of the performance data that was used to make each prediction ($\beta_1$) and the target blockage ratio ($\beta_2$). Note the different axes limits in \subref{fig:steirosBluffCT_30}-\subref{fig:steirosBluffCT_55}.}
     \label{fig:forecastBluffCT_steiros}
 \end{figure*}

 \begin{figure*}
     \centering
     \includegraphics[width=\textwidth]{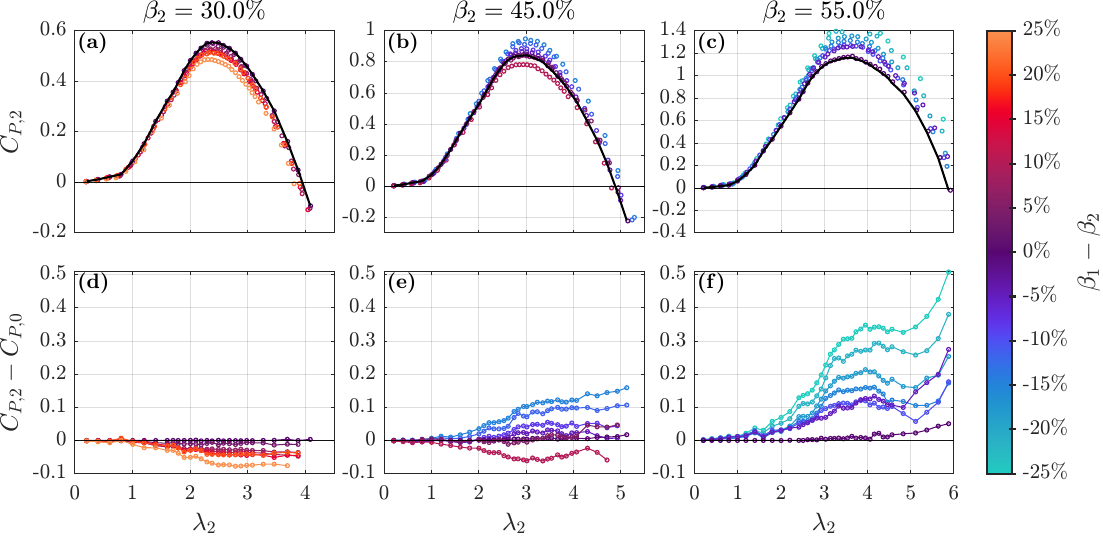}
     {\phantomsubcaption\label{fig:steirosBluffCP_30}
      \phantomsubcaption\label{fig:steirosBluffCP_45}
      \phantomsubcaption\label{fig:steirosBluffCP_55}
      \phantomsubcaption\label{fig:steirosBluffCP_30_error}
      \phantomsubcaption\label{fig:steirosBluffCP_45_error}
      \phantomsubcaption\label{fig:steirosBluffCP_55_error}
     }
     \caption{Confined $C_P$ predicted from the experimental data using a bluff-body method and the model of \citet{steiros_analytical_2022}. \subref{fig:steirosBluffCP_30}-\subref{fig:steirosBluffCP_55} Predicted thrust coefficient and \subref{fig:steirosBluffCP_30_error}-\subref{fig:steirosBluffCP_55_error} error between the predicted thrust coefficient and measured thrust coefficient ($C_{P,0}$, shown as the gray lines in \subref{fig:steirosBluffCP_30}-\subref{fig:steirosBluffCP_55}) for $\beta_2 = 30.0\%$, $45.0\%$, and $55.0\%$. Color indicates the difference between the blockage ratio of the performance data that was used to make each prediction ($\beta_1$) and the target blockage ratio ($\beta_2$). Note the different axes limits in \subref{fig:steirosBluffCP_30}-\subref{fig:steirosBluffCP_55}.}
     \label{fig:forecastBluffCP_steiros}
 \end{figure*}

\Cref{fig:forecastBluffCT_steiros,fig:forecastBluffCP_steiros} show the thrust coefficients and power coefficients, respectively, predicted at $\beta_2 = 30.0\%$, $45.0\%$, and $55.0\%$ when a bluff-body based blockage adjustment using \citeauthor{steiros_analytical_2022}'s model is applied to the experimental data.
As for the bluff-body based blockage adjustments using \citeauthor {houlsby_application_2008}'s model presented in \Cref{fig:forecastCT,fig:forecastCP}, trends in prediction accuracy follow trends in the ${C_T}^b-\lambda^b$ and ${C_P}^b-\lambda^b$ curves, which serve as the basis for blockage adjustments inspired by bluff-body theory.
Bluff-body prediction accuracy for $C_T$ using \citeauthor{steiros_analytical_2022}'s model is comparable to that of \citeauthor{houlsby_application_2008}'s model, owing to the similar collapse in the ${C_T}^b - \lambda^b$ curves obtained by both models (\Cref{fig:compare_ct_bypass}).
However, bluff-body prediction accuracy for $C_{P}$ for \citeauthor{steiros_analytical_2022}'s model is worse than for \citeauthor{houlsby_application_2008}'s model---especially at high $\lambda$---due to the worse collapse in the ${C_P}^b-\lambda^b$ observed (\Cref{fig:compare_cp_bypass}).
As for all other blockage adjustment methods considered in this study, prediction accuracy for both $C_T$ and $C_P$ becomes worse for a larger difference between $\beta_1$ and $\beta_2$, as well as at higher $\beta_2$ for a given difference between $\beta_1$ and $\beta_2$.

Comparing the two blockage adjustment methods based on \citeauthor{steiros_analytical_2022}'s model examined here, a Glauert-derived prediction yields accurate predictions of $C_T$ at lower $\lambda$, whereas a bluff-body based prediction for $C_T$ is more accurate at higher $\lambda$.
Therefore, a hybrid approach that utilizes either a Glauert-derived prediction or a bluff-body based prediction depending on the regime could provide adequate prediction accuracy over all $\lambda$.
However, such a hybrid approach would not be effective for predicting $C_P$, as both Glauert-derived predictions and bluff-body predictions of $C_P$ using \citeauthor{steiros_analytical_2022}'s model become less accurate as $\lambda$ increases.

\newpage
\section*{Solution Space for Generalized Blockage Adjustment}

\begin{figure}[hbt!]
    \centering
    \includegraphics[width=\textwidth]{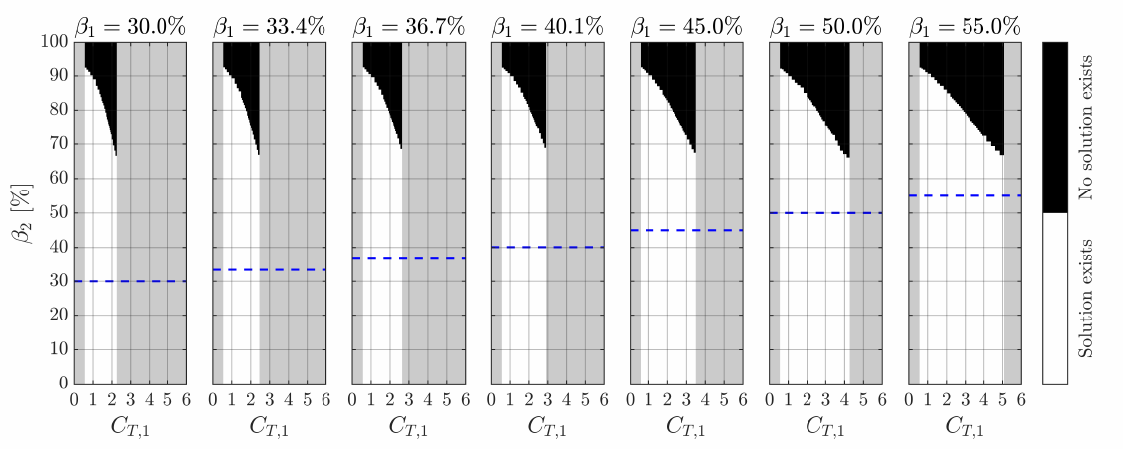}
    \caption{Conditions under which physically meaningful predictions for performance at $\beta_2$ can be obtained using the generalized blockage adjustment method described in \Cref{sec:lmadForecasting} and the collected performance data (\cref{fig:arrayPerf}).
    Here, it is assumed that $Fr_h$ is constant between $\beta_1$ and $\beta_2$.
    The white regions in each panel represent combinations of $C_{T,1}$ and $\beta_2$ that yield physical, sub-critical solutions, whereas the black regions indicate combinations that yield only non-physical solutions. 
    The gray regions in each panel indicate $C_{T,1}$ outside of the range of values measured at each $\beta_1$, and the dashed blue line in each panel indicates $\beta_1$ relative to the range of $\beta_2$ at which performance can be predicted.}
    \label{fig:forecastingLimits}
\end{figure}

\newpage
\section*{PIV Sensitivity Analysis}

\FloatBarrier

\begin{figure}[t]
    \centering
    \includegraphics[width=1.0\textwidth]{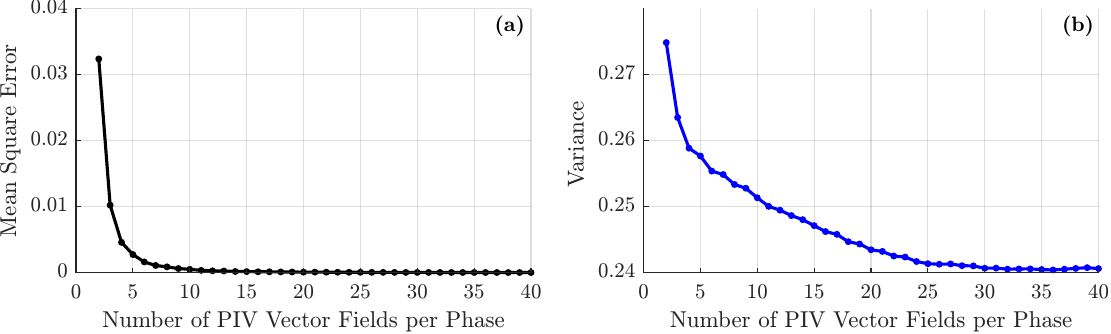}
    {\phantomsubcaption\label{fig:piv_phaseConv_mean}
    \phantomsubcaption\label{fig:piv_phaseConv_std}}
    \caption{Convergence of the phase-averaged $U_x/U_{\infty}$ field as a function of the number of PIV vector fields included in the average for a representative case at $\beta = 45.0\%$, $\lambda = 2.9$ and $\theta = 135^{\circ}$ in a plane $0.6D$ downstream of the array. \subref{fig:piv_phaseConv_mean} Mean squared error between successive phase-averaged $U_x/U_{\infty}$ fields as additional PIV vector fields are added. \subref{fig:piv_phaseConv_mean} Variance of the population of PIV vector fields used to calculate the phase-average.}
    \label{fig:PhaseAvgConvergence}
\end{figure}

To evaluate the statistical convergence of a given phase-averaged velocity field, we examine how each additional vector field included influences the resulting phase-averages.
\Cref{fig:piv_phaseConv_mean} shows the mean squared error between successive phase-averaged $U_x/U_\infty$  fields for a representative case at $\beta = 45.0\%$, $\lambda = 2.9$, $\theta =135^{\circ}$ in the plane $0.6D$ downstream of the array.
It is observed that the resulting mean square error in the streamwise velocity (normalized by the freestream velocity) decreases sharply between 1-10 vector fields before stabilizing below $3\times10^{-5}$ after the inclusion of 28 vector fields, indicating good convergence of the phase-averaged field.
Similarly, the variance of the vector field population for this case (\cref{fig:piv_phaseConv_std}) plateaus after the inclusion of 30 vector fields.
These results indicate that 40 vector fields are sufficient to provide a representative mean flow field at a given phase.
While shown here for a representative case, similar results are observed across all of the tested $\beta$, $\lambda$, and $\theta$.

\begin{figure}[t]
    \centering
    \includegraphics[width=1.0\textwidth]{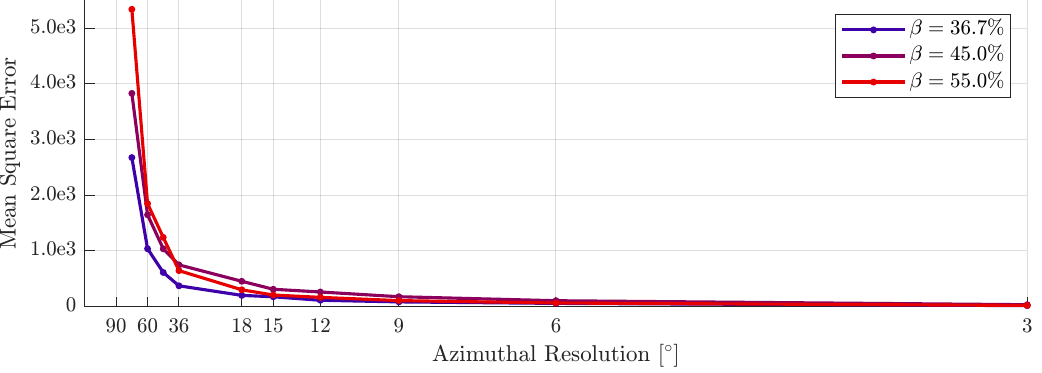}
    \caption{Mean square error of the time-averaged $U_x/U_\infty$ field as a function of the azimuthal separation between equally-spaced phase-average vector fields in a plane $0.6D$ downstream of the array at each blockage and corresponding optimal $\lambda$. The spacing between entries on the x-axis denotes the increase in the number of unique azimuthal positions that are sampled with successive increases in azimuthal resolution.}
    \label{fig:PIV_timeConvergence}
\end{figure}

\begin{figure}
    \centering
    \includegraphics[width=1.0\textwidth]{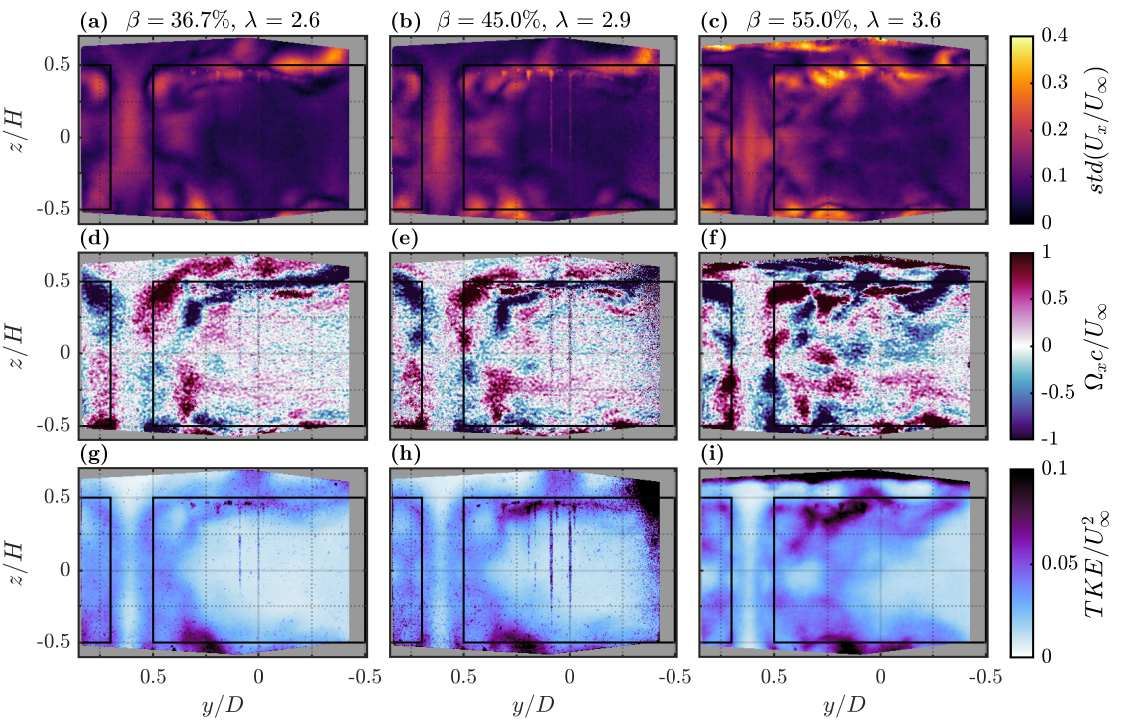}
    {\phantomsubcaption\label{fig:piv_stdev_36_12deg}
    \phantomsubcaption\label{fig:piv_stdev_45_12deg}
    \phantomsubcaption\label{fig:piv_stdev_55_12deg}
    \phantomsubcaption\label{fig:piv_vorticity_36_12deg}
    \phantomsubcaption\label{fig:piv_vorticity_45_12deg}
    \phantomsubcaption\label{fig:piv_vorticity_55_12deg}
    \phantomsubcaption\label{fig:piv_tke_36_12deg}
    \phantomsubcaption\label{fig:piv_tke_45_12deg}
    \phantomsubcaption\label{fig:piv_tke_55_12deg}
    }
    \caption{Examples of non-idealities in the measured flow fields that are not present in the flow fields assumed by linear momentum theory, shown for the streamwise flow through a plane $0.6D$ downstream of the array operating at $\beta = 36.7\%$, $45.0\%$, and $55.0\%$ and optimal $\lambda$.
    The fields shown are \ul{calculated using $12^{\circ}$ azimuthal resolution}, whereas the fields in \Cref{fig:pivVariability} are calculated using a $3^{\circ}$ azimuthal resolution.
    \subref{fig:piv_stdev_36}--\subref{fig:piv_stdev_55} Standard deviation of the measured streamwise velocity field.
    \subref{fig:piv_vorticity_36}--\subref{fig:piv_vorticity_55} Time-averaged streamwise vorticity at the same location and conditions.
    \subref{fig:piv_tke_36}--\subref{fig:piv_tke_55} Time-averaged turbulent kinetic energy (TKE) measured at the same location and conditions.}
    \label{fig:PIVVariability_12deg}
\end{figure}

Additionally, to evaluate the effect of azimuthal resolution (i.e., number of phases sampled per rotation) on the time-averaged field, we utilize the data at higher azimuthal resolution ($3^{\circ}$) for the optimal tip-speed ratio .
\Cref{fig:PIV_timeConvergence} shows the mean square error between successive time-averaged $U_x/U_{\infty}$ fields as azimuthal resolution is increased for representative cases at the optimal $\lambda$ for each $\beta$.
As the azimuthal resolution increases, the mean square error initially decreases sharply, then decreases more slowly for resolutions below $15^{\circ}$.
While the mean square error continues to decrease slowly up to the maximum resolution tested ($3^{\circ}$), we observe that using a phase separation between $3^{\circ}$ and $36^\circ$ (or in other words, a minimum of 5 equally distributed individual phase-averaged fields) produces a visually indistinguishable time-averaged velocity field.
For secondary data products (standard deviation, vorticity, and TKE), calculation using the higher resolution ($3^{\circ}$) data reduces small-scale noise in the resulting time-averaged fields, but results in the same large-scale structure as for $12^{\circ}$ resolution (shown for comparison in \cref{fig:PIVVariability_12deg}).
In summary, a phase separation of $12^{\circ}$ is sufficient to achieve convergence in the time-averaged flow fields.
Furthermore, validation cases where PIV acquisition was unsynchronized with the turbine rotation (400 frames acquired at 15 Hz) yield equivalent time-averaged fields.

\newpage
\section*{Additional PIV Flow Field Figures}


\begin{figure*}[hbt!]
      \centering
      \includegraphics[width=\textwidth]{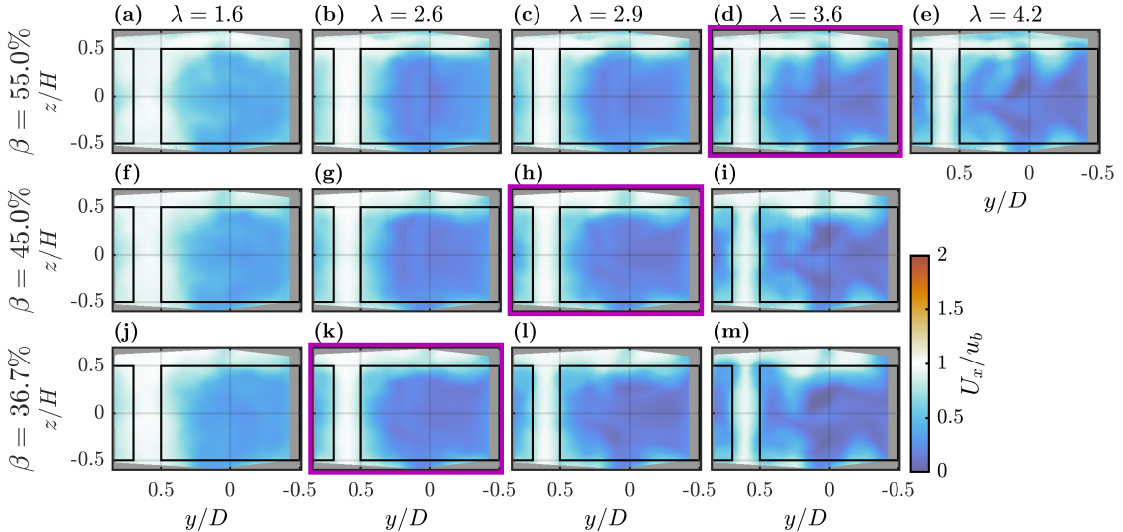}
      \caption{Time-averaged streamwise velocity $0.6D$ downstream of the array as function of $\beta$ and $\lambda$, normalized by the corresponding bypass velocity, $u_b$, predicted by \ul{the model of Houlsby \textit{et al.}} \citep{houlsby_application_2008}. The black rectangles in each tile indicate the projected area of the turbines in the field of view. Tiles corresponding to the optimal $\lambda$ for each blockage are outlined in purple (tiles (d), (h), and (k)), and correspond to \cref{fig:pivScaled_bypass_36,fig:pivScaled_bypass_45,fig:pivScaled_bypass_55}.}
      \label{fig:pivD06Ub}
\end{figure*}

\begin{figure*}
      \centering
      \includegraphics[width=\textwidth]{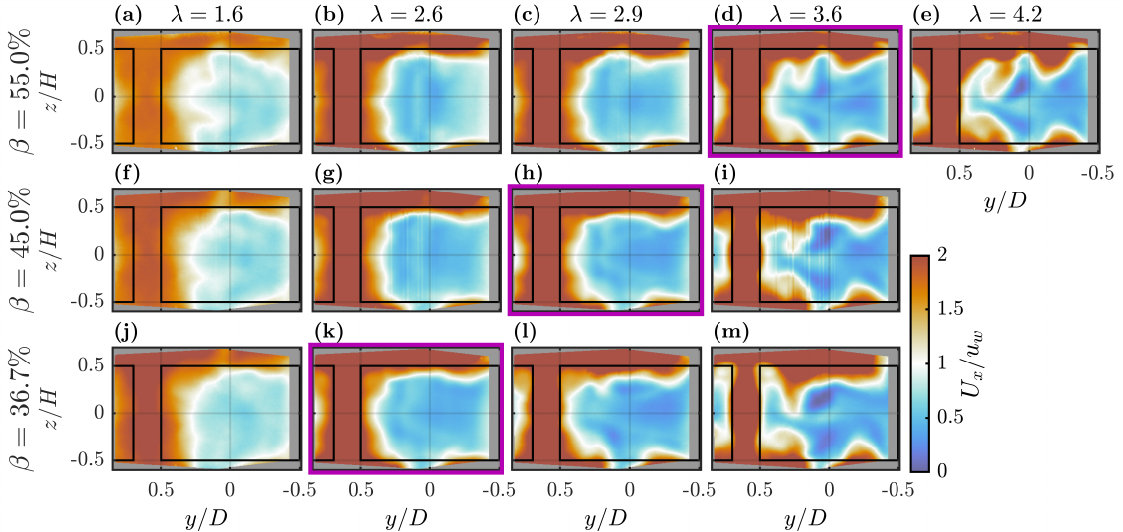}
      \caption{Time-averaged streamwise velocity $0.6D$ downstream of the array as function of $\beta$ and $\lambda$, normalized by the corresponding wake velocity, $u_w$, predicted by \ul{the model of Houlsby \textit{et al.}} \citep{houlsby_application_2008}. The black rectangles in each tile indicate the projected area of the turbines in the field of view. Flow fields corresponding to the optimal $\lambda$ for each blockage are outlined in purple (tiles (d), (h), and (k)), and correspond to \cref{fig:pivScaled_wake_36,fig:pivScaled_wake_45,fig:pivScaled_wake_55}.}
      \label{fig:pivD06Uw}
\end{figure*}


\begin{figure*}
      \centering
      \includegraphics[width=\textwidth]{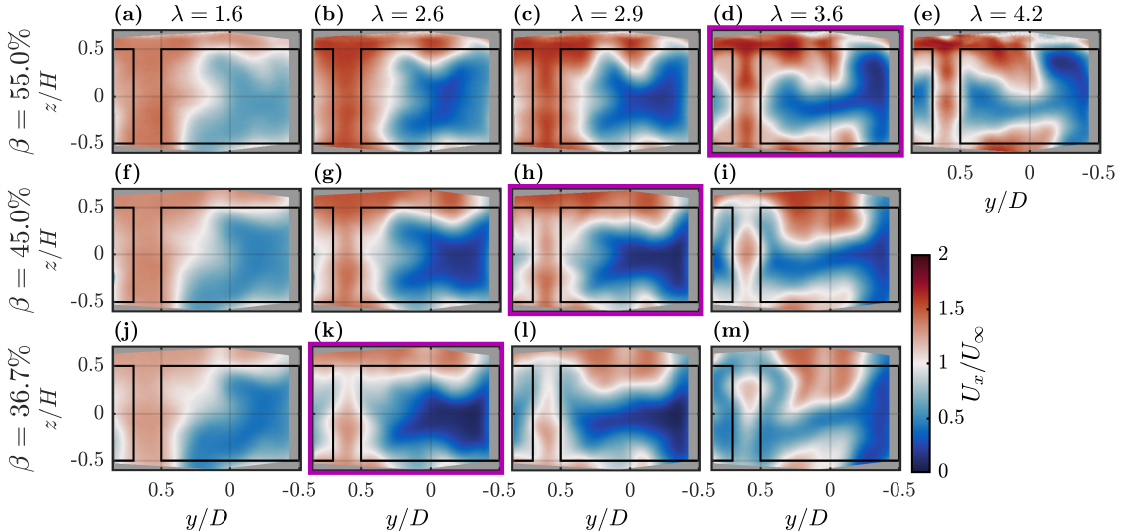}
      \caption{Time-averaged streamwise velocity $1.5D$ downstream of the array as function of $\beta$ and $\lambda$. Velocity measurements are normalized by the freestream velocity measured by the ADV. The black rectangles in each tile indicate the projected area of the turbines in the field of view. Flow fields corresponding to the optimal $\lambda$ for each blockage are outlined in purple (tiles (d), (h), and (k)).}
      \label{fig:pivD15Uinf}
\end{figure*}


\begin{figure*}
      \centering
      \includegraphics[width=\textwidth]{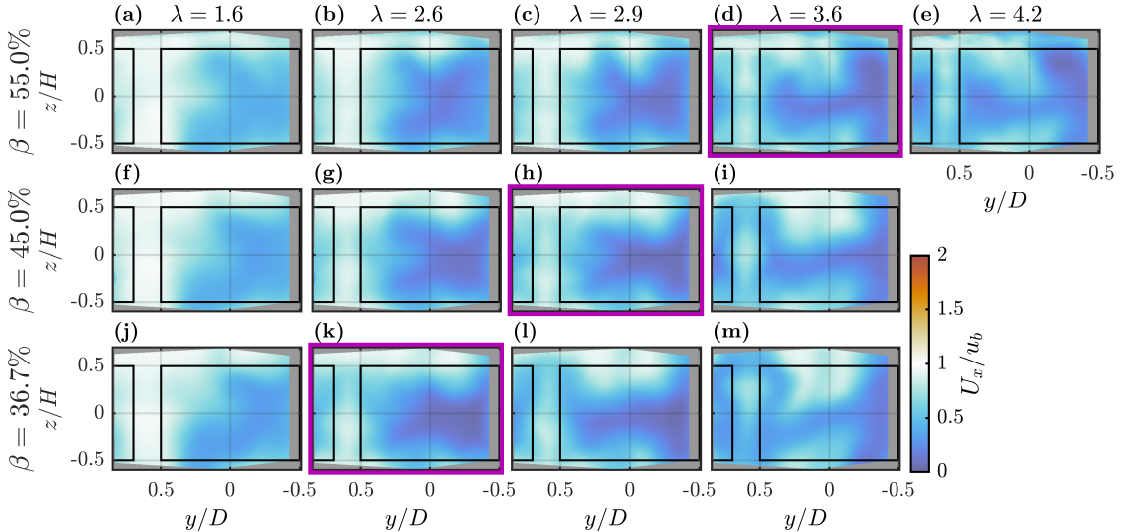}
      \caption{Time-averaged streamwise velocity $1.5D$ downstream of the array as function of $\beta$ and $\lambda$, normalized by the corresponding bypass velocity, $u_b$, predicted by \ul{the model of Houlsby \textit{et al.}}  \citep{houlsby_application_2008}. The black rectangles in each tile indicate the projected area of the turbines in the field of view. Tiles corresponding to the optimal $\lambda$ for each blockage are outlined in purple (tiles (d), (h), and (k)).}
      \label{fig:pivD15Ub}
\end{figure*}

\begin{figure*}
      \centering
      \includegraphics[width=\textwidth]{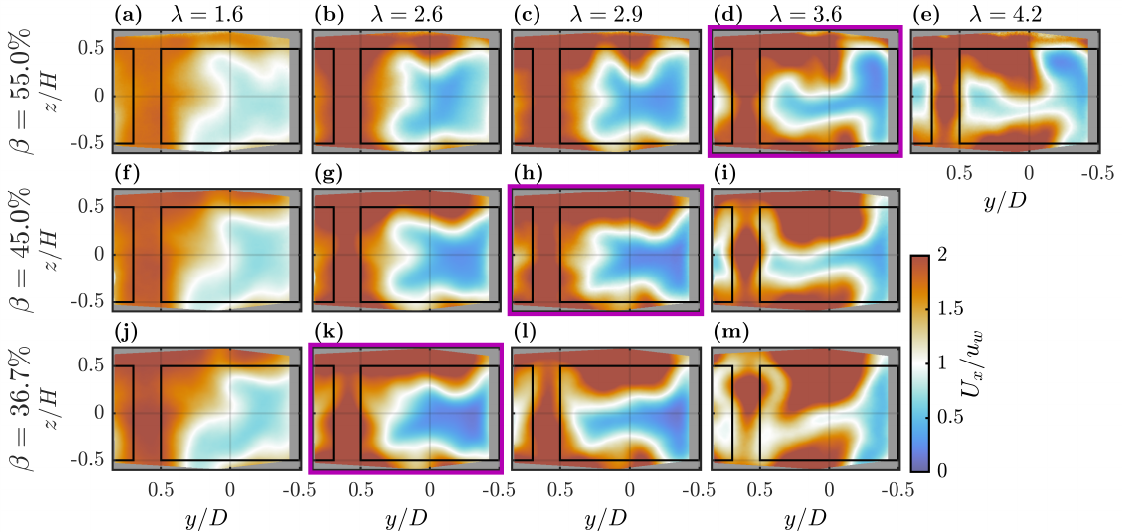}
      \caption{Time-averaged streamwise velocity $1.5D$ downstream of the array as function of $\beta$ and $\lambda$, normalized by the corresponding wake velocity, $u_w$, predicted by \ul{the model of Houlsby \textit{et al.}} \citep{houlsby_application_2008}. The black rectangles in each tile indicate the projected area of the turbines in the field of view. Flow fields corresponding to the optimal $\lambda$ for each blockage are outlined in purple (tiles (d), (h), and (k)).}
      \label{fig:pivD15Uw}
\end{figure*}


\begin{figure*}
    \centering
    \includegraphics[width=\textwidth]{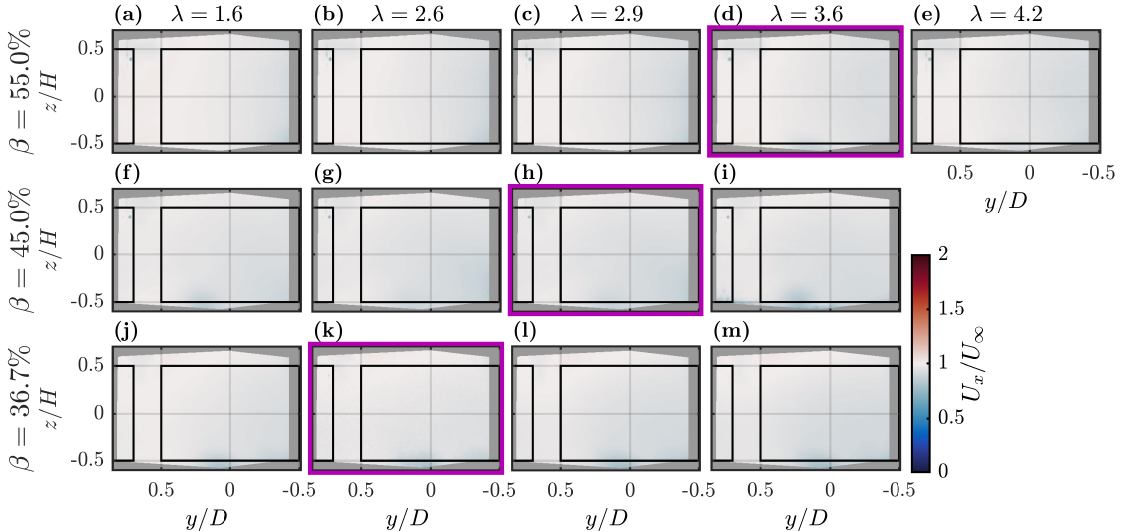}
    {\phantomsubcaption\label{fig:Up_B55_TSR1.6}
    \phantomsubcaption\label{fig:Up_B55_TSR2.6}
    \phantomsubcaption\label{fig:Up_B55_TSR2.9}
    \phantomsubcaption\label{fig:Up_B55_TSR3.6}
    \phantomsubcaption\label{fig:Up_B55_TSR4.2}
    
    \phantomsubcaption\label{fig:Up_B45_TSR1.6}
    \phantomsubcaption\label{fig:Up_B45_TSR2.6}
    \phantomsubcaption\label{fig:Up_B45_TSR2.9}
    \phantomsubcaption\label{fig:Up_B45_TSR3.6}
    
    \phantomsubcaption\label{fig:Up_B36_TSR1.6}
    \phantomsubcaption\label{fig:Up_B36_TSR2.6}
    \phantomsubcaption\label{fig:Up_B36_TSR2.9}
    \phantomsubcaption\label{fig:Up_B36_TSR3.6}
    }
    \caption{Time-averaged streamwise velocity $1.0D$ upstream of the array as function of $\beta$ and $\lambda$. Velocity measurements are normalized by the freestream velocity measured by the ADV. The black rectangles in each tile indicate the projected area of the turbines in the field of view. Flow fields corresponding to the optimal $\lambda$ for each blockage are outlined in purple (tiles (d), (h), and (k)).}
    \label{fig:pivUpstream1D}
\end{figure*} 

\begin{figure*}
    \centering
    \includegraphics[width=\textwidth]{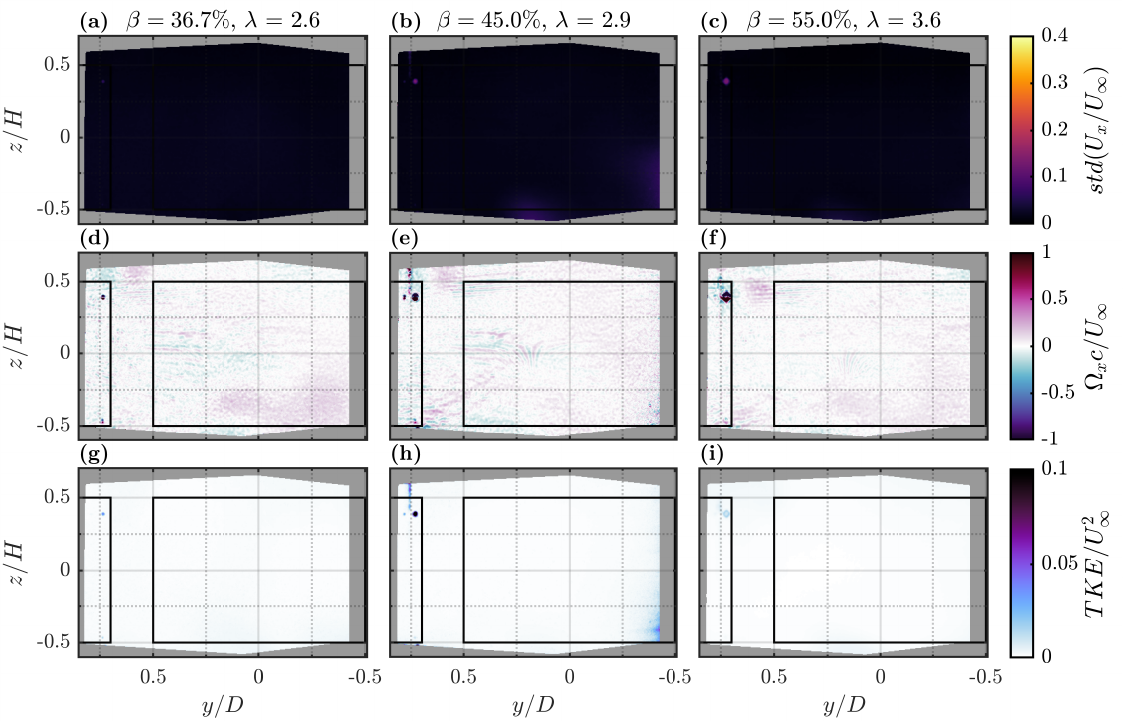}
    {\phantomsubcaption\label{fig:Up_stdev_36}
    \phantomsubcaption\label{fig:Up_stdev_45}
    \phantomsubcaption\label{fig:Up_stdev_55}
    \phantomsubcaption\label{fig:Up_vorticity_36}
    \phantomsubcaption\label{fig:Up_vorticity_45}
    \phantomsubcaption\label{fig:Up_vorticity_55}
    \phantomsubcaption\label{fig:Up_tke_36}
    \phantomsubcaption\label{fig:Up_tke_45}
    \phantomsubcaption\label{fig:Up_tke_55}
    }
    \caption{Examples of overall uniformity in the measured flow fields shown for the streamwise flow through a plane $1.0D$ upstream of the array operating at $\beta = 36.7\%$, $45.0\%$, and $55.0\%$ and optimal $\lambda$.
    \subref{fig:Up_stdev_36}--\subref{fig:Up_stdev_55} Standard deviation of the measured streamwise velocity field, calculated as the standard deviation of the phase-average streamwise velocity fields across all phases.
    \subref{fig:Up_vorticity_36}--\subref{fig:Up_vorticity_55} Time-averaged streamwise vorticity at the same location and conditions, calculated from the phase-averaged velocity fields and averaged across all phases.
    \subref{fig:Up_tke_36}--\subref{fig:Up_tke_55} Time-averaged turbulent kinetic energy (TKE) measured at the same location and conditions.
    To account for the inherent periodicity of cross-flow turbine wakes, the TKE field shown is obtained by calculating the TKE at each phase from the individual frames collected at that phase, then averaging the resulting fields across all phases. Note measurement noise near (0.75,0.4) in all frames due to reflections off the turbine shaft during PIV acquisition.}
    \label{fig:UpVariability}
\end{figure*}

\end{document}